\documentclass{aa}
\usepackage{graphicx}
\usepackage{txfonts}
\usepackage{hyperref}
\usepackage{cleveref}[2012/02/15]% v0.18.4; 
\usepackage[normalem]{ulem}
\hypersetup{colorlinks = true, citecolor = {blue}, urlcolor= {blue}}
\def\PGPU{$\varphi-$GPU}

\def\gapprox{\;\rlap{\lower 3.0pt                       % approximately smaller
        \hbox{$\sim$}}\raise 2.5pt\hbox{$>$}\;}
\def\lapprox{\;\rlap{\lower 3.1pt                       % approximately smaller
        \hbox{$\sim$}}\raise 2.7pt\hbox{$<$}\;}

\newcommand{\be}{ \begin{equation} }
\newcommand{\ee}{\end{equation}}

\newcommand{\ben}{\begin{enumerate}}
\newcommand{\een}{\end{enumerate}}

\renewenvironment{thebibliography}[1]{\begin{oldthebibliography}{#1}\setlength{\parskip}{0ex}\setlength{\itemsep}{0ex}}{\end{oldthebibliography}}

\usepackage{diagbox}
\usepackage{siunitx}
\usepackage{enumerate}
\newcommand{\orcid}[1]{\href{https://orcid.org/#1}{\protect\includegraphics[width=8pt]{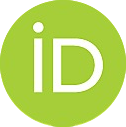}}}
\makeatletter
\renewcommand*\aa@pageof{, page \thepage{} of \pageref*{LastPage}}
\makeatother

\crefformat{footnote}{#2\footnotemark[#1]#3}

\begin{document}

\title{Milky Way globular clusters on cosmological timescales. I. Evolution of the orbital parameters in time-varying potentials}

\author{Maryna~Ishchenko
\inst{1,2}\orcid{0000-0002-6961-8170 }
\and
Margaryta~Sobolenko
\inst{1}\orcid{0000-0003-0553-7301}
\and
Peter~Berczik
\inst{3,4,1,2}\orcid{0000-0003-4176-152X}
\and
Sergey~Khoperskov
\inst{5,6}\orcid{0000-0003-2105-0763}
\and
Chingis~Omarov
\inst{2}\orcid{0000-0002-1672-894X}
\and
Olexander Sobodar
\inst{1}\orcid{0000-0001-5788-9996}
\and
Maxim~Makukov
\inst{2}\orcid{0000-0003-3643-9368}
}

\institute{Main Astronomical Observatory, National  
           Academy of Sciences of Ukraine,
           27 Akademika Zabolotnoho St, 03143 Kyiv, Ukraine
           \email{\href{mailto:marina@mao.kiev.ua}{marina@mao.kiev.ua}}
           \and
           Fesenkov Astrophysical Institute, 050020, Almaty, Kazakhstan
           \and
           Astronomisches Rechen-Institut, Zentrum f\"ur Astronomie, University of Heidelberg, M\"onchhofstrasse 12-14, 69120 Heidelberg, Germany
           \and
           Konkoly Observatory, Research Centre for Astronomy and Earth Sciences, E\"otv\"os Lor\'and Research Network (ELKH), MTA Centre of Excellence, Konkoly Thege Mikl\'os \'ut 15-17, 1121 Budapest, Hungary
           \and
           Leibniz Institut f\"{u}r Astrophysik Potsdam (AIP), An der Sternwarte 16, D-14482, Potsdam, Germany
           \and
           GEPI, Observatoire de Paris, Université PSL, CNRS, 5 Place Jules Janssen, 92190 Meudon, France
           }

\date{Received xxx / Accepted xxx}
    
\abstract    
% context heading (optional)
{Recent observational data show that the Milky Way galaxy contains about $170$ globular clusters. A fraction of them is likely formed in dwarf galaxies accreted onto the Milky Way in the past, while the remaining of clusters are formed in-situ. Therefore, different parameters, including orbits, of the globular clusters is a valuable tool for studying the Milky Way evolution. However, since the evolution of the 3D mass distribution of the Milky Way is poorly constrained, the orbits of the clusters are usually calculated in static potentials.}
% aims heading (mandatory)
{In this work, we study the evolution of the globular clusters in several external potentials, where we aim to quantify the effects of the evolving galaxy potential on the orbits of the globular clusters.}
% methods heading (mandatory)
{For the orbits calculation we used five Milky Way-like potentials from IllustrisTNG-100 simulation. The orbits of 159 globular clusters were integrated using a high-order $N$-body parallel dynamic code \PGPU, with initial conditions obtained from recent \textit{Gaia} Data Release 3 catalogues.}
{
We provide a classification of the globular clusters orbits according to their 3D shapes and association with different components of the Milky Way (disk, halo, bulge). We also found that the globular clusters in the external potentials have roughly similar energy-angular momentum distributions at the present time. However, both total energy and total angular momentum of the globular clusters are not conserved due to time-varying nature of the potentials. In some extreme cases, the total energy can change up to 40\%~(18 objects) over the last 5 Gyr of evolution. We found that the in-situ formed globular clusters are less affected by the evolution of the TNG potentials as compared to the clusters which are likely formed ex-situ. Therefore, our results suggest that time-varying potentials significantly affect the orbits of the GC, thus making it vital for understanding the formation of the Milky Way.
}
{}


\titlerunning{Orbital evolution of the globular clusters in external potentials}
\authorrunning{M.~Ishchenko et al.}
\maketitle

%%%%%%%%%%%%%%%%%%%%%%%%%%%%%%%%%%%%%%%%%%%%%%%%%%%%%%%%%%%%%%%%%%%%
\section{Introduction}\label{sec:Intr}
%%%%%%%%%%%%%%%%%%%%%%%%%%%%%%%%%%%%%%%%%%%%%%%%%%%%%%%%%%%%%%%%%%%%
According to the $\Lambda$CDM model, the Milky Way (MW) globular clusters (GCs) are the first stellar associations formed in the early Universe as gravitationally bound systems \citep{2019A&ARv..27....8G}. As the result their typical ages are $>10-12$~Gyr \citep{2009ApJ...694.1498M,VandenBerg2013, Valcin2020} and the present masses $\gtrsim10^{5}\rm M_{\odot}$ \citep{2013ApJ...772...82H,Kharchenko2013, Baumgardt2019, Baumgardt2021}. Many GCs' survived over long galactic history and, thus, became relics from the earliest era of galaxy formation \citep{Garro2022}. Recent observations reveal the picture where the MW halo contains a large number of GCs' \citep{Harris2010, VasBaum2021}, stellar streams \citep{Ibata2021} and satellite galaxies \citep{2012AJ....144....4M,McConnachie2021}. Most of the discovered and studied streams in the halo are the remnants of either GCs or dwarf galaxies destroyed in the process of ``merging history'' of our Galaxy~\citep{Mateu2023, Ibata2021,2023arXiv230105166F}. They undergo complete destruction (or mass loss) under the influence of the complex time-dependent gravitational field of our Galaxy. Study of kinematics and chemical abundances of the GCs captured by the MW is a very good tool for studying the past history of our Galaxy \citep{Xiang2022}, as well as searching possible progenitors for these GCs \citep{Massari2019,Malhan2022}.

Thanks to the publication of the \textit{Gaia}~(ESA) catalogues~\citep{Gaia2021, 2018A&A...616A...1G} full phase-space information is available for almost the entire population of the GCs, which makes possible to investigate their orbits. Several recent works studied the orbits of the GCs in the MW potential assuming its being axisymmetric and non-evolving in time~\citep{Massari2019, Myeong2019,Malhan2022}. These two assumptions result in the conservation of the total energy and the $z$-component of the angular momentum~($L_{\rm z}$) of the GCs, which is being used to group some of the GCs and link these kinematically coherent groups with their galaxy-progenitors~(either MW or accreted satellite). However, the above-mentioned assumptions are not fully correct because the MW galaxy contains the bar and boxy-peanut bulge in the center~\citep{1995ApJ...445..716D,2013MNRAS.435.1874W,2016AJ....152...14N}. In the outer parts, the axisymmetry of the MW potential is believed to be broken by the massive orbiting satellites~\citep{2018MNRAS.481..286L,2020MNRAS.492L..61L}, including the LMC/SMC systems~\citep{2015ApJ...802..128G,2021NatAs...5..251P,2021Natur.592..534C}. In the current picture, most of the mass of the disk is assembled from the smooth accretion of gas combined with the accretion along cold filaments~\citep{1991ApJ...377..365K,2003MNRAS.345..349B,2005MNRAS.363....2K,2009MNRAS.397L..64A}. Although the mass evolution of the MW is still uncertain, different models suggest its rapid increase at early times~(until $\approx$8 Gyr) followed by a slower evolution~\citep{2014ApJ...781L..31S} which is in agreement with the mass growth of the MW-like galaxies constrained via halo abundance matching~\citep{2013ApJ...771L..35V, Fattahi2019}. While some models suggest that the mass-growth of the MW-type galaxies affects the coherency of accreted and in-situ stellar populations in the kinematic space~\citep[e.g., in $E-L_{z}$][]{2019MNRAS.487L..72G,2021ApJ...920...10P,2022arXiv220604522K,2022arXiv221004245P}, the impact on the motions of the GCs remains unclear. 

The main feature of our study of the MW GC subsystem phase space distributions is a time variable external potential. This feature significantly differs our investigation from other similar works \citep[see for example][]{Massari2019, Vasiliev2019, Bajkova2021, Bajkova2021AstL, Malhan2022, Armstrong2021, Haghi2015}, where the authors use the fixed or the analytically time-evolving MW potential.
In series of articles about evolution of GC, and in particular this one, we aim to investigate how the evolution of the mass and spatial scales of both stellar and dark matter components of the MW affect the orbital parameters of the GCs. Since the evolution of the MW parameters is poorly constrained, we analyse the dynamics of the GCs in the MW-like evolving potentials directly obtained from IllustrisTNG cosmological simulations~\citep{2018MNRAS.473.4077P,2018MNRAS.475..624N}. 

The paper is organised as follows. In Section~\ref{subsec:data} we present the GCs selection with the error distribution from the \textit{Gaia}~data release 3 (DR3) catalogue. In Sections~\ref{subsec:tng} and \ref{subsec:integr}, we introduce the time-dependent gravitational potentials of the MW-type galaxies that were selected from IllustrisTNG-100 and shortly describe the numerical integrator. In Section~\ref{subsec:char-orb}, we present the individual orbits of the GCs in the evolving TNG potentials. In Section~\ref{sec:phase-space} we describe GCs phase space evolution. In Section~\ref{sec:con} we summarise our results.

%%%%%%%%%%%%%%%%%%%%%%%%%%%%%%%%%%%%%%%%%%%%%%%%%%%%%%%%%%%%%%%%%%%%%
\section{Methods}\label{sec:met}
%%%%%%%%%%%%%%%%%%%%%%%%%%%%%%%%%%%%%%%%%%%%%%%%%%%%%%%%%%%%%%%%%%%%

%%%%%%%%%%%%%%%%%%%%%%%%%%%%%%%%%%%%%%%%%%%%%%%%%%%%%%%%%%%%%%%%%%%%%
\subsection{Globular Clusters sample}\label{subsec:data}
%%%%%%%%%%%%%%%%%%%%%%%%%%%%%%%%%%%%%%%%%%%%%%%%%%%%%%%%%%%%%%%%%%%%

In our study, we used the recent catalogues of the GCs released by \cite{VasBaum2021} and \cite{Baumgardt2021}, which contain information about more than 160~individual objects. The catalogues include the complete 6D phase-space information required for the initial conditions of our simulations: right ascension~(RA), declination~(DEC), heliocentric distance~(D$_{\odot}$), proper motion in right ascension (PMRA), proper motion in declination (PMDEC), and the radial velocity~(RV). 
Because of the dynamical simulation of the GCs, we use also the self-gravity of the clusters together with the Galaxy external potential. We exclude the Mercer~5 object since there is no mass information for this GC from the catalogues above. 

\begin{figure*}[htbp!]
\centering
\includegraphics[width=0.80\linewidth]{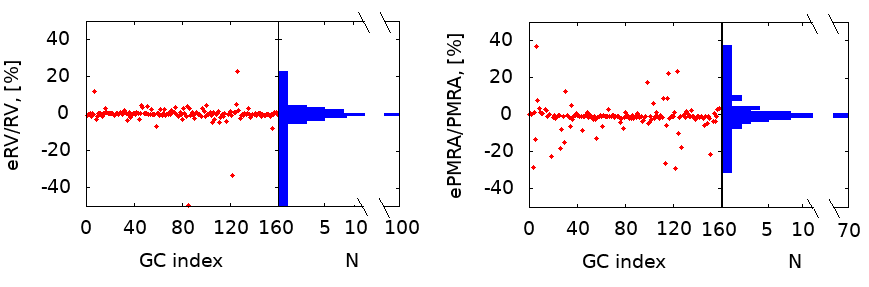}
\includegraphics[width=0.80\linewidth]{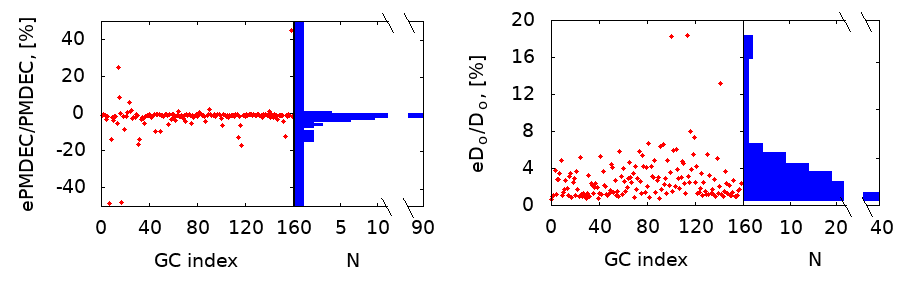}
\caption{Distribution of the relative errors \textit{(from left to right)}: for the radial velocity (eRV), proper motions in right ascension (ePMRA) and declination (ePMDEC), and in the heliocentric distance eD$_{\odot}$. Blue histograms represent the error distributions. The following GCs are not shown due to the large values of these errors: Crater, Lae~3, NGC~6760, BH~261, NGC~6553, and Pal~13, except the graph on the bottom right panel.}
\label{fig:errors}
\end{figure*}

Before the orbital integration, we analysed the GCs \textit{Gaia} measurement errors influence which may change the orbital shape during the orbital integration \citep[see][]{Bajkova2021}. We analysed the errors in PMRA, PMDEC, RV and D$_{\odot}$ for 159~GCs from the catalogue above and in Fig.~\ref{fig:errors} presented the distribution of the relative errors of listed parameters. As we see, the vast majority (up to 95\%) of the errors are well below 5\% but some GCs (Crater and Lae 3) have a quite large value in proper motion. Analysing the errors in the heliocentric distance we found only 18 objects that have relative errors larger than 5\% and only three GCs with errors larger than 10\% (namely: 2MASS-GC01, VVV-CL001, and Pal 10). The influence of the measurement errors on the orbital integration is presented in subsection \ref{subsec:err-orbits}.

%-------------------------------------------------------------------------%
%-------------------------------------------------------------------------%

For the transformation of the positions and velocities to the Cartesian Galactocentric rest-frame \citep[see][for the coordinate transformation equations]{Johnson1987, Bovy2011}, we assumed the Galactocentric distance of the Sun $R_{\odot}=8.178$ kpc \citep{Gravity2019, Reid2004}, a height above the Galactic plane $Z_{\odot}=20.8$~pc \citep{Bennett2019} and the velocity of the Local Standard of Rest (LSR) $V_{\rm LSR}=234.737$~km~s$^{-1}$ \citep{Bovy2012, Drimmel2018}. Accordingly, the Sun is located at $X_{\odot} = -8178$~pc, $Y_{\odot} = 0$~pc and $Z_{\odot} = 20.8$~pc in our Cartesian Galactocentric coordinate system. For the peculiar velocity of the Sun we used the following values with respect to the LSR: $U_{\odot}=11.1$~km~s$^{-1}$, $V_{\odot}=12.24$~km~s$^{-1}$, $W_{\odot}=7.25$~km~s$^{-1}$~\citep{Schonrich2010}. 

%%%%%%%%%%%%%%%%%%%%%%%%%%%%%%%%%%%%%%%%%%%%%%%%%%%%%%%%%%%%%%%%%%%%
\subsection{Time-variable gravitational potentials from Illustris simulation}\label{subsec:tng}
%%%%%%%%%%%%%%%%%%%%%%%%%%%%%%%%%%%%%%%%%%%%%%%%%%%%%%%%%%%%%%%%%%%%

Since the time evolution of the MW potential is poorly constrained from observations, for the GCs orbits integration, we have selected the MW-like galaxies from the most recent Illustris cosmological numerical simulations. For our purpose, we used the publicly available snapshot data from the IllustrisTNG-100~\citep{2018MNRAS.473.4077P,2018MNRAS.475..624N, 2018MNRAS.475..676S, 2018MNRAS.480.5113M, 2018MNRAS.477.1206N,2019ComAC...6....2N}. 
These simulations successfully reproduce the main scaling relations and evolution of galaxy sizes~\citep{2018MNRAS.474.3976G}, including the formation of realistic disc galaxies~\citep{2019MNRAS.490.3196P}, the gas-phase mass-metallicity relation~\citep{2019MNRAS.484.5587T}, and the range of MW-type galaxies observed kinematic properties~\citep{2018MNRAS.481.1950L}. The Illustris simulations were used to investigate various processes of galactic evolution, including the gas-stripping phenomena~\citep{2019MNRAS.483.1042Y,2020A&A...638A.133L}, central black hole activity~\citep{2019MNRAS.484.4413H,2021MNRAS.500.4004D}, star formation quenching~\citep{2018MNRAS.474.3976G,2020MNRAS.491.4462D}, and many others. These results ensure that the investigation of properties of different types of galaxies from the Illustris cosmological simulations allows us to study the GCs dynamics in an environment similar to the MW.

The IllustrisTNG-100 is characterised by a simulation box approximately 100~Mpc$^3$. In a box of such size, one can resolve a sufficient number of the MW-mass disk galaxies (in our sample 54 object\footnote{MW-like potentials are presented on a web page of the project \url{https://bit.ly/3b0lafw}}) with the mass resolution of $7.5\times10^{6}\rm\;M_{\odot}$ for dark matter and $1.4\times10^{6}\rm\;M_{\odot}$ for the baryonic particles, respectively. For our analysis, in the Illustris simulations, we identified the MW-like galaxy candidates with at least $10^{5}$ dark matter particles and at least $10^{3}$ baryonic particles (stars and gas) at redshift zero.

From the simulated galaxies in the IllustrisTNG-100, we selected our five galaxies~({\tt \#411321, \#441327, \#451323, \#462077}, and {\tt \#474170}) that at redshift zero well reproduce MW parameters~(disk and halo masses with the spatial scales) at present the best~(see Table~\ref{tab:pot-val}). 

Due to the limited particle number of our selected TNG-TVPs, we cannot resolve a separate bulge component. This is mainly caused by the limited mass resolution of the IllustrisTNG-100, that is, simply we do not have enough particles to determine such a component as a separate bulge in our galaxy mass model.

We also used a circular velocity value at the solar distance~($\approx$8 kpc) in the model as an extra parameter to select the best TNG galaxies which represent the MW-type systems. This value indicates the position of the Sun at present. According to the age and chemical compositions of the stars in the solar neighbourhood, we know that over the past few billion years, there were no large changes in the radial motion of the masses. This means that the circular velocity at the distance of the Sun in the Galactic disk should remain approximately constant during the last few billion years near the $V_{\odot}\approx235$~km~s$^{-1}$ \citep{Mardini2020}. In Appendix~\ref{app:tng-pot-vel}, we present the circular velocity evolution for the selected five TNG external potentials. In the same figures, we also present the time evolution of the selected TNG-TVP circular velocity radial distribution. These figures were obtained directly from the mass distribution of the corresponding IllustrisTNG-100 simulation snapshots for three different redshifts z = 0, -5 Gyr (z = 0.48) and -10 Gyr (z = 1.74). Due to the galaxy mass decrease as a function of lookback time, we clearly see the circular velocity decrease at all galactocentric radii.

To obtain the spatial scales of the disks and dark matter haloes, we decomposed the mass distribution using the Miyamoto-Nagai (MN)~$\Phi_{\rm d} (R,z)$ \citep{Miyamoto1975} and Navarro–Frenk–White (NFW) $\Phi_{\rm h} (R,z)$ \citep[][]{NFW1997} potentials:
\begin{equation}
\begin{split}
\Phi_{\rm tot} &= \Phi_{\rm d} (R,z) + \Phi_{\rm h} (R,z) = \\
&= - \frac{GM_{\rm d}}{\sqrt{R^{2}+\Bigl(a_{\rm d}+\sqrt{z^{2}+b^{2}_{\rm d}}\Bigr)^{2}}} - 
\frac{GM_{\rm h}\cdot{\rm ln}\Bigr(1+\frac{\sqrt{R^{2}+z^{2}}}{b_{\rm h}}\Bigl)}{\sqrt{R^{2}+z^{2}}},
\end{split}
\end{equation}
where 
$R=\sqrt{x^{2}+y^{2}}$ is the planar Galactocentric radius, 
$z$ is the distance above the plane of the disc, 
$G$ is the gravitational constant, 
$a_{\rm d}$ is the disk scale length, 
$b_{\rm d,h}$ are the disk and halo scale heights, respectively, 
$M_{\rm d}$ and 
$M_{\rm h}=4\pi\rho_{0}b^{3}_{\rm h}$ ($\rho_{0}$ is the central mass density of the halo) are masses of the disk and halo, respectively. 

In our case all these mass distribution parameters are functions of lookback time. We performed such a fit for each available snapshot in each of our selected galaxies. The code, which we use to find these parameters, is also publicly available at GitHub\footnote{The ORIENT: \\~\url{ https://github.com/Mohammad-Mardini/The-ORIENT}}.

\begin{figure*}[htbp]
\centering
\includegraphics[width=0.95\linewidth]{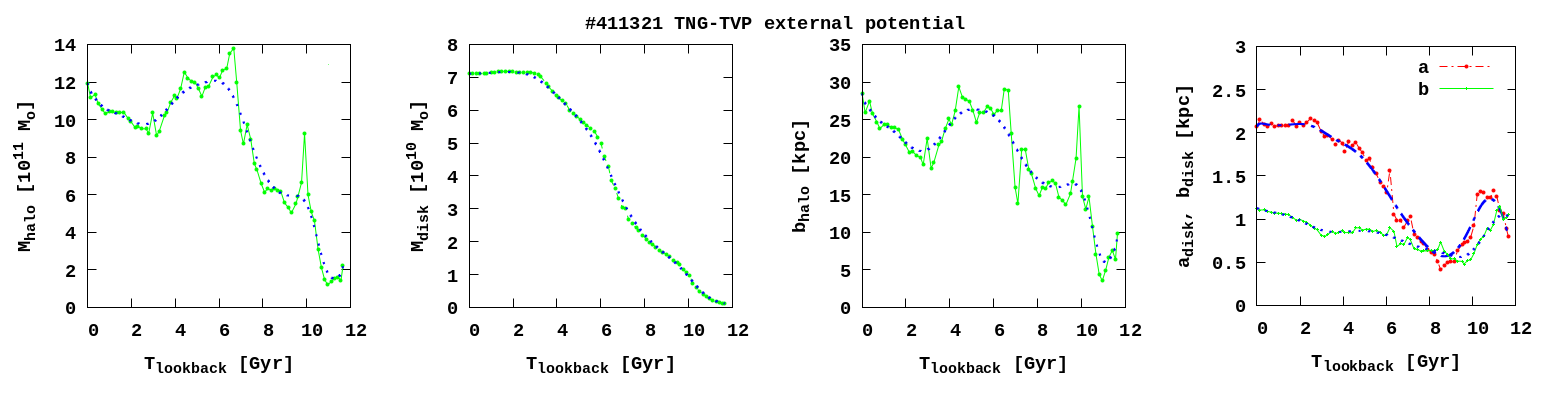}
\includegraphics[width=0.95\linewidth]{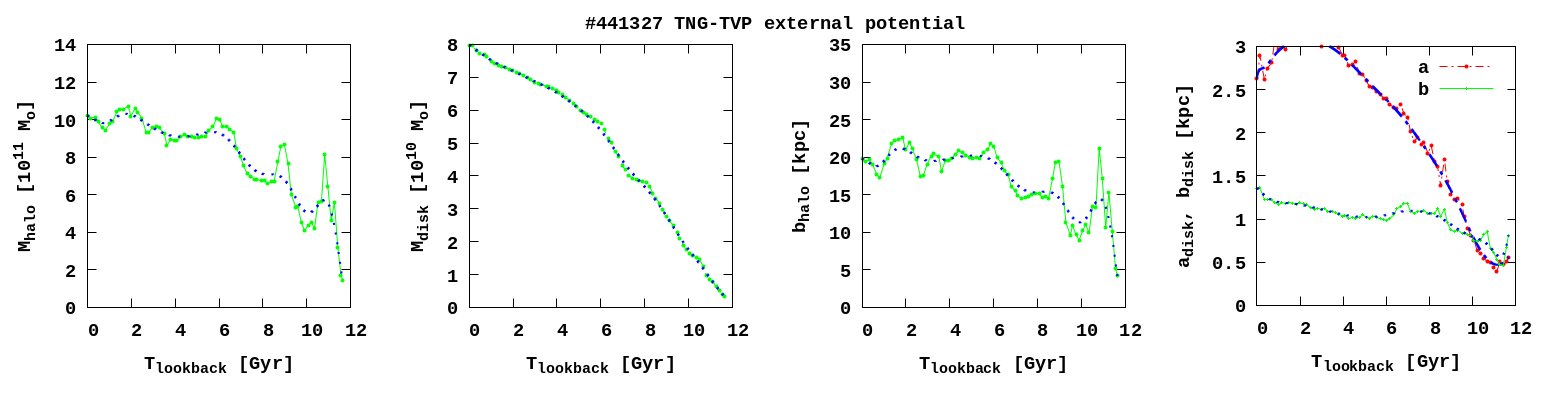}
\includegraphics[width=0.95\linewidth]{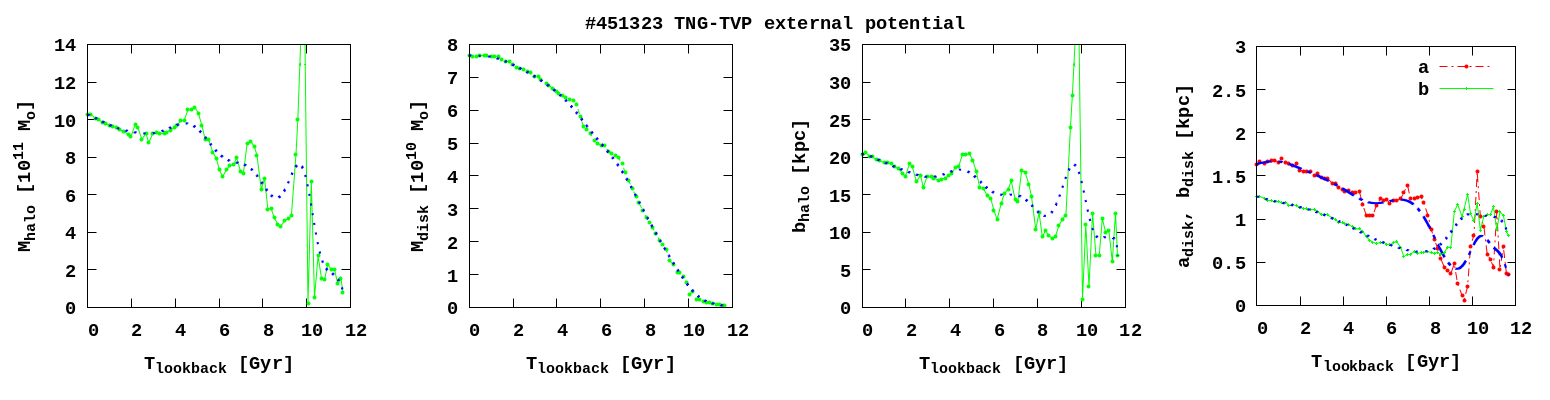}
\includegraphics[width=0.95\linewidth]{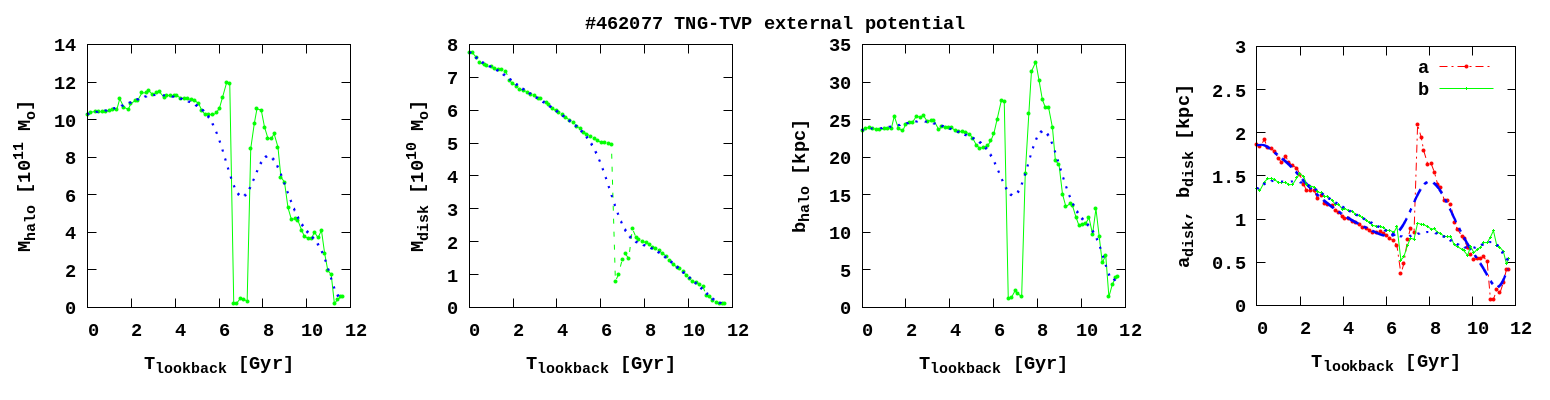}
\includegraphics[width=0.95\linewidth]{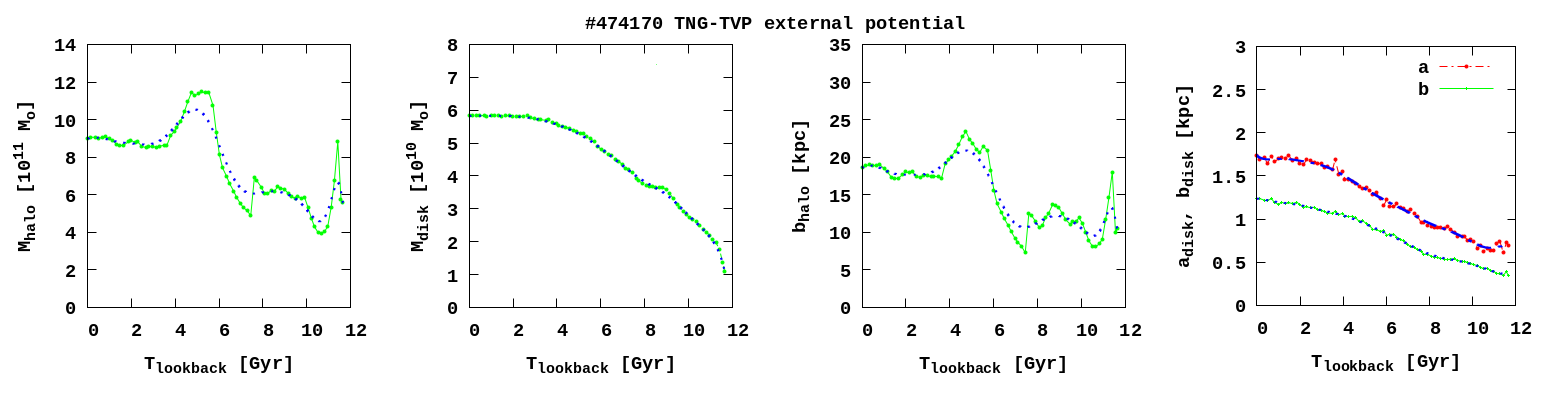}
\caption{Evolution of halo and disk masses, and their characteristic scales for \textit{(from top to bottom):} {\tt \#411321, \#441327, \#451323, \#462077}, and {\tt \#474170} TNG-TVPs in time. Halo mass $M_{\rm h}$, disk mass $M_{\rm d}$, NFW halo scale parameter $b_{\rm h}$, MN disk scale parameters $a_{\rm d}$ and $b_{\rm d}$ are presents \textit{from left to right}. Solid green lines with dots show the parameters recovered from the IllustrisTNG-100 data. Dotted and dash-dotted blue lines correspond to the values after the interpolation and smoothing with a 1~Myr time step that was used in the orbital integration.}
\label{fig:MW-TNG}
\end{figure*}

For each snapshot available in the IllustrisTNG database, we decomposed the mass distribution of each selected galaxy. Next, we interpolated the variation in masses and spatial scales of dark matter and stellar disk among the snapshots with the timestep of $1$~Myr. Finally, to avoid some short-time scale perturbations, likely linked to the mergers or errors in the galaxy identification, we additionally smoothed the evolution of the main galaxy parameters. 
The above procedure results are shown for our five time-variable potentials~(TNG-TVPs) in Fig.~\ref{fig:MW-TNG}. 

%-------------------------------------------------------------------------%
\begin{table*}[htbp]
\caption{Parameters of the time-varying potentials selected from the IllustrisTNG-100 simulation at redshift zero. The last column shows the parameters of the corresponding MW components according to \cite{BBH2022} at present.}
\centering
\begin{tabular}{llccccccc}
\hline
\hline
\multicolumn{1}{c}{Parameter} & Unit & {\tt \#411321} & {\tt \#441327} & {\tt \#451323} & {\tt \#462077} & {\tt \#474170} & Milky Way \\
\hline
\hline
Disk mass, $M_{\rm d}$         & $10^{10}~\rm M_{\odot}$ & 7.110 & 7.970 & 7.670 & 7.758 & 5.825 & 6.788 \\
Halo mass, $M_{\rm h}$         & $10^{12}~\rm M_{\odot}$ & 1.190 & 1.020 & 1.024 & 1.028 & 0.898 & 1.000 \\
Disk scale length, $a_{\rm d}$ & 1~kpc                     & 2.073 & 2.630 & 2.630 & 1.859 & 1.738 & 3.410 \\
Disk scale height, $b_{\rm d}$ & 1~kpc                     & 1.126 & 1.356 & 1.258 & 1.359 & 1.359 & 0.320 \\
Halo scale height, $b_{\rm h}$ & 10 kpc                  & 2.848 & 1.981 & 2.035 & 2.356 & 1.858 & 2.770 \\
\hline
\end{tabular}
\label{tab:pot-val}
\end{table*} 
%-------------------------------------------------------------------------%

%-------------------------------------------------------------------------%
%-------------------------------------------------------------------------%

%%%%%%%%%%%%%%%%%%%%%%%%%%%%%%%%%%%%%%%%%%%%%%%%%%%%%%%%%%%%%%%%%%%%%
\subsection{Integration procedure}\label{subsec:integr}
%%%%%%%%%%%%%%%%%%%%%%%%%%%%%%%%%%%%%%%%%%%%%%%%%%%%%%%%%%%%%%%%%%%%

For the GCs' orbits integration we used a high-order parallel dynamical $N$-body code \PGPU \footnote{$N$-body code \PGPU: \\~\url{ https://github.com/berczik/phi-GPU-mole}} \citep{Berczik2011,BSW2013}. The code is based on the fourth-order Hermite integration scheme with hierarchical individual block time steps. 
For the force calculation, acting on the particles during the integration, we combine the particles self gravity with the time variable external potential described in the previous subsection. For the self gravity calculation (as a first order simplification) we apply the current mass of the individual GCs.
We applied the integration timestep parameter $\eta$=0.01 \citep{MA1992}, which gives for our investigation the required integration accuracy. 
As a test case, we run a typical simulation with the fixed for today (that is, for $t = 0$) values of the external potential {\tt \#411321} with up-to-date halo and disk masses and scale lengths. During this test case the total relative energy drift ($\Delta E/ E_{t=0}$) over a 10 Gyr backward integration was below $\approx2.5\times10^{-13}$. For the production runs, we embedded our 159 GC into the selected five external TNG-TVPs as a point masses (set the masses of GC as for today). For each GC-point we set for position and velocity the central values from \cite{Baumgardt2021} and \cite{VasBaum2021}. 

%%%%%%%%%%%%%%%%%%%%%%%%%%%%%%%%%%%%%%%%%%%%%%%%%%%%%%%%%%%%%%%%%%%%%%%%%%%%%%%%
\section{Characteristics of the individual Globular Clusters orbital evolution in five time-varying potentials}\label{subsec:char-orb}
%%%%%%%%%%%%%%%%%%%%%%%%%%%%%%%%%%%%%%%%%%%%%%%%%%%%%%%%%%%%%%%%%%%%%%%%%%%%%%%%

%%%%%%%%%%%%%%%%%%%%%%%%%%%%%%%%%%%%%%%%%%%%%%%%%%%%%%%%%%%%%%%%%%%%%%%%%%%%%%%%
\subsection{Influence of the measurement errors}\label{subsec:err-orbits}
%%%%%%%%%%%%%%%%%%%%%%%%%%%%%%%%%%%%%%%%%%%%%%%%%%%%%%%%%%%%%%%%%%%%%%%%%%%%%%%%

\begin{figure*}[htbp!]
\centering
\includegraphics[width=0.99\linewidth]{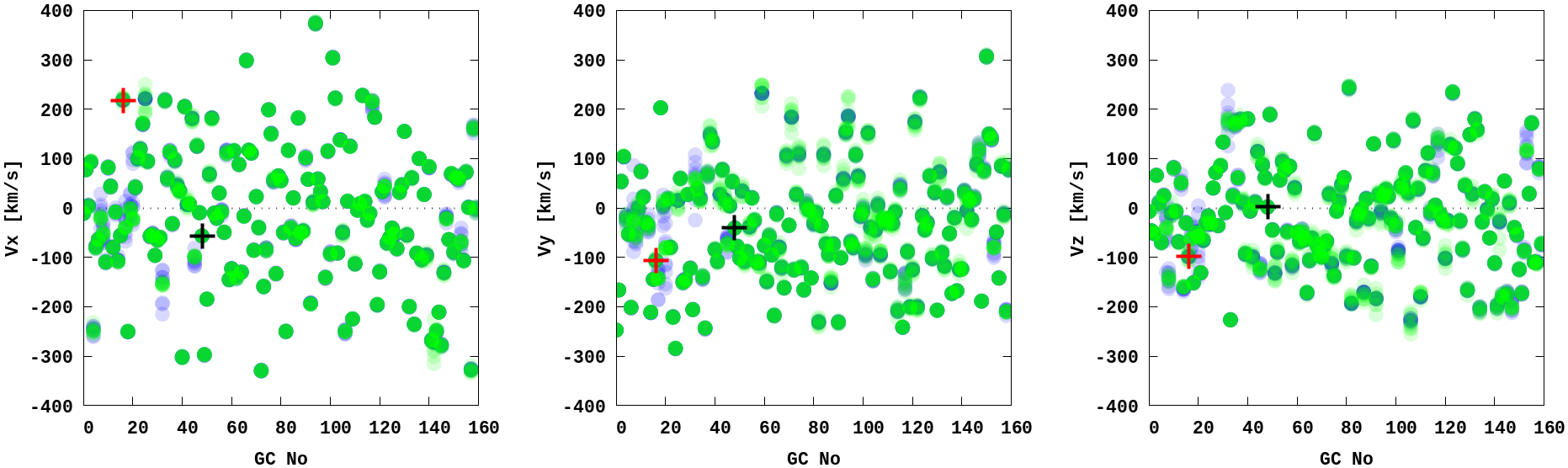}
\caption{Influence of error randomisation to the initial velocity components in the Galactocentric Cartesian reference frame. Distance determination -- green circles and velocity determination -- blue circles. Red cross marked Crater, black -- NGC 6121.}
\label{fig:ini-err}
\end{figure*}

Prior to the analysis of the orbital types of the GCs, we investigate the impact of the initial conditions uncertainties on the initial velocities values and structure of the orbits. For each GC in our sample, we generated extra $10$ initial conditions where the initial velocities take into account the normal distribution of the errors for PMRA, PMDEC, RV (blue circles), and D$_{\odot}$ (green circles) from \cite{Baumgardt2021}\footnote{\label{note1}Error values for PMRA, PMDEC, RV, and D$_{\odot}$ from \url{https://people.smp.uq.edu.au/HolgerBaumgardt/globular/orbits_table.txt}}, see Fig. \ref{fig:ini-err}. As we see, both randomisations have a similar effect on the GCs' initial velocities. In general, the errors by velocities are even slightly dominated over the errors due to the distance determination. The effect from the velocity determination errors is even slightly larger for a few objects (six) with indexes below 30.  

%-------------------------------------------------------------------------%
%-------------------------------------------------------------------------%

\begin{figure*}[htbp!]
\centering
\includegraphics[width=0.99\linewidth]{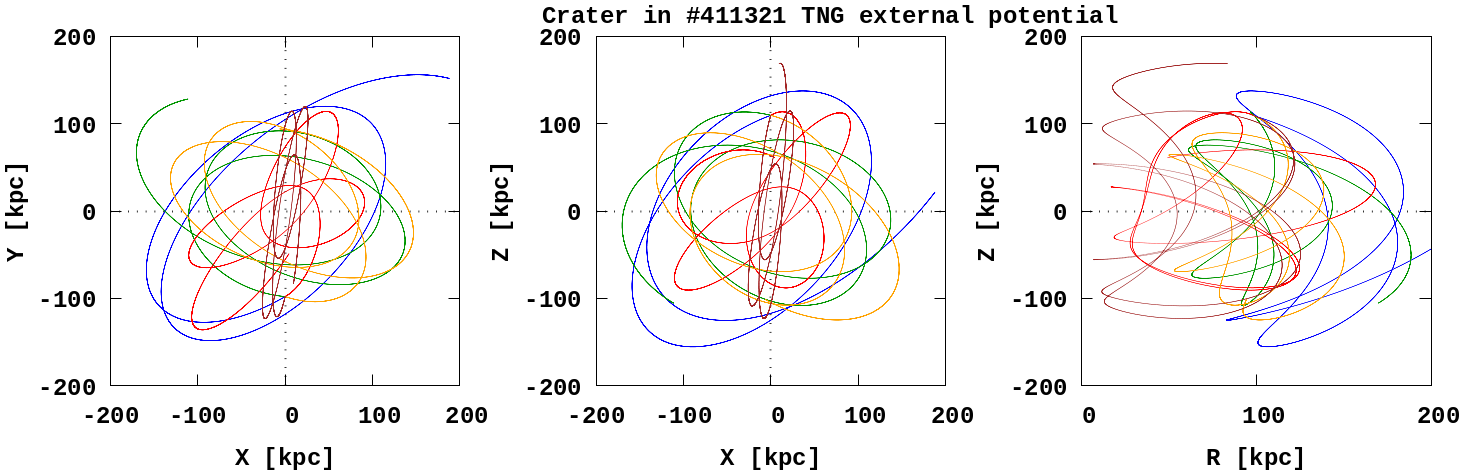}
\includegraphics[width=0.99\linewidth]{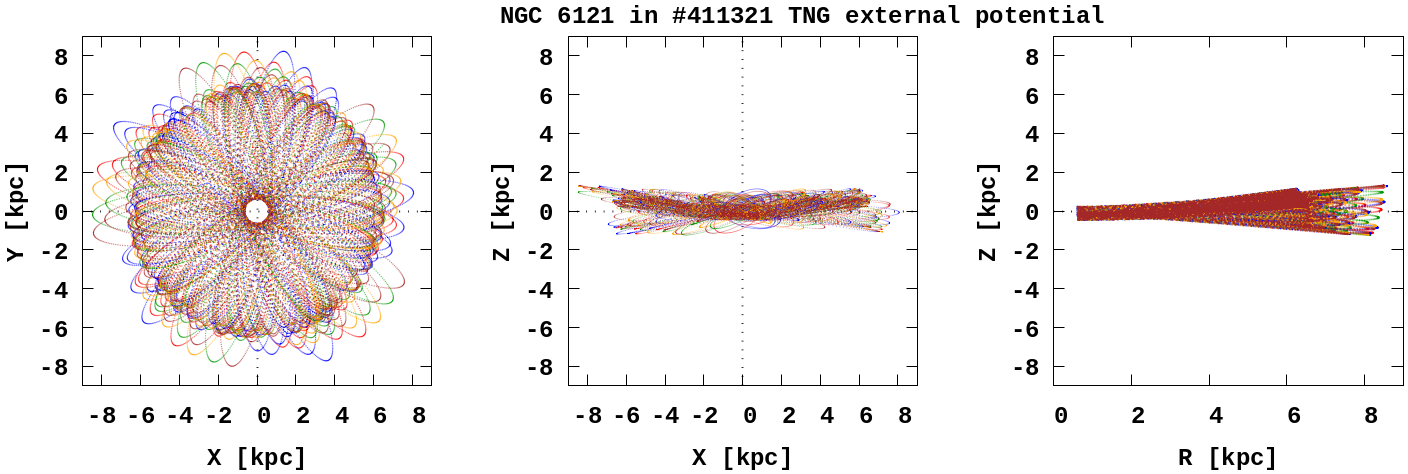}
\caption{Changes in the orbits caused by the uncertainties in the radial velocity and proper motions\cref{note1} for Crater and NGC 6121. The presented orbits were integrated for 10 Gyr lookback time. The different colours with transparency are represent the orbits for five random realisations of the initial conditions in the {\tt \#411321} TNG-TVP.}
\label{fig:orb-err}
\end{figure*}

In Fig.~\ref{fig:orb-err} we illustrate the impact of the initial conditions uncertainties on the orbital evolution of two selected GCs: Crater~(belongs to the $5\%$ with the largest velocity uncertainties: 211\%, 100\% and 0.4\% for proper motions and radial velocity) as for the NGC~6121 (with the one of the smallest values: 0.1\%, 0.12\%, and 2\% respectively). These two GCs are marked by crosses in Fig. \ref{fig:ini-err}. The velocity uncertainties significantly change the orbital characteristics in the case of Crater. In particular, the extension of the orbit along the principal axes, especially in the disk plane, is very different in various GCs realisations. At the same time, the vertical motions are less affected. In contrast, for the NGC~6121, which parameters are well measured, the orbits are less affected by the velocity uncertainties, as we can see on the bottom row of Fig.~\ref{fig:orb-err}. Thus, since the velocity uncertainties for most of our GCs are similar to the NGC~6121, we assume that the orbits of the majority of the GCs are robust to the variations of the initial condition. 

%-------------------------------------------------------------------------%
%-------------------------------------------------------------------------%

As we can see in Fig.~\ref{fig:orb-err} the influence of the relative errors of the heliocentric distances on the orbital shape of our GC sample is indeed less significant compared to the other relative errors. As an example, we present the orbits of two selected GC: Carter (largest heliocentric distance) and NGC 6121 (smallest heliocentric distance). As we can see, the effect from the relative distance error is significantly smaller compared to the effect from the proper motions and radial velocity errors, presented on the same plot. Next, we focus our attention only on the GC-point orbits obtained using the central values for positions and velocities from \cite{Baumgardt2021} and \cite{VasBaum2021}. 

%%%%%%%%%%%%%%%%%%%%%%%%%%%%%%%%%%%%%%%%%%%%%%%%%%%%%%%%%%%%%%%%%%%%%%%%%%%%%%%%
\subsection{The evolution of the Globular Clusters orbital elements}\label{subsec:evol-orb}
%%%%%%%%%%%%%%%%%%%%%%%%%%%%%%%%%%%%%%%%%%%%%%%%%%%%%%%%%%%%%%%%%%%%%%%%%%%%%%%%

\begin{figure*}[htbp]
\centering
\includegraphics[width=0.49\linewidth]{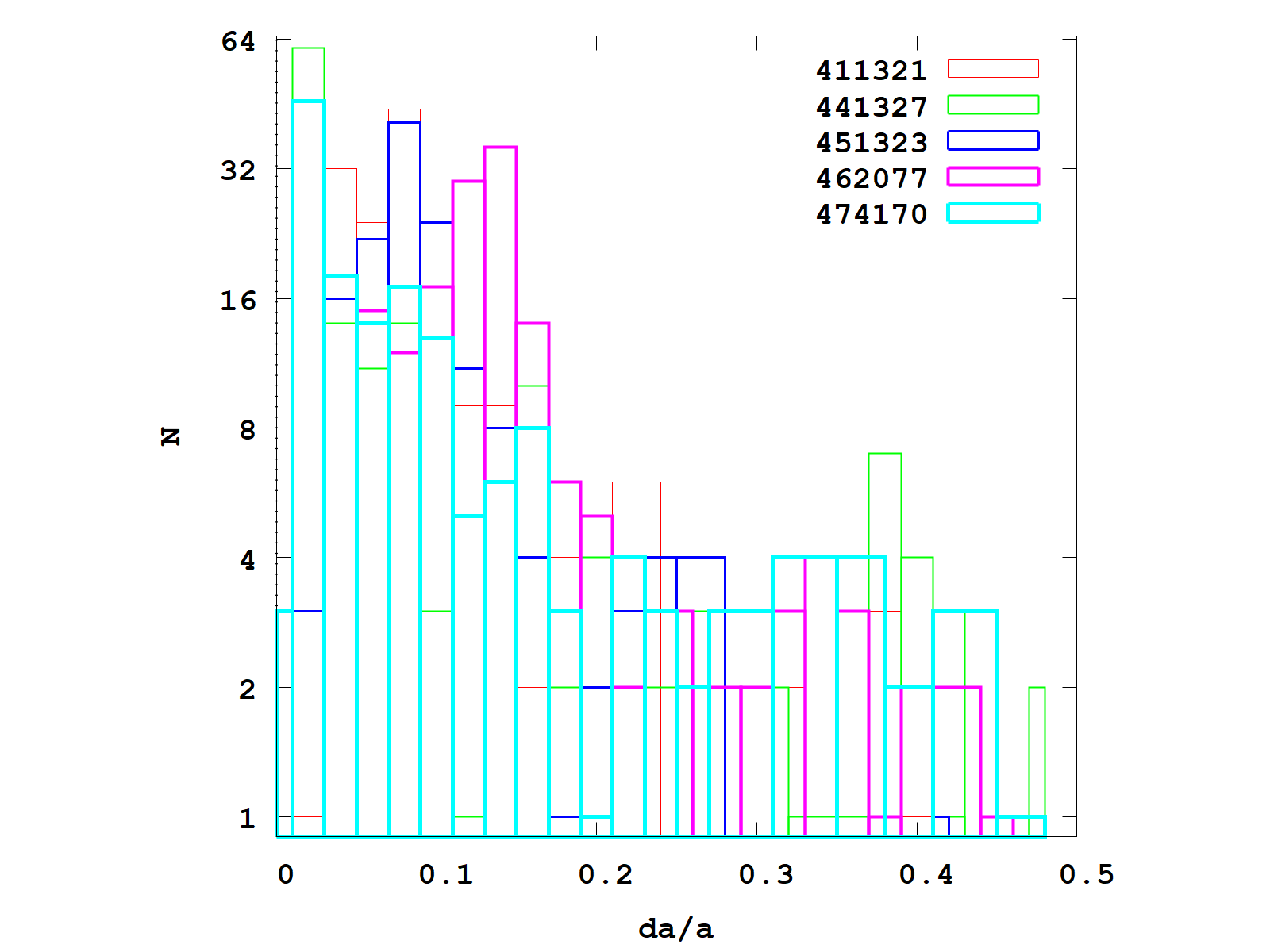}
\includegraphics[width=0.49\linewidth]{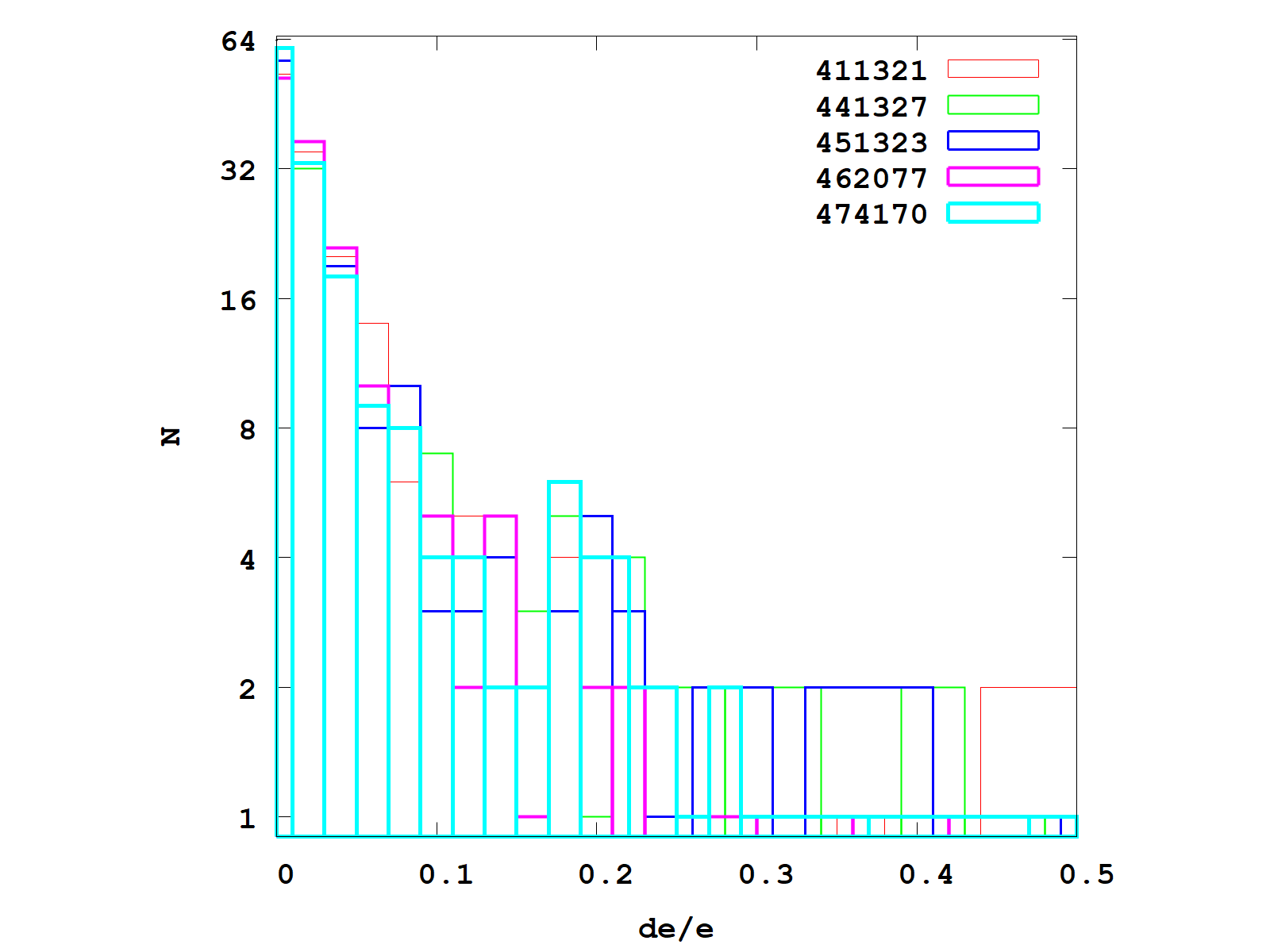}
\caption{Relative orbital semi-major $\Delta a/a$ (left) and eccentricity $\Delta e/e$ (right) changes during the GCs' orbital evolution for all five TNG-TVPs.}
\label{fig:da_de}
\end{figure*}

\begin{figure}[htbp]
\centering
\includegraphics[width=0.99\linewidth]{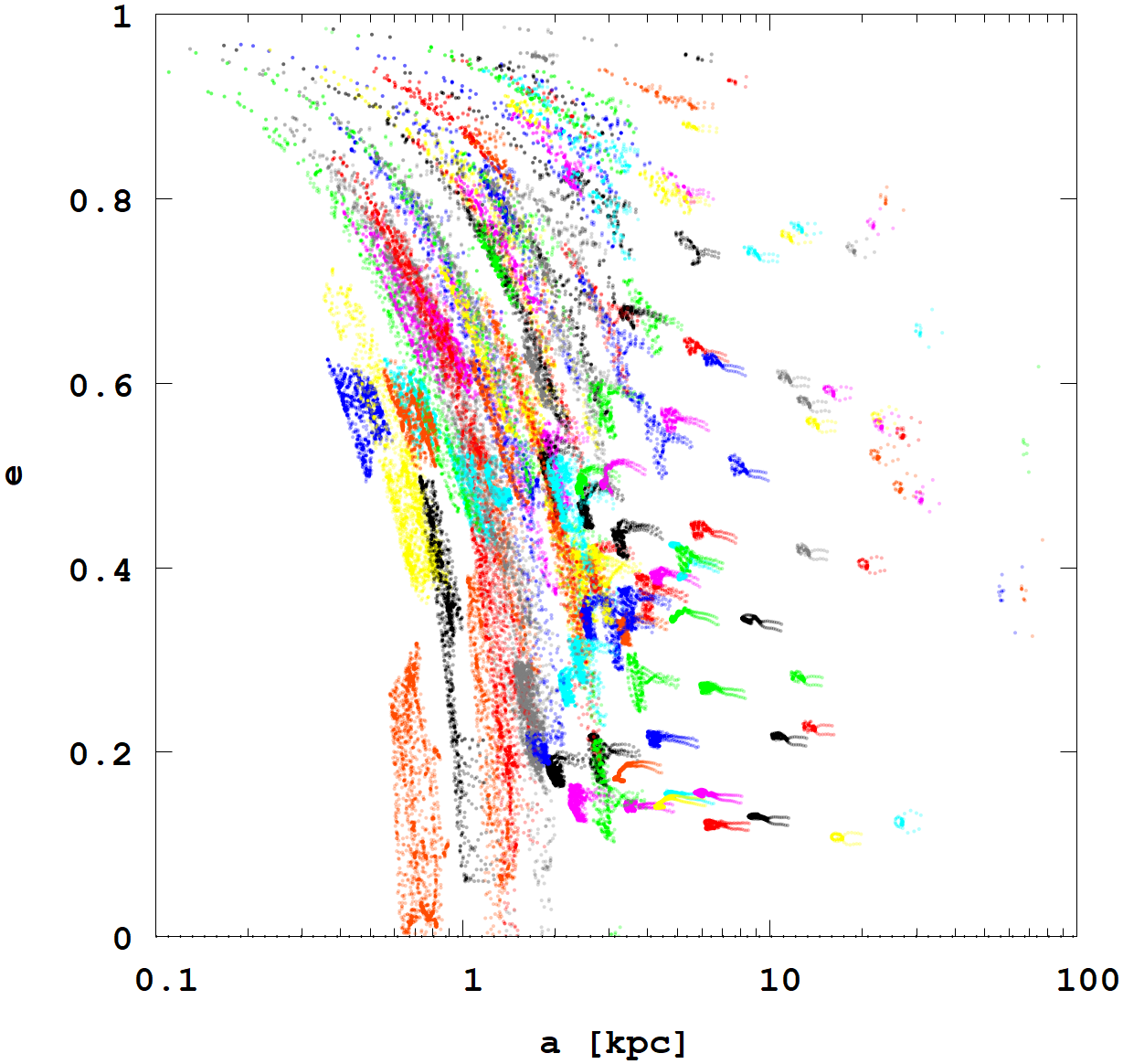}
\caption{Evolution of the GCs' orbital semi-major and eccentricity during the whole backward integration time in the case of {\tt \#411321} TNG-TVP. We can clearly see the separation between the `inner' and `outer' GCs. The `inner' GCs ($a\leq$3 kpc) have a more regular and larger eccentricity changes during the evolution. The `outer' GCs ($a>$3 kpc) have a much smaller eccentricity changes during the whole backward integration time.} 
\label{fig:orb-evol}
\end{figure}

Assuming the cosmologically motivated time variable potential we automatically taking into account the possible influence of the Milky Way satellite galaxies on to the GCs subsystem. If we compare the individual GCs' orbital elements (semi-major axis - {\tt a} and eccentricity - {\tt ecc}) variation over the whole integration time (sometimes over the hundreds or thousands of GCs' orbital revelations). In Fig.~\ref{fig:da_de} we present the orbital elements relative changes in five TNG-TVPs. In Fig.~\ref{fig:orb-evol} we have shown the time evolution of the {\tt a} and {\tt ecc}. The strong influence on the GCs' orbits have already discussed in details in the papers \cite{Garrow2020} and \cite{Boldrini2022}. Our Fig.~\ref{fig:da_de} and Fig.~\ref{fig:orb-evol} are directly comparable with the Figures 1, 2 and 3 from the \cite{Garrow2020}. As we can see the relative changes of {\tt a} and {\tt ecc} behave in a similar range. In our case the {\tt a} evolution is even larger in a factor of two. We also notice a very similar distribution of the relative quantities for the different TNG-TVP potentials.

As a illustration of the time evolution of orbital elements of all GCs we have shown such a data for the {\tt \#411321} TNG-TVP in Fig.~\ref{fig:orb-evol}. Here we can clearly see the separation between the `inner' and `outer' GCs. The `inner' GCs (a $\lesssim$3 kpc) have a more regular and larger eccentricity changes during the evolution. The `outer' GCs (a $\gtrsim$3 kpc) have much smaller eccentricity changes during the whole backward integration time. This Fig.~\ref{fig:orb-evol} can be directly compared with the Figure 3 from the \cite{Garrow2020}. As we can see, our potential keep bound the individual GCs, even with large {\tt a} during the whole time of integration -- 10 Gyr.

%-------------------------------------------------------------------------%
%-------------------------------------------------------------------------%

%-------------------------------------------------------------------------%
%-------------------------------------------------------------------------%

For example, in the works of \cite{Garrow2020} and \cite{Boldrini2022} during their GCs orbit investigations the MW major satellites were added as an extra gravitational perturbation in the fixed MW potential approximation. In contrast in our case, the IllustrisTNG-100 mass assembling history of our theoretical MW-like galaxies (see Fig. \ref{fig:MW-TNG}) provides us with the same or even a larger scale of orbital perturbations for individual GCs during their time evolution.

\begin{figure*}[htbp!]
\centering
\includegraphics[width=0.49\linewidth]{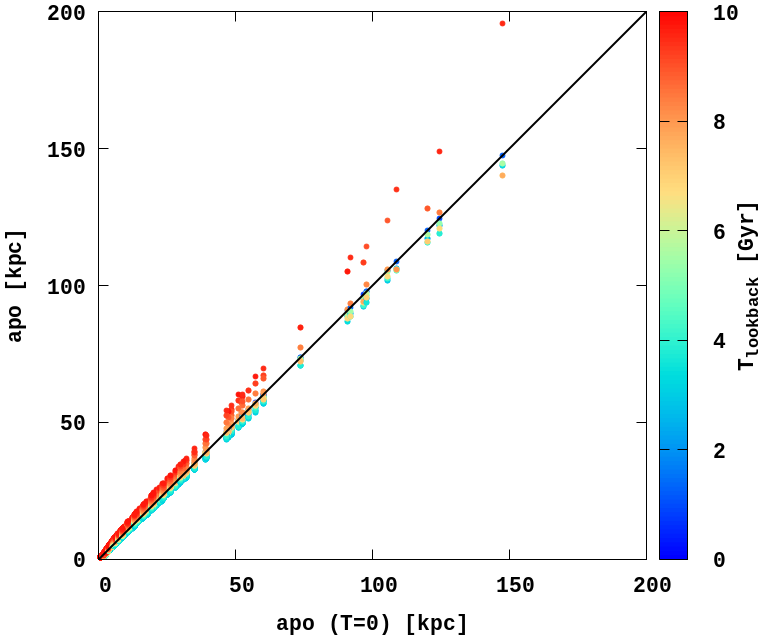}
\includegraphics[width=0.49\linewidth]{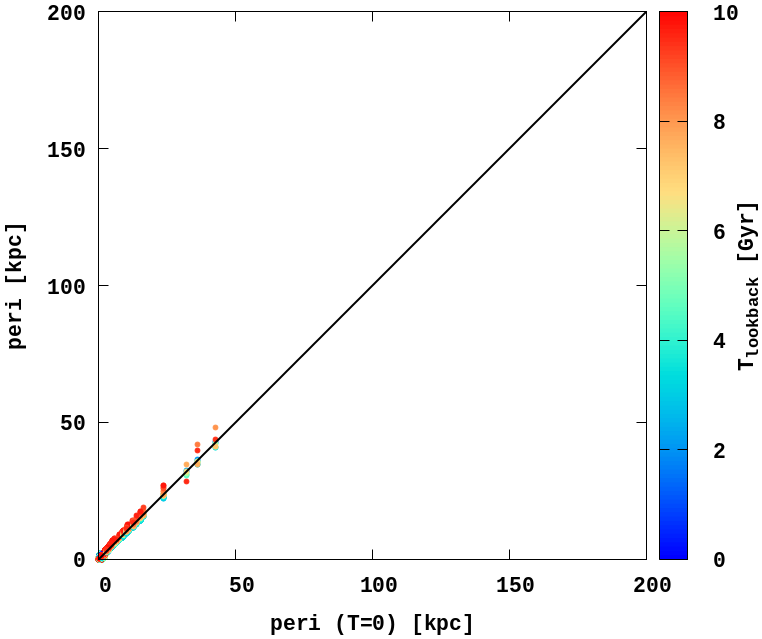}
\caption{Evolution of the apocenters (\textit{left panel}) and pericenters (\textit{right panel}) for the 159 GCs in the {\tt \#411321} TNG-TVP. The colour code corresponds to the lookback time.} 
\label{fig:apo-peri-all}
\end{figure*}

In Fig.~\ref{fig:apo-peri-all} we present the GCs' main orbital parameters evolution: apocenters and pericenters (as for a similar plot one can look of the work \citealt[see Figure 9]{Bajkova2021AstL}). On the $X$-axis we plotted the {\tt apo} and {\tt peri} values of each GCs in the static (fixed) potential, that is, using the initial values of the {\tt \#411321} IllustrisTNG-100 potential. The general behaviour of these quantities in our Fig.~\ref{fig:apo-peri-all} and in the work of \cite{Bajkova2021AstL} are quite similar. Up to roughly 5 Gyr lookback time we we did not notice significant changes in these quantities. But after approximately 8 Gyr the differences (especially in apocenters values) are become more significant (up to $\sim$25\%).

%-------------------------------------------------------------------------%
%-------------------------------------------------------------------------%

\begin{figure*}[htbp!]
\centering
\includegraphics[width=0.99\linewidth]{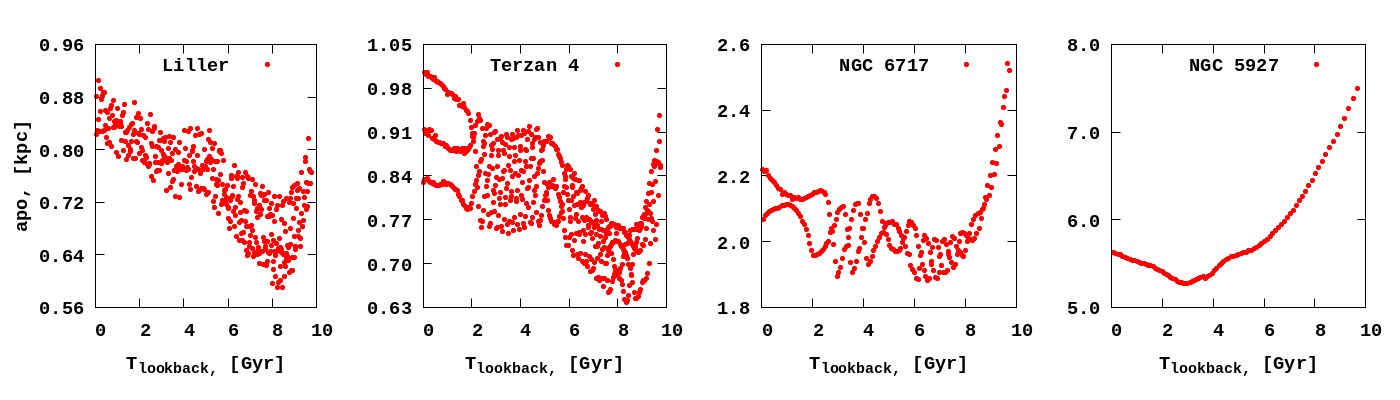}
\includegraphics[width=0.99\linewidth]{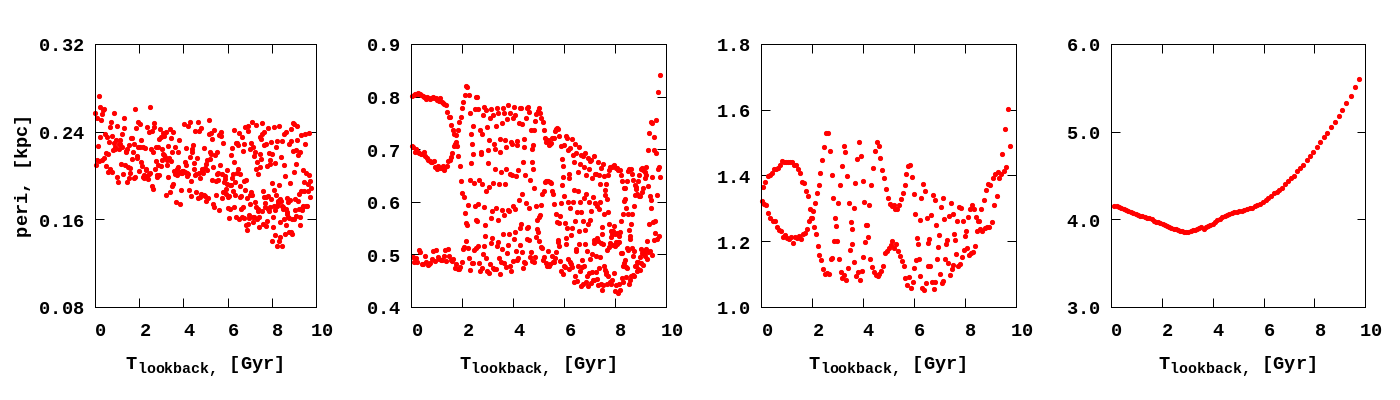}
\caption{Evolution of the apocenters (\textit{upper panel}) and pericenters (\textit{bottom panel}) for the (\textit{from left to right}) Liller, Terzan 4, NGC 6717, and NGC 5927 GCs. The presented orbits were integrated for 10 Gyr lookback time. Each point represents the individual orbital passages of the GC in the {\tt \#411321} TNG-TVP}.
\label{fig:apo}
\end{figure*}

In Fig.~\ref{fig:apo} we present the evolution of the apocenters and pericenters during the full 10 Gyr lookback time integration in the {\tt \#411321} TNG-TVP. For the visualisation, we selected several GCs with different heliocentric distances. In our representative sample, the Liller and Terzan 4 are the closest GCs to the Galactic centre.

%-------------------------------------------------------------------------%
%-------------------------------------------------------------------------%

As we see in Fig.~\ref{fig:apo}, the Liller and Terzan 4 as closest GCs have a clear decreasement of the apocenter with the minimum values at the 8 Gyr with further increasement. The NGC 6717 has a slow decline with the minimum at the same time. Meanwhile, we don't see such a clear picture of pericenter evolution. Such a behaviour can be understood taking into account the more strong dependence of the apocenter value (compared to the pericenter) from the mass and size changes of our TNG-TVPs. At the same time, the GC 5927 apo- and pericenter changes are `synchronised' and have a minimum value at 3 Gyr and after it starts to increase. The evolution of the apocenters and pericenters for all GCs we present on the web-site\cref{note2}, for all five TNG-TVPs and for each GC separately.

\begin{figure*}[htbp!]
\centering
\includegraphics[width=0.99\linewidth]{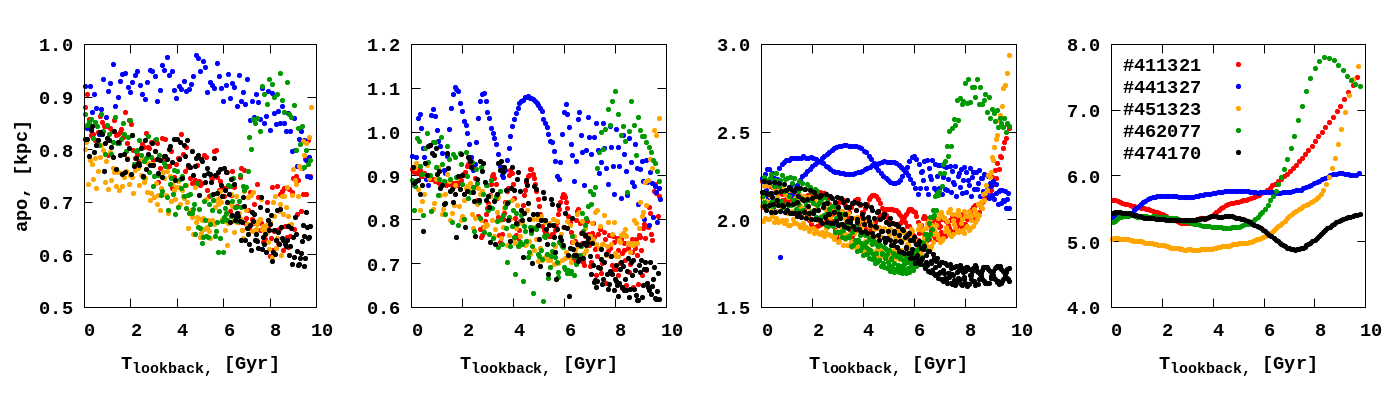}
\includegraphics[width=0.99\linewidth]{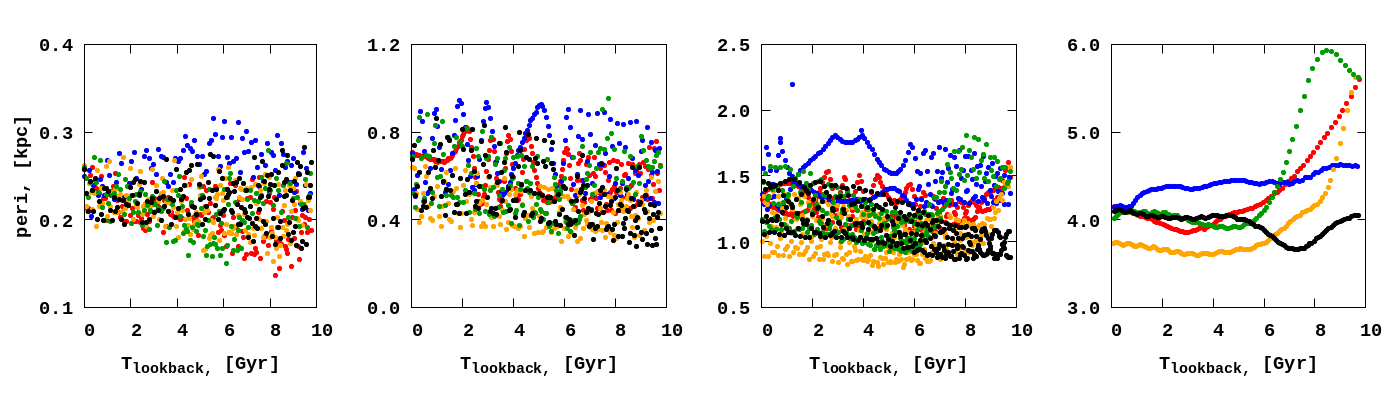}
\caption{Evolution of the apocenters (\textit{upper panel}) and pericenters (\textit{bottom panel}) for the (\textit{from left to right}) Liller, Terzan 4, NGC 6717, and NGC 5927 GCs in five TNG-TVPs, where colours correspond: red -- {\tt \#411321}, blue -- {\tt \#441327}, orange -- {\tt \#451323}, green -- {\tt \#462077}, and black -- {\tt \#474170} respectively. The presented orbits were integrated for 10 Gyr lookback time.}
\label{fig:apo-pot}
\end{figure*}

Also, we present in Fig.~\ref{fig:apo-pot} the comparison of apocenters and pericenters of evolution for selected GCs in the five TNG-TVPs simultaneously. The pericenter evolution in all the five TNG-TVPs looks very similar. The apocenter evolution shows some differences for different potentials, which we can understand as a more strong influence of the time variable masses and sizes of MW-like potentials to the orbits.

%-------------------------------------------------------------------------%
%-------------------------------------------------------------------------%

%%%%%%%%%%%%%%%%%%%%%%%%%%%%%%%%%%%%%%%%%%%%%%%%%%%%%%%%%%%%%%%%%%%%%%%%%%%%%%%%
\subsection{Types of orbits}\label{subsec:type-orb}
%%%%%%%%%%%%%%%%%%%%%%%%%%%%%%%%%%%%%%%%%%%%%%%%%%%%%%%%%%%%%%%%%%%%%%%%%%%%%%%%

We calculated the orbits of 159 GCs lookback time for 10 Gyr in each of our TNG-TVP potentials. The orbital evolution for our GC sample is presented online\footnote{\label{note2} We also presented orbits for all 159 GCs in TNG-TVP potentials on the web site of the project \url{https://bit.ly/3b0lafw}}. The visualisation was carried out in three projections of the Cartesian Galactocentric rest-frame: ($X$, $Y$), ($X$, $Z$), and ($R$, $Z$), where $R$ is the planar Galactocentric radius. As we see, the orbits change in different potentials, but their general shape remains similar. 

The shape filled by an orbit in an axisymmetric potential might be classified as a short-axis tube, thin tube, etc., depending on specifics of the potential (e.g., oblate/prolate) and the particular combination of integrals of motion the orbit possesses \citep{carpintero1998,merritt_1999}. For our purposes, after a visual analysis of 159 orbits (in each of our TNG-TVP potentials), we divide them into four main types (categories): 
\begin{itemize}
    \item Tube orbit (TB) -- $110$~(69\%) orbits. The orbit is mainly in the ($X$, $Y$) Galactic plane with a hole in the centre. The orbit in the ($X$, $Z$) plane has a boxy shape and in the ($R,Z$) has a trapezoidal shape. 
    \item Perpendicular tube orbit (PT) -- $8$~(5\%) orbits. The orbital plane  is close to the meridional, also contains a hole in the centre and in the ($R$, $Z$) plane has a trapezoidal shape. 
    \item Long radial orbit (LR) -- $19$~(12\%) orbits.  These are long-period, near-hyperbolic orbits characterised by large eccentricity with no prominent hole in the centre.
    \item Irregular orbit (IR) -- $22$~(14\%) orbits. The orbit can not be classified as one of the described above. 
\end{itemize}

We listed the orbit types for each GC in Table~\ref{tab:GC3}. In Fig.~\ref{fig:to} we show some typical examples of the orbits used in the classification above.

%%%%%%%%%%%%%%%%%%%%%%%%%%%%%%%%%%%%%%%%%%%%%%%%%%%%%%%%%%%%%%%%%%%%%%%%%%%%%%%%
\subsection{Association with the regions of the Galaxy}\label{subsec:gal-reg}
%%%%%%%%%%%%%%%%%%%%%%%%%%%%%%%%%%%%%%%%%%%%%%%%%%%%%%%%%%%%%%%%%%%%%%%%%%%%%%%%

We assumed that Galaxy has several spatial components: bulge, thin disk, thick disk and main halo. Each GCs' orbit can be associated predominantly with one of them (based on the distance criteria). Such associations weakly depend on orbits shape. We analysed the orbits of all 159 GCs in each potential to sort them by ownership to the different regions of the Galaxy at the past billion years. For this classification we used the following characteristic scales according to \cite{Bland-Hawthorn2016} for orbital apocenter and pericenter of each GCs:
\begin{itemize}
    \item Bulge (BL). Scale height in $Z$ directions - $\sim$0.2 kpc. Scale length in $R$ plane - $\sim$0.7 kpc. With such a scales we determine the boundary dimensions for the bulge as $\sim$2.1 $\times$ $\sim$0.6 kpc. We associated nine GCs. 
    
    \item Thin disk (TN). Scale height in $Z$ directions - $\sim$0.3 kpc. Scale length in $R$ plane - $\sim$2.6 kpc. With such a scales we determine the boundary dimensions for the thin disk as $\sim$7.8 $\times$ $\sim$0.9 kpc. We associated ten GCs. 
    
    \item Thick disc (TH). Scale height in $Z$ directions - $\sim$0.9 kpc. Scale length in $R$ plane - $\sim$2 kpc. With such a scales we determine the boundary dimensions for the thick disc as $\sim$6 $\times$ $\sim$2.7 kpc. We associated 47 GCs. 
    
    \item Halo (HL). Over then $\sim$2.7 kpc in $Z$ direction and more than $\sim$7.8 kpc in $R$ plane. We associated 94 GCs. 
\end{itemize}

%-------------------------------------------------------------------------%
\begin{figure}[htbp]
\centering
\includegraphics[width=0.99\linewidth]{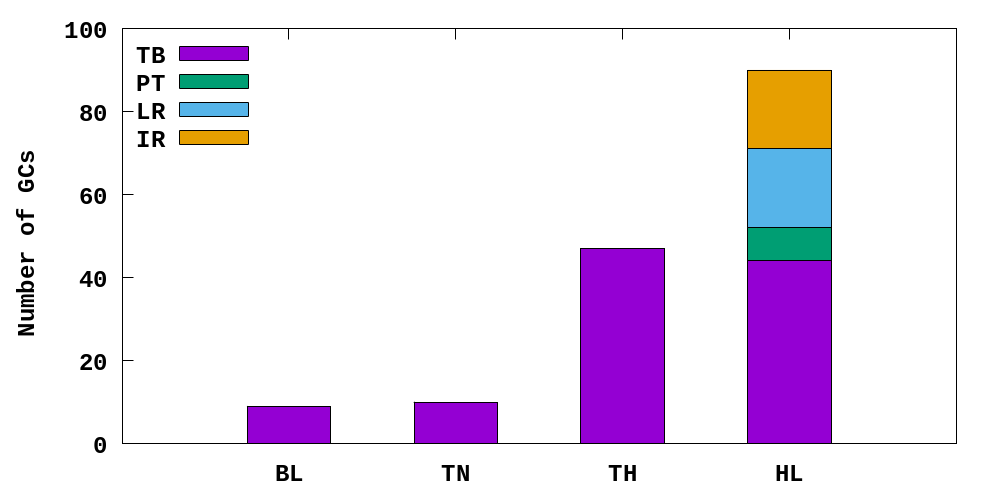}
\caption{Distribution of the GCs by their orbit types in four Galactic components: bulge (BL), thin disk (TN), thick disk (TH), and halo (HL). Colour coding shows different orbit types: tube orbit (TB), perpendicular tube orbit (PT), long radial orbit (LR) and irregular orbit (IR).} 
\label{fig:gisogal-orb}
\end{figure}
%-------------------------------------------------------------------------%

As we see from Fig.~\ref{fig:gisogal-orb} most GCs at the present stage (last billion year) belong to the halo and thick disk, at 59\% and 30\%, respectively. The bulge and thin disk contain just 5\% and 6\% GCs. But in the distant past there is a high probability that some of the GCs belonged to other regions of the Galaxy \cite[e.g.][]{Bajkova2020, Sun2023}. A possible explanation is that GCs survive better when they are far from the Galactic centre. We listed the associations with different Galaxy components for each GC in Table~\ref{tab:GC3}. This classification is also stable for each of the five different TNG-TVP, which made, in our point of view, these potentials quite representative.

The majority of the GCs from the bulge, thin and thick have tube orbits, see Table \ref{app:table}. Possibly, it underwent extreme encounter(s) with another GC, which changed its orbit. In contrast, halo GCs display all types of orbits, with the most common being tube, irregular and long radial orbits. Also, the halo harbours objects with all GCs that have perpendicular tube orbits.

%%%%%%%%%%%%%%%%%%%%%%%%%%%%%%%%%%%%%%%%%%%%%%%%%%%%%%%%%%%%%%%%%%%%%%%%%%%%%%%%
\subsection{Energy changes of Globular Clusters during evolution}\label{subsec:phase-space-evol}
%%%%%%%%%%%%%%%%%%%%%%%%%%%%%%%%%%%%%%%%%%%%%%%%%%%%%%%%%%%%%%%%%%%%%%%%%%%%%%%%

\begin{figure*}[htbp]
\centering
\includegraphics[width=0.49\linewidth]{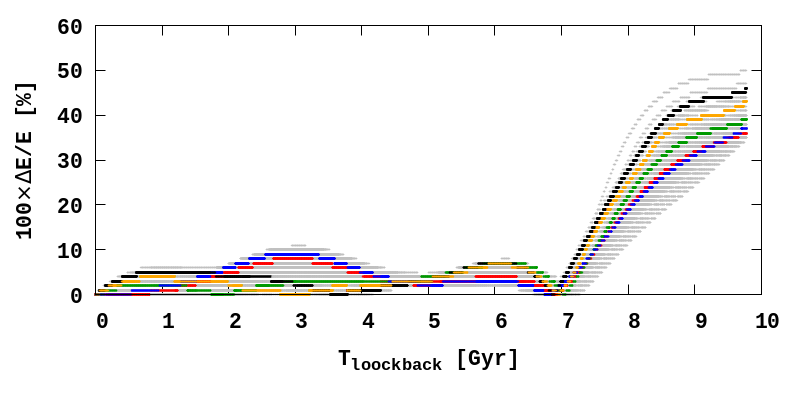}
\includegraphics[width=0.49\linewidth]{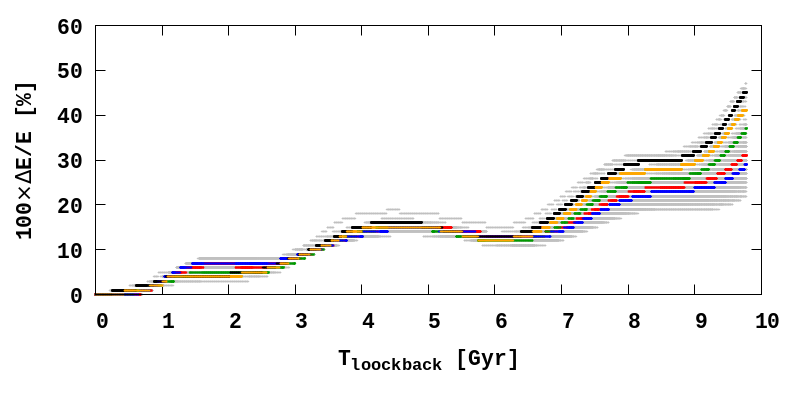}
\includegraphics[width=0.49\linewidth]{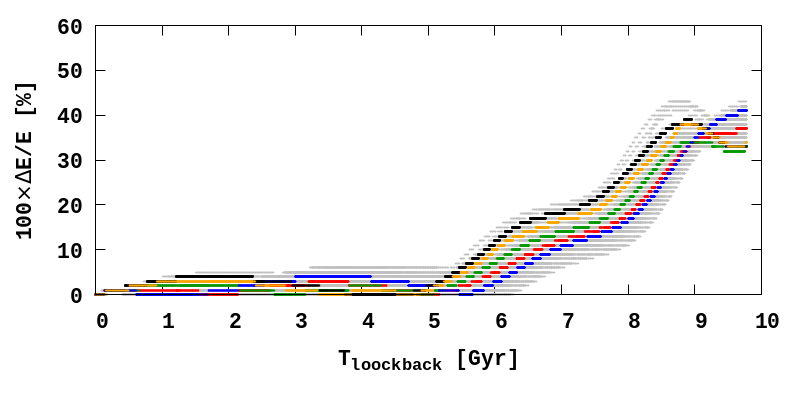}
\includegraphics[width=0.49\linewidth]{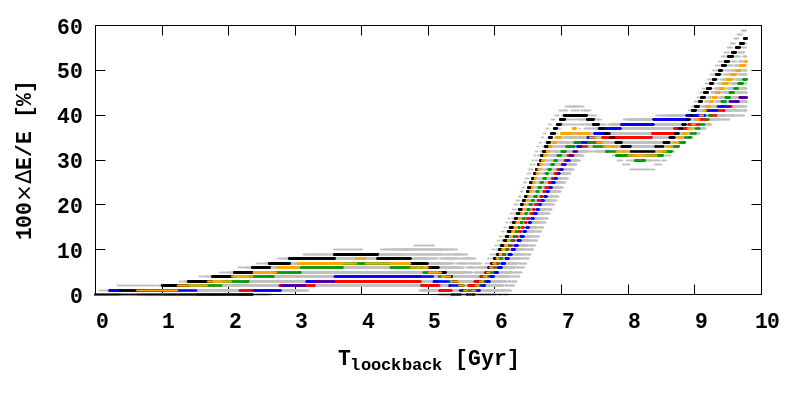}
\includegraphics[width=0.49\linewidth]{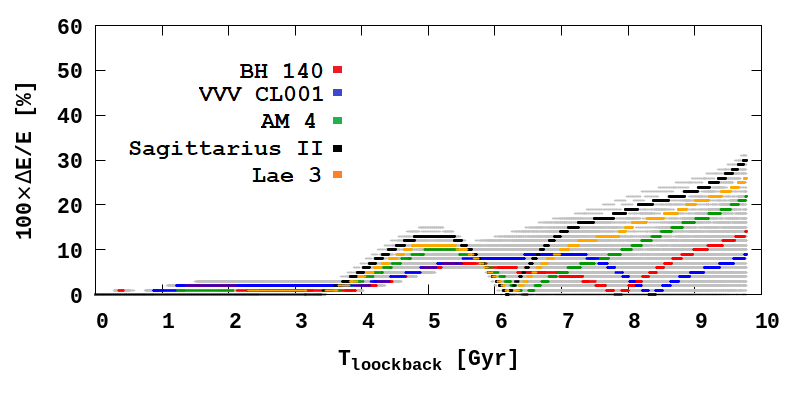}
\caption{Relative energy $\Delta E/E$ changes during the GCs' orbital evolution in per cent.  {\it From left to right from top to bottom}: {\tt \#411321, \#441327, \#451323, \#462077}, and {\tt \#474170} TNG-TVPs. Colour-coding: red -- BH~140, green -- AM~4, blue -- VVV~CL001, black -- Sagittarius~II, orange -- Lae~3, and grey -- for all other GCs.}
\label{fig:de}
\end{figure*}

Since we consider the dynamics of the MW GCs in the evolving potentials, the integrals of motions~(energy and angular momentum) are not conserved over time. In order to quantify this, we calculated the relative energy changes $\Delta E/E$ during the evolution of GCs over 10 Gyr. In Fig.~\ref{fig:de} we present the relative energy changes of individual GCs during the orbit integration. In general, we see the effect of Galaxy mass changes in our selected TNG-TVPs (Fig.~\ref{fig:MW-TNG}). We found that, the maximum relative energy changes are $\sim$45-50\%. As an example, in Fig.~\ref{fig:de} we highlighted five GCs in colour: BH~140 and VVV~CL001 (red and blue colours) associated with the Galactic disk and likely formed in-situ; AM~4, Sagittarius~II, and Lae~3 associated with the halo and likely born outside the Galaxy \citep[see the definition of association][]{Malhan2022, Fernandez-Trincado2021}. All other 154 GCs are marked with a grey colour. As we can see from Fig.~\ref{fig:de}, the in-situ GCs are less affected by the changes in the gravitational potential compared to the accreted clusters. This trend is especially visible on the last panel of the Fig.~\ref{fig:de} for the {\tt \#474170 TNG-TVP}. 

%-------------------------------------------------------------------------%
%-------------------------------------------------------------------------%

We also identified the list of GCs whose relative energy has changed by more than 40\%. For simplicity, we present the list of GCs from the {\tt \#411321} TNG-TVP only: 

\begin{itemize}
    \item $\Delta E/E=45\%\ldots50\%$: Crater, Pal~3, Sagittarius~II (black colour in Fig.~\ref{fig:de}) Eridanus, AM~1, and Pal~4;
    
    \item $\Delta E/E=42\%\ldots44\%$: NGC~2419, Pyxis, Pal~14, Lae~3 (orange colour in Fig.~\ref{fig:de}), and Whiting~1;
    
    \item $\Delta E/E=40\%\ldots41\%$: NGC~5694, Arp~2, Terzan~8, Terzan~7, NGC~7006, Pal~12, and Pal~13.
\end{itemize}

All of them belong to the halo (HL) and, in most cases they have either irregular (IR) or perpendicular tube-type (PT) of orbits, see Table~\ref{tab:GC3}.

%-------------------------------------------------------------------------%
\begin{figure*}[htbp!]
\centering
\includegraphics[width=0.33\linewidth]{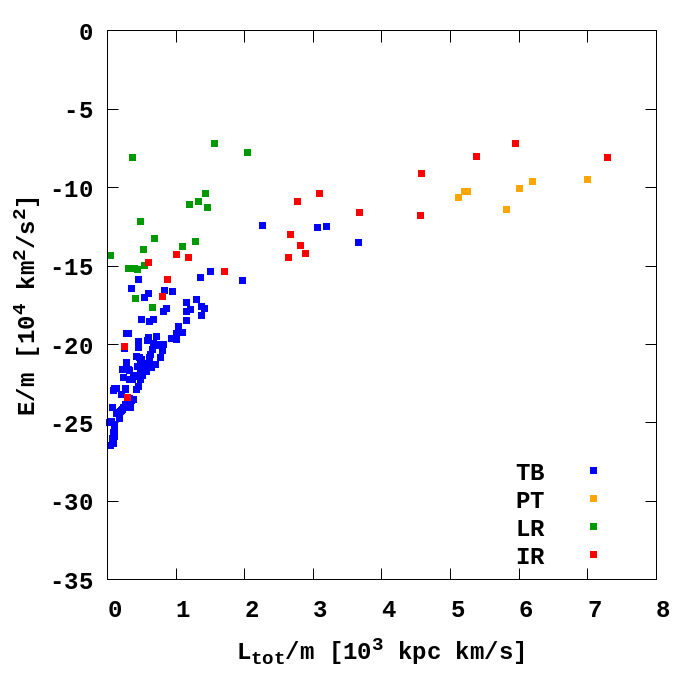}
\includegraphics[width=0.33\linewidth]{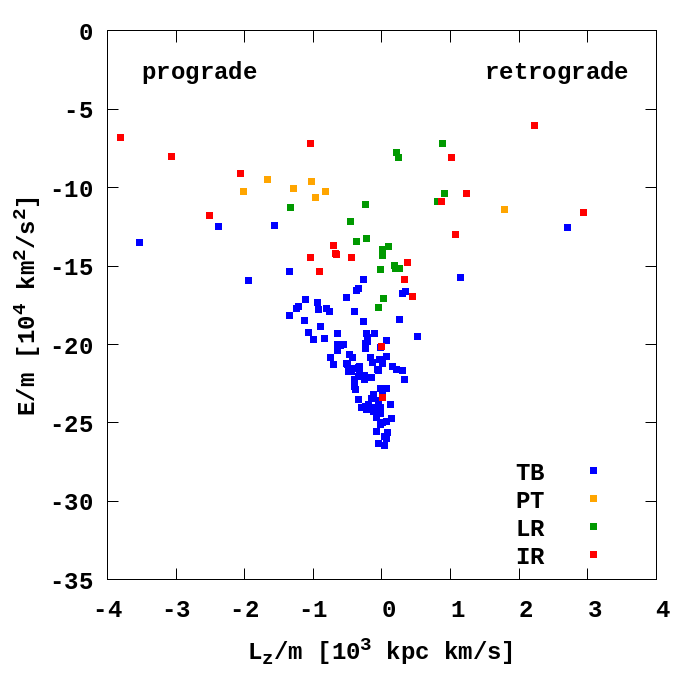}
\includegraphics[width=0.33\linewidth]{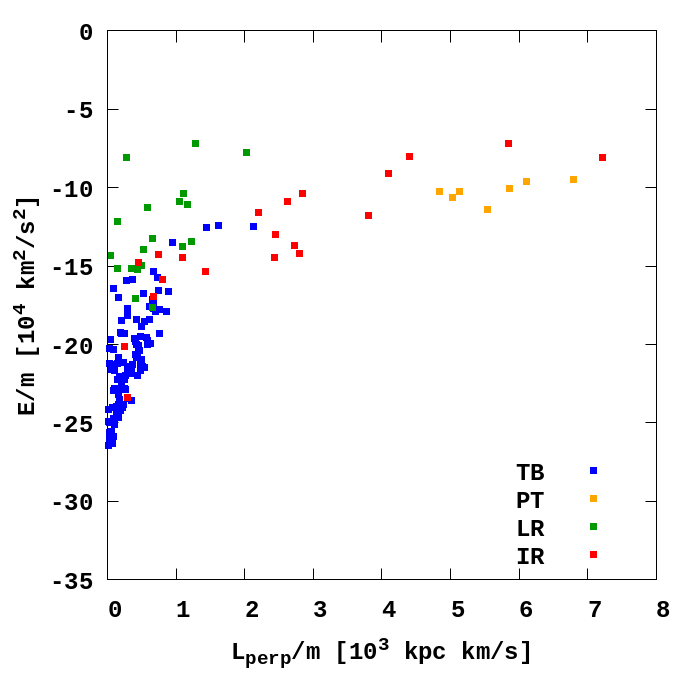}
\caption{Distribution of orbital types at redshift zero for the {\tt \#411321} TNG-TVP. {\it From left to right:} total angular momentum versus energy ($L_{\rm tot}$, $E$), $z$-th component of the angular momentum versus total energy ($L_{\rm z}$, $E$), the perpendicular component of the angular momentum versus total energy ($L_{\rm perp}$, $E$). The colour corresponds to the type of the GCs' orbit: blue -- tube orbit (TB), orange -- perpendicular tube orbit (PT), green -- long radial orbit (LR), and red - irregular orbit (IR). Pal~3, Crater, and Sagittarius~II are not shown in ($L_{\rm tot}$, $E$) and ($L_{\rm perp}$, $E$) phase spaces, also Sagittarius~II is not shown in the ($L_{\rm z}$, $E$) phase space due to their extremely large values.}
\label{fig:to1}
\end{figure*}
%-------------------------------------------------------------------------%

%%%%%%%%%%%%%%%%%%%%%%%%%%%%%%%%%%%%%%%%%%%%%%%%%%%%%%%%%%%%%%%%%%%%%%%%%%%%%%%%
\section{Globular Clusters in phase spaces}\label{sec:phase-space}
%%%%%%%%%%%%%%%%%%%%%%%%%%%%%%%%%%%%%%%%%%%%%%%%%%%%%%%%%%%%%%%%%%%%%%%%%%%%%%%%
%%%%%%%%%%%%%%%%%%%%%%%%%%%%%%%%%%%%%%%%%%%%%%%%%%%%%%%%%%%%%%%%%%%%%%%%%%%%%%%%
\subsection{Present-day properties of the Globular Clusters}\label{subsec:phase-space-cur}
%%%%%%%%%%%%%%%%%%%%%%%%%%%%%%%%%%%%%%%%%%%%%%%%%%%%%%%%%%%%%%%%%%%%%%%%%%%%%%%%

In our study we analyse the GCs' phase space distributions in the energy-angular momentum coordinates for each of the TNG-TVPs. Below, we consider specific values of energy and angular momentum normalised by the GCs' mass. In particular, we analyse the phase-space distribution of the GCs at present in three phase space coordinates combinations: total angular momentum~($L_{\rm tot}$) versus total energy ($E$), $z$-component of the angular momentum~($L_{\rm z}$) versus total energy ($E$), and the perpendicular component of the angular momentum~($L_{\rm perp}$) versus total energy ($E$), where $L_{\rm tot}=\sqrt{L_{x}^{2}+L_{y}^{2}+L_{z}^{2}}$ and $L_{\rm perp}=\sqrt{L_{x}^{2}+L_{y}^{2}}$. The orientation of the $Z$-axis in our Cartesian Galactocentric coordinates is directed towards to the Galactic South Pole. By this definition the angular momentum of the Sun is negative.

Next we discuss the main relations in the energy-angular momentum space for the GCs of different orbital types. In Fig.~\ref{fig:to1}, the GCs are colour-coded according to the type of orbit in the {\tt \#411321} potential. GCs with TB orbits have better GCs bounds, orbiting close to the Galactic center since their total energies are largest (by absolute value). Most of the GCs with TB obits show from zero to some prograde rotation with the smallest perpendicular angular momentum component. It can indicate their common dynamical origin or possible orbital evolution as a consequence of their similar angular momentum losses timescales. 

Globular clusters with IR orbits show a wide range of the total energy and the angular momentum components and overlap with the GCs of other types of orbits. Their origin is unlikely the same based on the quite wide range of their angular momentum. 

The GCs with LR orbits have a wide range of the total energies but tend to have smaller absolute values of the angular momentum. Interestingly these GCs have a relatively narrow range of  $L_{\rm perp}$ with a weak trend to have a retrograde rotation. 

The smallest group of the GCs with PT are grouped together in all the energy-angular momentum phase coordinates presented in Fig.~\ref{fig:to1}. They have roughly the same total energy, substantial net rotation, and strong motions perpendicular to the disk plane. The large values of $L_{\rm perp}$ might suggest their common origin. As it was mentioned above, the GCs have the same type of orbits in different TNG-TVPs. Therefore, the described picture is quite similar in all external potentials considered in the paper. 

\begin{figure*}[htbp!]
\centering
\includegraphics[width=0.33\linewidth]{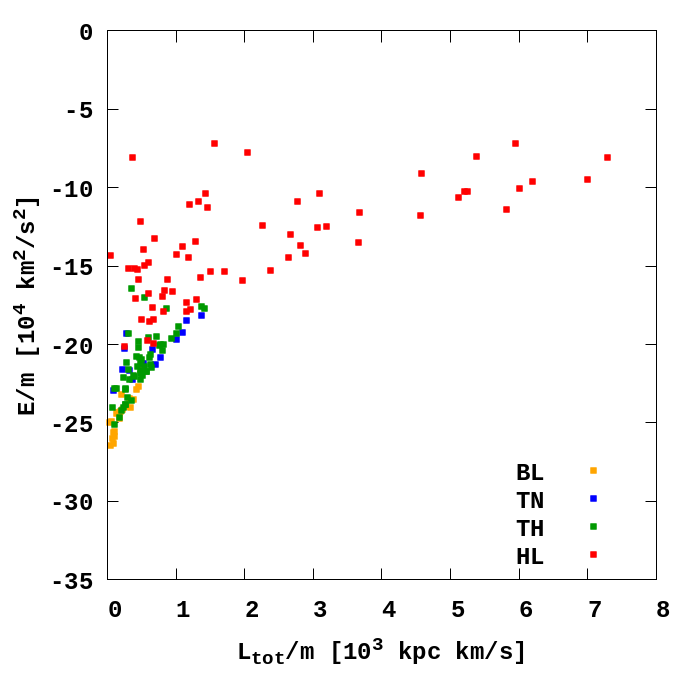}
\includegraphics[width=0.33\linewidth]{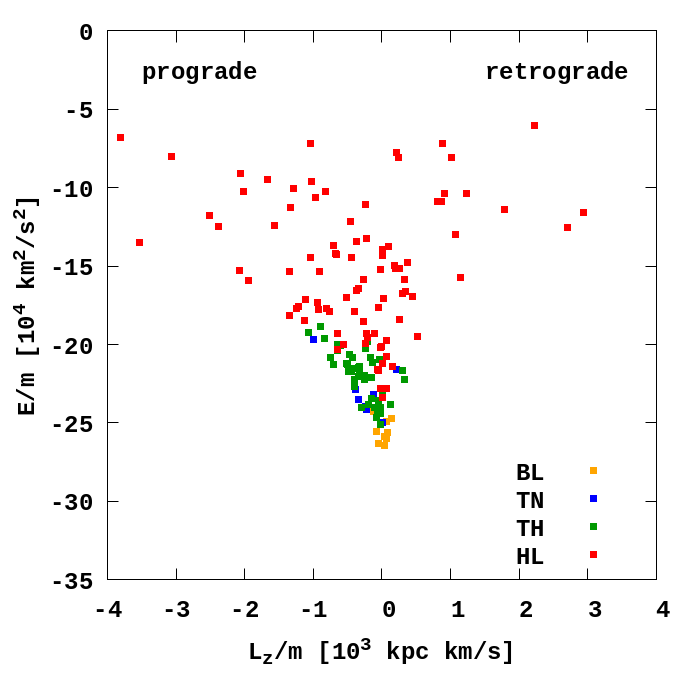}
\includegraphics[width=0.33\linewidth]{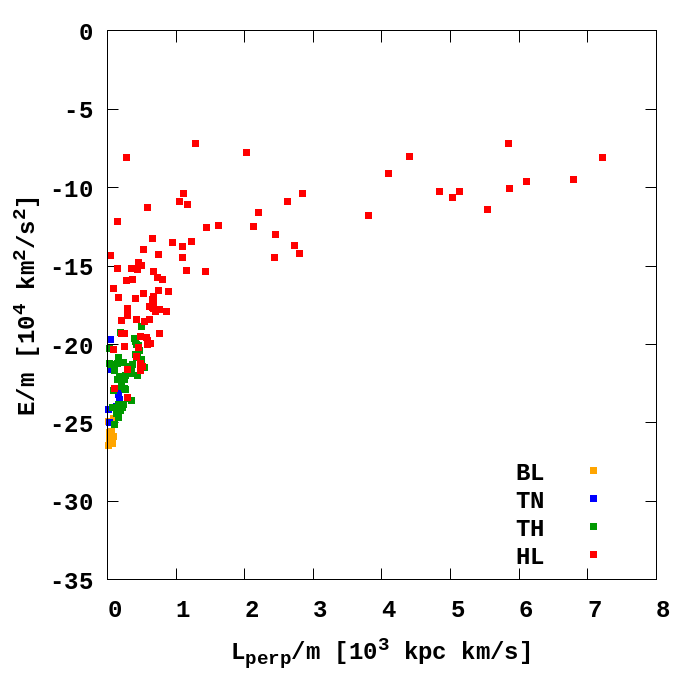}
\includegraphics[width=0.33\linewidth]{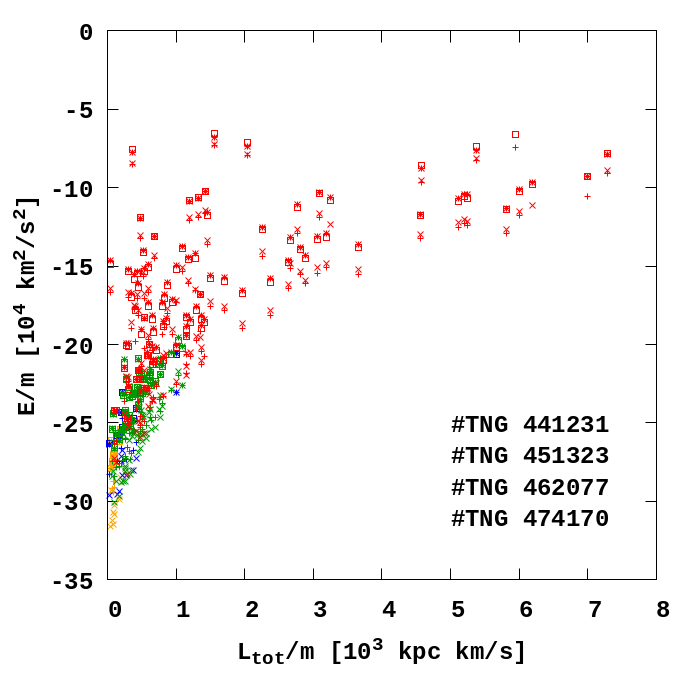}
\includegraphics[width=0.33\linewidth]{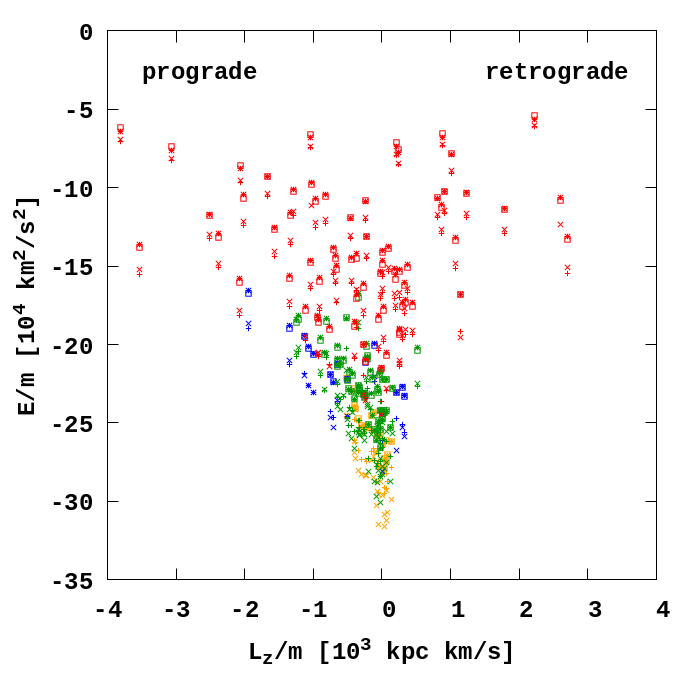}
\includegraphics[width=0.33\linewidth]{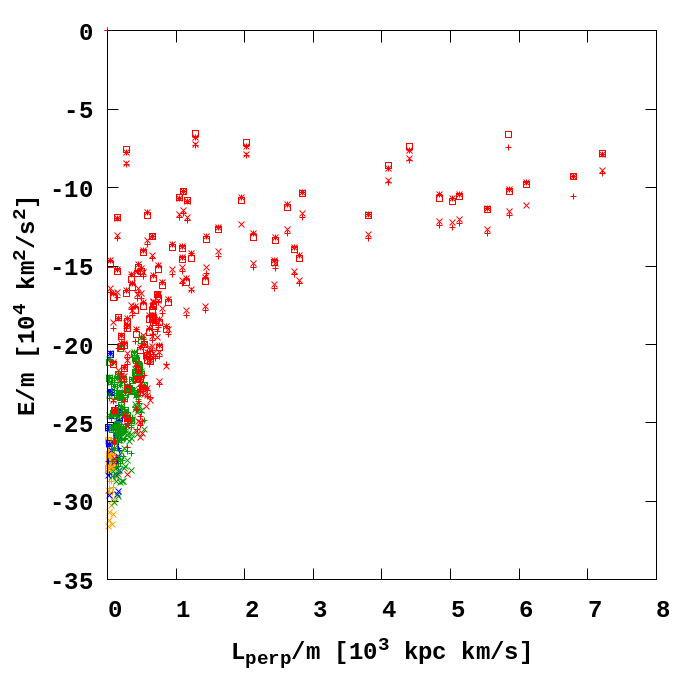}
\caption{Position of GCs in different regions of the Galaxy at the present time for the {\tt \#411321} TNG-TVP, {\it top panel}. {\it From left to right}: total angular momentum, $z$-th component of the angular momentum and perpendicular component of the angular momentum vs energy. Colour coding represents the GCs that belong to the different regions in the Galaxy: orange -- bulge (BL), blue -- thin disk (TN), green -- thick disk (TH), and red -- halo (HL). {\it Bottom panel}: the same values but for the {\tt \#441327} (plus), {\tt \#451323} (cross), {\tt \#462077} (star), and {\tt \#474170} (empty square) TNG-TVPs. The GCs Pal~3, Crater, and Sagittarius~II are not shown in ($L_{\rm tot}$, $E$) and ($L_{\rm perp}$, $E$) phase spaces and also Sagittarius~II is not shown in ($L_{\rm z}$, $E$) phase space due to of their extremely high values.}
\label{fig:tp-orb}
\end{figure*}

Next, we show the association of the GCs with different Galaxy components presented in the energy-angular momentum phase coordinates. In Fig.~\ref{fig:tp-orb},~(top row), we colour-codded the GCs associated with bulge~(BL, yellow), thin disk~(TN, blue), thick disk~(TH, green), and halo~(HL, red) Galactic regions. For simplicity, we present the GCs' parameters only in the {\tt \#411321} TNG-TVP. As we can see from the {\it top panel} in Fig.~\ref{fig:tp-orb}, as expected, the GCs that belong to the bulge (BL) and thick disk (TH) have the maximum negative energy values in the entire sample. But the bulge GCs have a relatively narrow energy range compared to the thick disk GCs. The GCs associated with the halo form a system which is less bound with the Galaxy. They have wide energy and angular momentum ranges.

In Fig.~\ref{fig:tp-orb} we present a comparison of the GCs' positions in the phase space for other four {\tt \#441327, \#451323, \#462077}, and {\tt \#474170} TNG-TVPs, {\it bottom panel}. The colours (as on the {\it top panel}) represent the associations with the different regions of the Galaxy. Different symbols correspond to different potentials. Comparing the GCs in different potentials at the present we can conclude that there are no significant differences in their values in the angular momentum space. The visible differences in energy values can be easily explained with different parameters of the galactic potentials (see Table~\ref{tab:pot-val}) at the present. At the same time, we can note that our sample of external potentials does not give any strong mutual discrepancies. 

%%%%%%%%%%%%%%%%%%%%%%%%%%%%%%%%%%%%%%%%%%%%%%%%%%%%%%%%%%%%%%%%%%%%%%%%%%%%%%%%
\subsection{The Globular Cluster subsystem phase space evolution in five time-varying potentials}\label{subsec:lz-e-TNG}
%%%%%%%%%%%%%%%%%%%%%%%%%%%%%%%%%%%%%%%%%%%%%%%%%%%%%%%%%%%%%%%%%%%%%%%%%%%%%%%%

\begin{figure*}[htbp]
\centering
\includegraphics[width=0.33\linewidth]{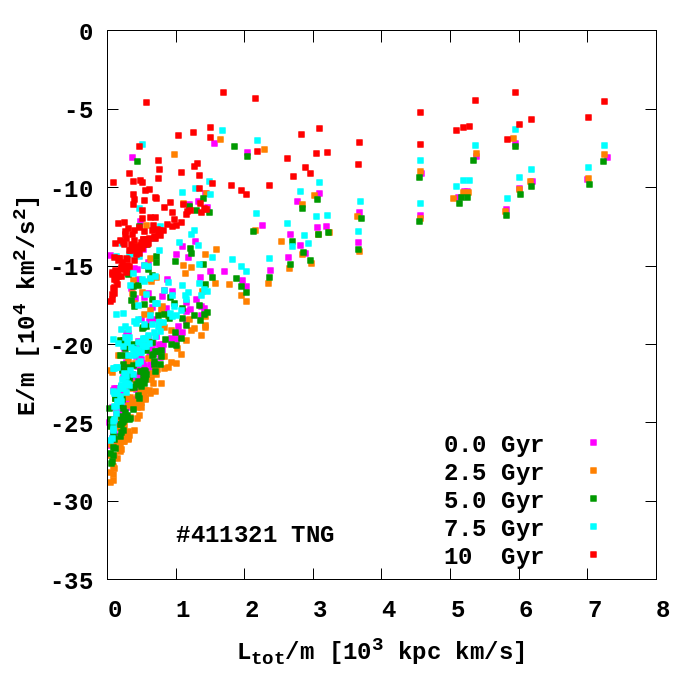}
\includegraphics[width=0.33\linewidth]{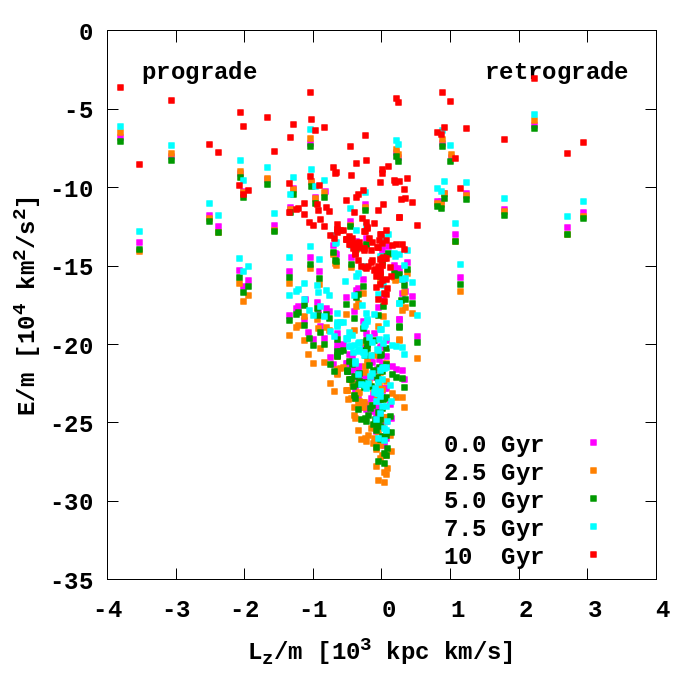}
\includegraphics[width=0.33\linewidth]{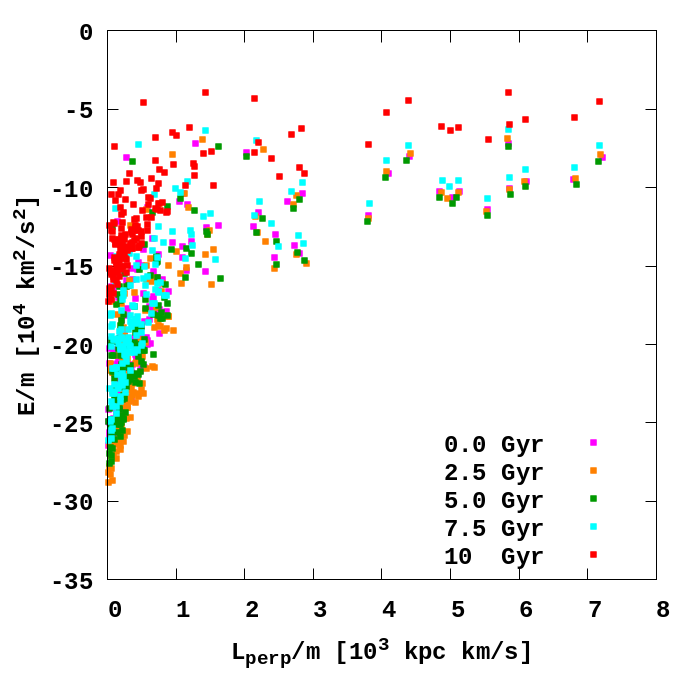}
\includegraphics[width=0.33\linewidth]{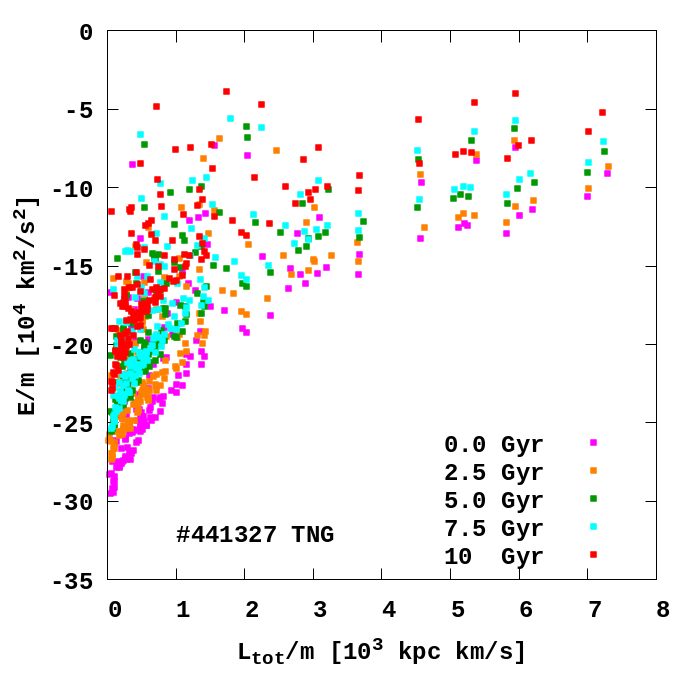}
\includegraphics[width=0.33\linewidth]{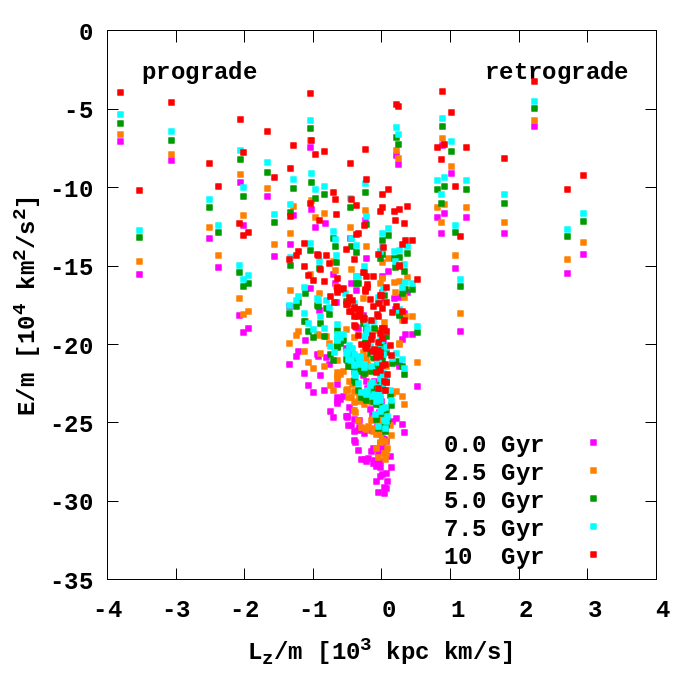}
\includegraphics[width=0.33\linewidth]{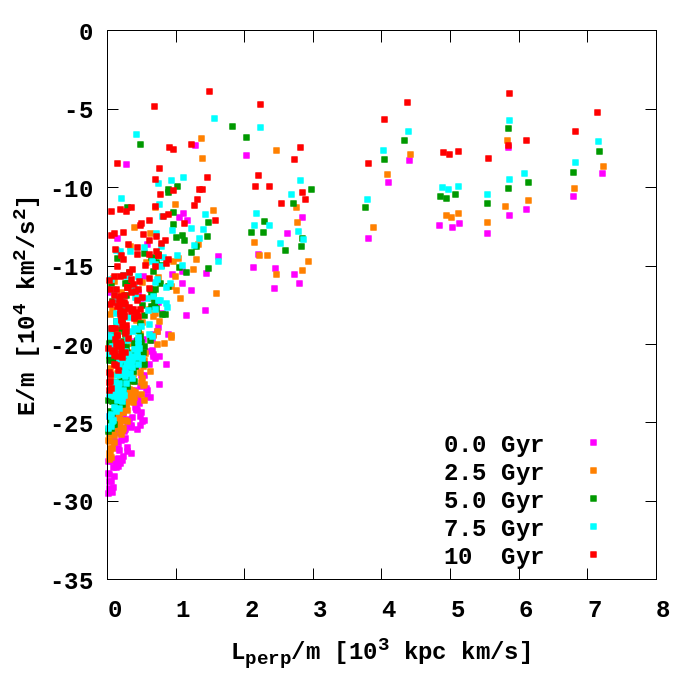}
\includegraphics[width=0.33\linewidth]{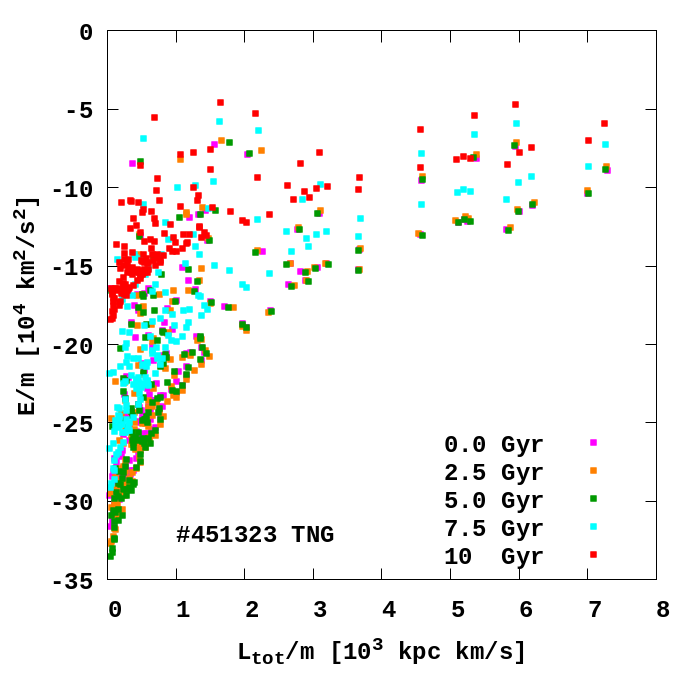}
\includegraphics[width=0.33\linewidth]{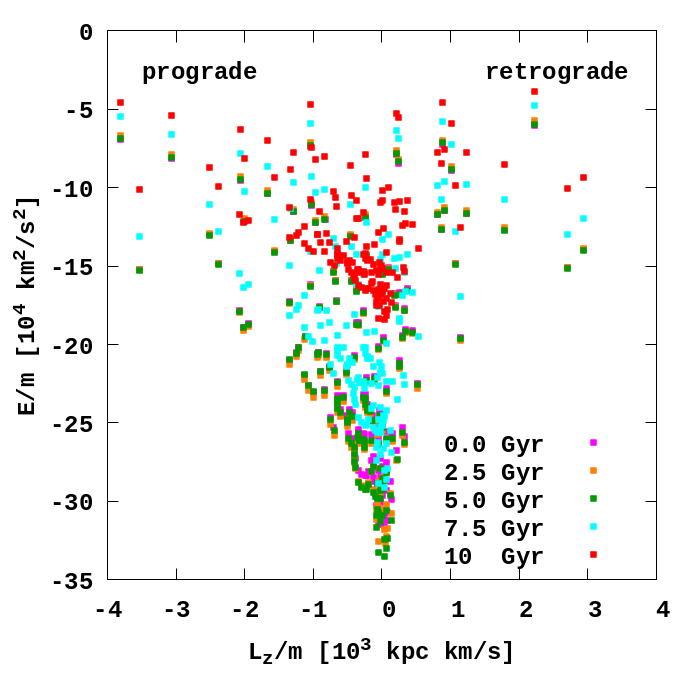}
\includegraphics[width=0.33\linewidth]{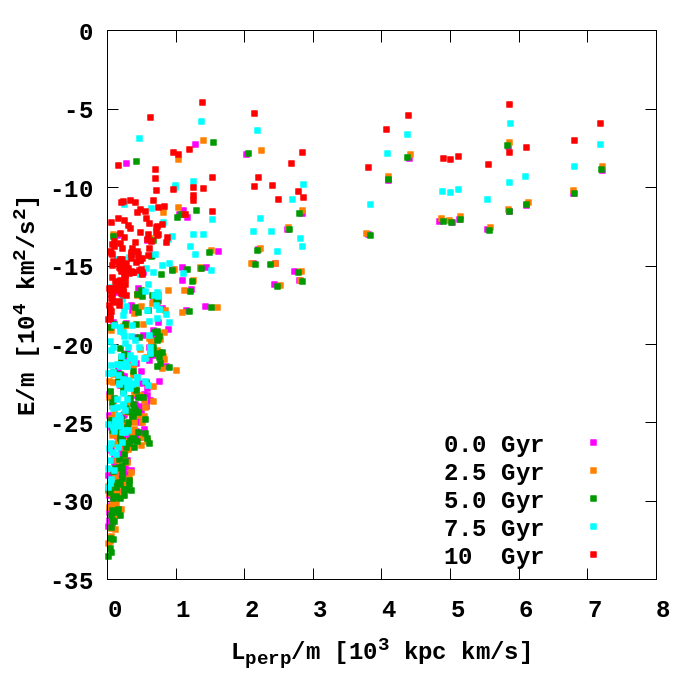}
\caption{Evolution of the total angular momentum vs energy phase space ($L_{\rm tot}$, $E$) {\it(left panels, top to bottom)}, $z$-th component of the angular momentum vs energy ($L_{\rm z}$, $E$) {\it(middle, top to bottom)} and the perpendicular component of the angular momentum vs energy ($L_{\rm perp}$, $E$) {\it(right, top to bottom)} for {\tt \#411321, \#441327}, and {\tt \#451323} TNG-TVPs. The colours represent data at different times in Gyr lookback time: magenta is $T=0$, orange is $T=2.5$, green is $T=5$, cyan is $T=7.5$ and red is $T=10$. The GCs Pal~3, Crater, and Sagittarius~II are not shown in the ($L_{\rm tot}$, $E$) and ($L_{\rm perp}$, $E$) phase spaces, and also Sagittarius~II is not shown in ($L_{\rm z}$, $E$) phase space due to of their extremely high values.}
\label{fig:Ltot1}
\end{figure*}

\begin{figure*}[htbp]
\centering
\includegraphics[width=0.33\linewidth]{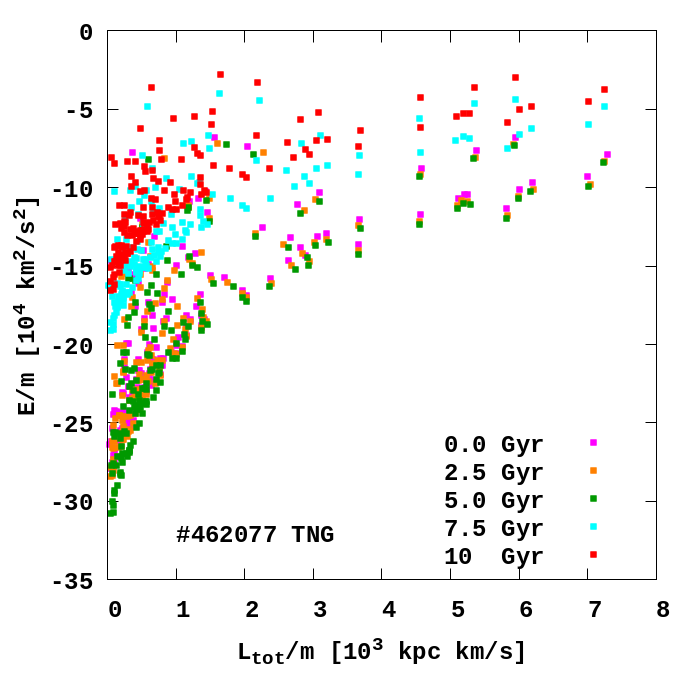}
\includegraphics[width=0.33\linewidth]{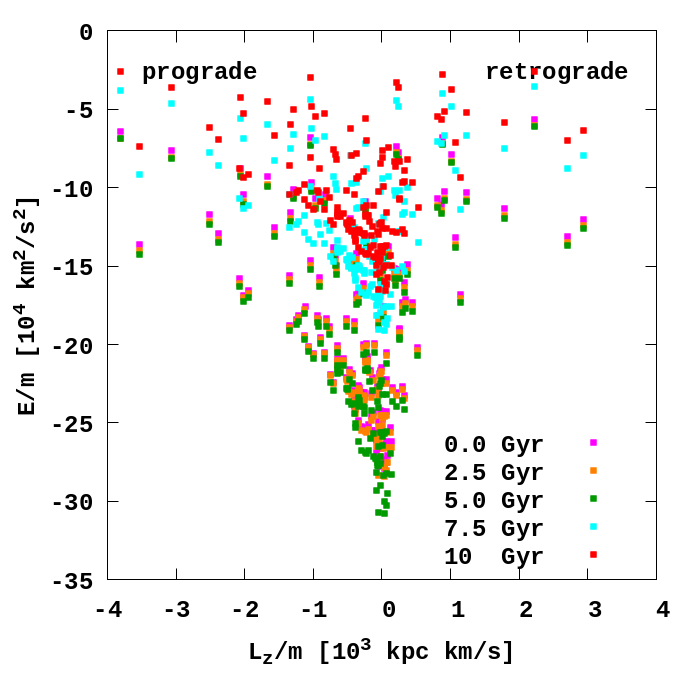}
\includegraphics[width=0.33\linewidth]{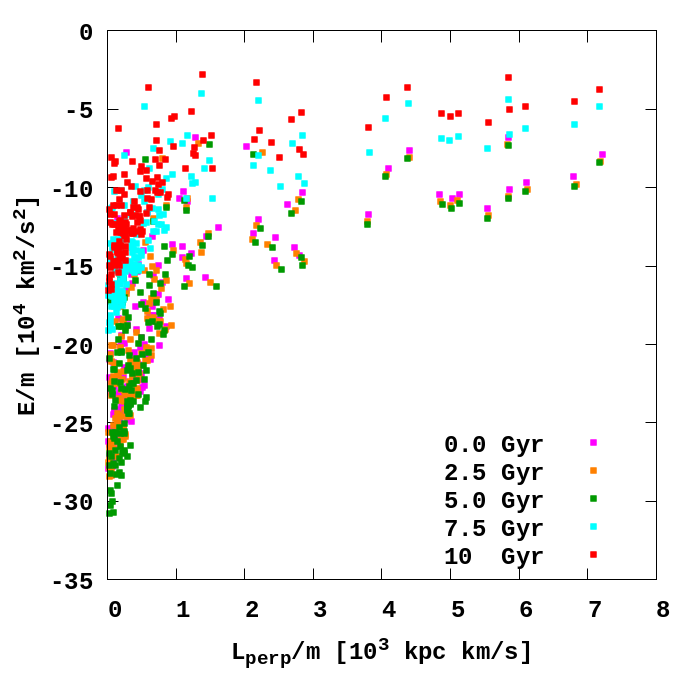}
\includegraphics[width=0.33\linewidth]{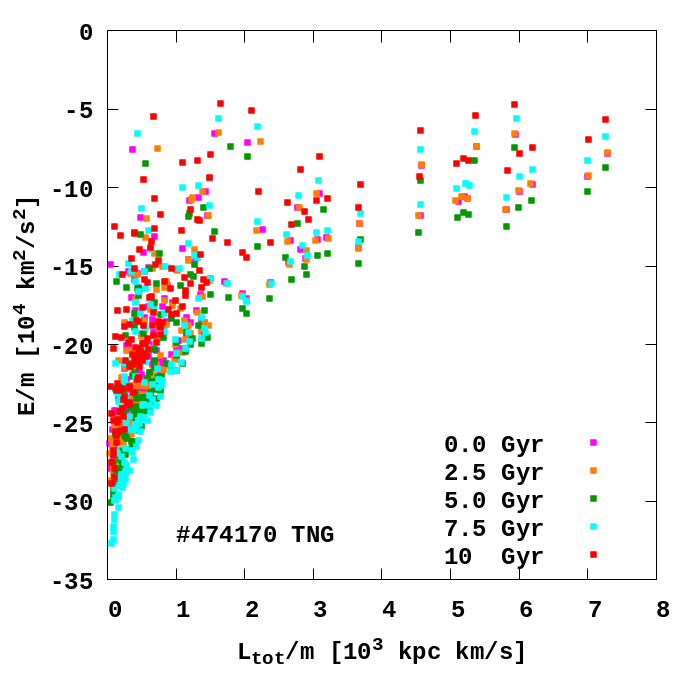}
\includegraphics[width=0.33\linewidth]{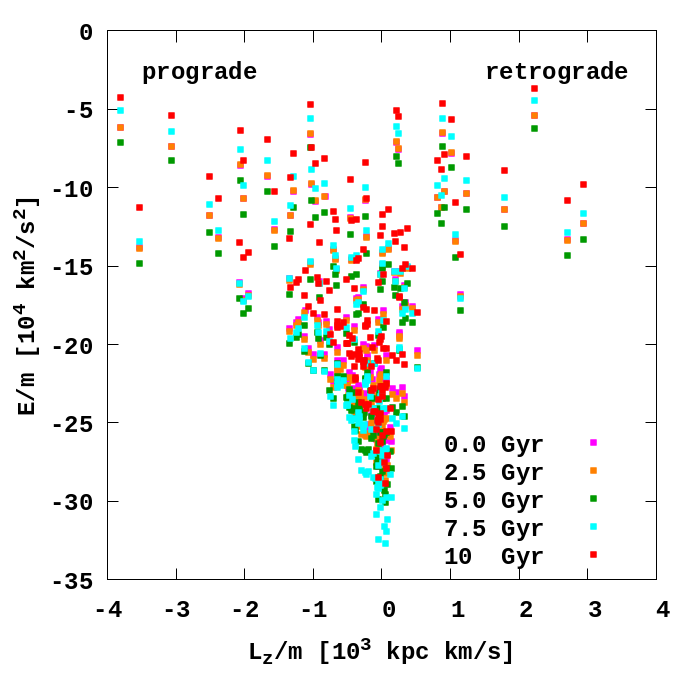}
\includegraphics[width=0.33\linewidth]{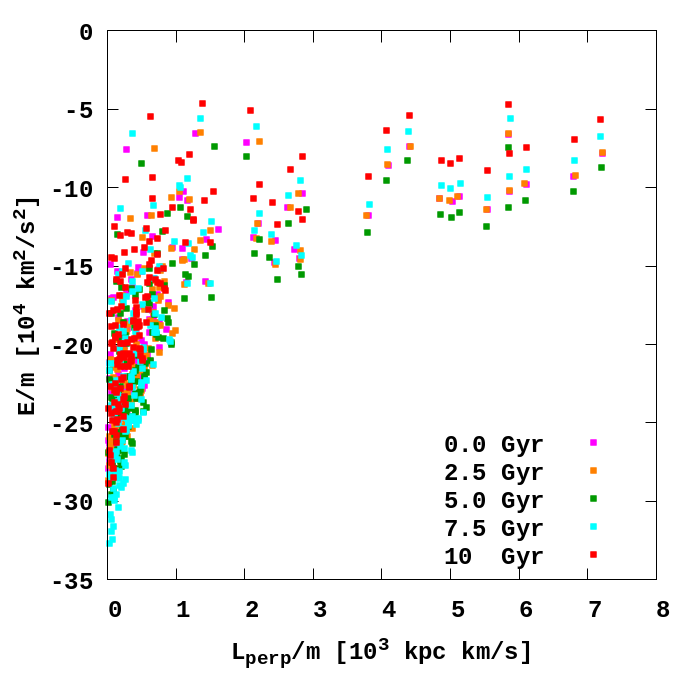}
\caption{Same as in Fig.~\ref{fig:Ltot1} but for {\tt \#462077, \#474170} TNG-TVPs.}
\label{fig:Ltot2}
\end{figure*}

Using the time-evolving potentials, we can explore the effect of the Galactic parameters time-variation on the GCs' phase space distributions in energy and angular momentum. In Figs.~\ref{fig:Ltot1} and~\ref{fig:Ltot2} we show the GC systems snapshots for different times (0.0, 2.5, 5.0, 7.5, and 10~Gyr lookback time) for each of our TNG-TVPs. 

As we can see from Figs.~\ref{fig:Ltot1} and \ref{fig:Ltot2}, all our external potentials provide roughly similar phase space energy values for individual GCs. The difference between potentials becomes more defined during evolution. In general, we can expect that at the beginning of the backwards integration~($t=0$~Gyr), the energies of the GCs are more negative (that is, their bounding energies are larger; magenta) and at the end of the simulation (10 Gyr, red) their energies are more positive -- this would mean GCs are less bounded which is not the case. Our expectations are fully met by the potential {\tt \#441327} (Fig.~\ref{fig:Ltot1}, \textit{second row}). Other TNG-TVP external potentials show slightly different behaviour. For example, the {\tt \#451323} -- (Fig.~\ref{fig:Ltot1}, \textit{third row}), {\tt \#462077} -- (Fig.~\ref{fig:Ltot2}, \textit{first row}) external potentials have maximum bounding energies at 5 Gyr (green). We can explain such behaviour if we check the evolution of the mass and characteristic scales from Fig.~\ref{fig:MW-TNG}. For these two MW-like potentials around 5 Gyr, we can observe higher mass and smaller scale sizes (as for the halo and disk). 

%-------------------------------------------------------------------------%
%-------------------------------------------------------------------------% 

%-------------------------------------------------------------------------%
%-------------------------------------------------------------------------%

%-------------------------------------------------------------------------%
%-------------------------------------------------------------------------%

For the potential {\tt \#411321} -- (Fig.~\ref{fig:Ltot1}, first row) the energy evolution shows a more complex behaviour. First the energy values for GCs are going down (approximately 3 Gyr; orange and approximately 5 Gyr; green) and only after $\approx$ 7 Gyr (cyan) they starts to raise. Such a complex behaviour might be also understood based on the evolution of the halo and disk masses visible in Fig.~\ref{fig:MW-TNG} for this TNG-TVP. 

In a case of the {\tt \#474170} TNG-TVP (Fig.~\ref{fig:Ltot2}, second row). In contrast to the previous external potentials here in the beginning (up to approximately 3 Gyr; magenta--orange) we have almost a constant halo mass and characteristic scale. 

In general, we conclude that the evolution of the GCs' phase space distributions weakly varies in the five selected TNG-TVPs. The differences are obviously caused by slightly different accretion history, which originally come from the IllustrisTNG-100 simulation. 

%%%%%%%%%%%%%%%%%%%%%%%%%%%%%%%%%%%%%%%%%%%%%%%%%%%%%%%%%%%%%%%%%%%%%%%%%%%%%%%
\section{Conclusions}\label{sec:con}
%%%%%%%%%%%%%%%%%%%%%%%%%%%%%%%%%%%%%%%%%%%%%%%%%%%%%%%%%%%%%%%%%%%%%%%%%%%%%%%

In this paper, we studied the orbital evolution of 159 Milky Way globular clusters up to 10 Gyr lookback time. For the initial positions, velocities and masses of each GCs, we used the values from the catalogues \citep{VasBaum2021}, which contain full 6D information from 
\textit{Gaia}~DR3. Analysing the errors and their distribution for the proper motions of right ascension and declination, heliocentric distances and radial velocities, we found that the vast majority (up to 95\%) of the GCs have relative errors below 5\% which allowed us to study the orbits with great precision.

For the orbital integration, we used a high-order parallel dynamical $N$-body code \PGPU. To mimic the evolution of the MW mass distribution over time, we used the data from the IllustrisTNG-100 cosmological simulation. From the simulated subhaloes, we selected galaxies that best reproduce the Milky Way parameters at present. To obtain the spatial scales of the disks and dark matter halos, we decompose the mass distribution using the MN and NFW potentials. Also, to verify the similarity of the chosen TNG potentials to our Galaxy, we added an extra parameter - the observed circular velocity of rotation at the solar radius. Finally, we selected the five TNG galaxies {\tt \#411321,  \#441327, \#451323, \#462077}, and {\tt \#474170} as time-variable external potentials for our study.

The main results of this study are as follows.

\begin{itemize}
    
    \item We integrated the orbits of 159 MW GCs up to 10 Gyr evolution in five external potentials. According to our results, we select a set of GCs which (at present time) belong to four main structural components of the Galaxy. We define nine GCs as currently belonging to the bulge (BL), ten -- to the thin disk (TN), 47 -- to the thick disc (TH), and 94 -- to the halo (HL). 
    
    \item We also proposed the orbit classification according to the GCs' orbital shapes in the $X-Y$, $X-Z$, and $R-Z$ coordinate projections. Based on this classification, we found 110 GCs with tube orbits (TB), eight GCs with perpendicular tube orbits (PT), 19 GCs with long radial orbits (LR), and 22 GCs with irregular orbits (IR).

    \item We analysed the evolution of orbital parameters of the GCs in all five external MW-like potentials. The relative individual changes of the semi-major axis and the eccentricities are quite significant (up to 50\%). The time evolution of these orbital elements clearly shows the separation between the `inner' and `outer' GCs. The `inner' GCs ($a\leq$3 kpc) have more regular and larger eccentricity changes during the evolution. The `outer' GCs ($a>$3 kpc) have much smaller eccentricity changes during the whole backward integration time.
    
    \item We analysed the evolution of the apocenters and pericenters for GCs and as a result, we can conclude the more prominent dependencies of the apocenter from the evolution of the MW-like potentials (masses and sizes). In case of pericenters, we don't see such strong variations.    
    
    \item All our external potentials provide similar energy distributions for individual GCs. Comparing the GCs in different potentials at present, we can also conclude that there are no significant differences in the angular momentum space. Some differences were found in the energy distribution, which can be explained by different parameters of the galactic potentials~(relative masses and spatial scales of stellar discs and DM halo). However, more differences (in GCs individual energies) appeared once we study the time evolution of the GC systems in the time-varying potentials. 
    
    \item We individually checked the relative energy changes $\Delta E/E$ during the orbital evolution of GCs up to 10 Gyr. Here we identified 18 GCs for which the change is more than 40\%. Such a large change in the relative energies of these clusters also suggests their possible non-local origin \citep[see also][]{Malhan2022} (energy changes from 50\% to 40\% accordingly): Crater, Pal~3, Sagittarius~II, Eridanus, AM~1, Pal~4, NGC~2419, Pyxis, Pal~14, Lae~3, Whiting~1, NGC~5694, Arp~2, Terzan~8, Terzan~7, NGC~7006, Pal~12, and Pal~13. We found that the in-situ formed GCs are less affected by the evolution of the TNG potentials as compared to the clusters, which are likely formed ex-situ.
\end{itemize}

Our study clearly shows that in order to constrain better the origin of the MW GCs it is vital to use time-variable Galactic potentials for the long-term ($>3--4$ Gyrs) analysis of the GC orbital evolution. 

%%%%%%%%%%%%%%%%%%%%%%%%%%%%%%%%%%%%%%%%%%%%%%%%%%%%%%%%%%%%%%%%%%%%%
\begin{acknowledgements}
%%%%%%%%%%%%%%%%%%%%%%%%%%%%%%%%%%%%%%%%%%%%%%%%%%%%%%%%%%%%%%%%%%%%%

The authors thank the anonymous referee for a very constructive report and suggestions that helped significantly improve the quality of the manuscript.

MI, PB, CO and MM acknowledge support within the grant No. AP14869395 of the Science Committee of the Ministry of Science and Higher Education of Kazakhstan (`Triune model of Galactic center dynamical evolution on cosmological time scale').

The work of MI was supported by the Grant of the National Academy of Sciences of Ukraine for young scientists. 

MS acknowledges the support under the Fellowship of the President of Ukraine for young scientists 2022-2024. 

The work of PB, MI and MS was also supported by the Volkswagen Foundation under the grant No. 97778.

The work of PB was supported by the Volkswagen Foundation under the special stipend No. 9B870 (2022).

PB and MI also acknowledge support from the National Academy of Sciences of Ukraine under the Main Astronomical Observatory GPU computing cluster project No.~13.2021.MM.

This work has made use of data from the European Space Agency (ESA) mission \textit{Gaia} (\url{https://www.cosmos.esa.int/gaia}), processed by the \textit{Gaia} Data Processing and Analysis Consortium (DPAC, \url{https://www.cosmos.esa.int/web/gaia/dpac/consortium}). Funding for the DPAC has been provided by national institutions, in particular the institutions participating in the \textit{Gaia} Multilateral Agreement. 
\end{acknowledgements}

%%%%%%%%%%%%%%%%%%%%%%%%%%%%%%%%%%%%%%%%%%%%%%%%%%%%%%%%%%%%%%%%%%%%%
\bibliographystyle{mnras}  % style aa.bst
\bibliography{gc-coll}   % your references Yourfile.bib

\begin{thebibliography}{}
\makeatletter
\relax
\def\mn@urlcharsother{\let\do\@makeother \do\$\do\&\do\#\do\^\do\_\do\%\do\~}
\def\mn@doi{\begingroup\mn@urlcharsother \@ifnextchar [ {\mn@doi@}
  {\mn@doi@[]}}
\def\mn@doi@[#1]#2{\def\@tempa{#1}\ifx\@tempa\@empty \href
  {http://dx.doi.org/#2} {doi:#2}\else \href {http://dx.doi.org/#2} {#1}\fi
  \endgroup}
\def\mn@eprint#1#2{\mn@eprint@#1:#2::\@nil}
\def\mn@eprint@arXiv#1{\href {http://arxiv.org/abs/#1} {{\tt arXiv:#1}}}
\def\mn@eprint@dblp#1{\href {http://dblp.uni-trier.de/rec/bibtex/#1.xml}
  {dblp:#1}}
\def\mn@eprint@#1:#2:#3:#4\@nil{\def\@tempa {#1}\def\@tempb {#2}\def\@tempc
  {#3}\ifx \@tempc \@empty \let \@tempc \@tempb \let \@tempb \@tempa \fi \ifx
  \@tempb \@empty \def\@tempb {arXiv}\fi \@ifundefined
  {mn@eprint@\@tempb}{\@tempb:\@tempc}{\expandafter \expandafter \csname
  mn@eprint@\@tempb\endcsname \expandafter{\@tempc}}}

\bibitem[\protect\citeauthoryear{{Agertz}, {Teyssier}  \& {Moore}}{{Agertz}
  et~al.}{2009}]{2009MNRAS.397L..64A}
{Agertz} O.,  {Teyssier} R.,   {Moore} B.,  2009, \mn@doi [\mnras]
  {10.1111/j.1745-3933.2009.00685.x}, \href
  {https://ui.adsabs.harvard.edu/abs/2009MNRAS.397L..64A} {397, L64}

\bibitem[\protect\citeauthoryear{{Armstrong}, {Bekki}  \& {Ludlow}}{{Armstrong}
  et~al.}{2021}]{Armstrong2021}
{Armstrong} B.~M.,  {Bekki} K.,   {Ludlow} A.~D.,  2021, \mn@doi [\mnras]
  {10.1093/mnras/staa3391}, \href
  {https://ui.adsabs.harvard.edu/abs/2021MNRAS.500.2937A} {500, 2937}

\bibitem[\protect\citeauthoryear{{Bajkova} \& {Bobylev}}{{Bajkova} \&
  {Bobylev}}{2021}]{Bajkova2021}
{Bajkova} A.~T.,  {Bobylev} V.~V.,  2021, \mn@doi [Research in Astronomy and
  Astrophysics] {10.1088/1674-4527/21/7/173}, \href
  {https://ui.adsabs.harvard.edu/abs/2021RAA....21..173B} {21, 173}

\bibitem[\protect\citeauthoryear{{Bajkova}, {Carraro}, {Korchagin}, {Budanova}
  \& {Bobylev}}{{Bajkova} et~al.}{2020}]{Bajkova2020}
{Bajkova} A.~T.,  {Carraro} G.,  {Korchagin} V.~I.,  {Budanova} N.~O.,
  {Bobylev} V.~V.,  2020, \mn@doi [\apj] {10.3847/1538-4357/ab8ea7}, \href
  {https://ui.adsabs.harvard.edu/abs/2020ApJ...895...69B} {895, 69}

\bibitem[\protect\citeauthoryear{{Bajkova}, {Smirnov}  \& {Bobylev}}{{Bajkova}
  et~al.}{2021}]{Bajkova2021AstL}
{Bajkova} A.~T.,  {Smirnov} A.~A.,   {Bobylev} V.~V.,  2021, \mn@doi [Astronomy
  Letters] {10.1134/S106377372107001X}, \href
  {https://ui.adsabs.harvard.edu/abs/2021AstL...47..454B} {47, 454}

\bibitem[\protect\citeauthoryear{{Baumgardt} \& {Vasiliev}}{{Baumgardt} \&
  {Vasiliev}}{2021}]{Baumgardt2021}
{Baumgardt} H.,  {Vasiliev} E.,  2021, \mn@doi [\mnras]
  {10.1093/mnras/stab1474}, \href
  {https://ui.adsabs.harvard.edu/abs/2021MNRAS.505.5957B} {505, 5957}

\bibitem[\protect\citeauthoryear{{Baumgardt}, {Hilker}, {Sollima}  \&
  {Bellini}}{{Baumgardt} et~al.}{2019}]{Baumgardt2019}
{Baumgardt} H.,  {Hilker} M.,  {Sollima} A.,   {Bellini} A.,  2019, \mn@doi
  [\mnras] {10.1093/mnras/sty2997}, \href
  {https://ui.adsabs.harvard.edu/abs/2019MNRAS.482.5138B} {482, 5138}

\bibitem[\protect\citeauthoryear{{Bennett} \& {Bovy}}{{Bennett} \&
  {Bovy}}{2019}]{Bennett2019}
{Bennett} M.,  {Bovy} J.,  2019, \mn@doi [\mnras] {10.1093/mnras/sty2813},
  \href {https://ui.adsabs.harvard.edu/abs/2019MNRAS.482.1417B} {482, 1417}

\bibitem[\protect\citeauthoryear{{Bennett}, {Bovy}  \& {Hunt}}{{Bennett}
  et~al.}{2022}]{BBH2022}
{Bennett} M.,  {Bovy} J.,   {Hunt} J. A.~S.,  2022, \mn@doi [\apj]
  {10.3847/1538-4357/ac5021}, \href
  {https://ui.adsabs.harvard.edu/abs/2022ApJ...927..131B} {927, 131}

\bibitem[\protect\citeauthoryear{{Berczik} et~al.,}{{Berczik}
  et~al.}{2011}]{Berczik2011}
{Berczik} P.,  et~al., 2011, in International conference on High Performance
  Computing, HPC-UA 2011. pp 8--18

\bibitem[\protect\citeauthoryear{{Berczik}, {Spurzem}  \& {Wang}}{{Berczik}
  et~al.}{2013}]{BSW2013}
{Berczik} P.,  {Spurzem} R.,   {Wang} L.,  2013, in Third International
  Conference on High Performance Computing, HPC-UA 2013. pp 52--59 (\mn@eprint
  {arXiv} {1312.1789})

\bibitem[\protect\citeauthoryear{{Birnboim} \& {Dekel}}{{Birnboim} \&
  {Dekel}}{2003}]{2003MNRAS.345..349B}
{Birnboim} Y.,  {Dekel} A.,  2003, \mn@doi [\mnras]
  {10.1046/j.1365-8711.2003.06955.x}, \href
  {https://ui.adsabs.harvard.edu/abs/2003MNRAS.345..349B} {345, 349}

\bibitem[\protect\citeauthoryear{{Bland-Hawthorn} \&
  {Gerhard}}{{Bland-Hawthorn} \& {Gerhard}}{2016}]{Bland-Hawthorn2016}
{Bland-Hawthorn} J.,  {Gerhard} O.,  2016, \mn@doi [\araa]
  {10.1146/annurev-astro-081915-023441}, \href
  {https://ui.adsabs.harvard.edu/abs/2016ARA&A..54..529B} {54, 529}

\bibitem[\protect\citeauthoryear{{Boldrini} \& {Bovy}}{{Boldrini} \&
  {Bovy}}{2022}]{Boldrini2022}
{Boldrini} P.,  {Bovy} J.,  2022, \mn@doi [\mnras] {10.1093/mnras/stac2578},
  \href {https://ui.adsabs.harvard.edu/abs/2022MNRAS.516.4560B} {516, 4560}

\bibitem[\protect\citeauthoryear{{Bovy}}{{Bovy}}{2011}]{Bovy2011}
{Bovy} J.,  2011, PhD thesis, New York University

\bibitem[\protect\citeauthoryear{{Bovy} et~al.,}{{Bovy}
  et~al.}{2012}]{Bovy2012}
{Bovy} J.,  et~al., 2012, \mn@doi [\apj] {10.1088/0004-637X/759/2/131}, \href
  {https://ui.adsabs.harvard.edu/abs/2012ApJ...759..131B} {759, 131}

\bibitem[\protect\citeauthoryear{Carpintero \& Aguilar}{Carpintero \&
  Aguilar}{1998}]{carpintero1998}
Carpintero D.~D.,  Aguilar L.~A.,  1998, \mn@doi [Monthly Notices of the Royal
  Astronomical Society] {10.1046/j.1365-8711.1998.01320.x}, 298, 1

\bibitem[\protect\citeauthoryear{{Chemerynska}, {Ishchenko}, {Sobolenko},
  {Khoperskov}  \& {Berczik}}{{Chemerynska} et~al.}{2022}]{Chemerynska2022}
{Chemerynska} I.~V.,  {Ishchenko} M.~V.,  {Sobolenko} M.~O.,  {Khoperskov}
  S.~A.,   {Berczik} P.~P.,  2022, \mn@doi [Advances in Astronomy and Space
  Physics] {10.17721/2227-1481.18-24}, \href
  {https://ui.adsabs.harvard.edu/abs/2022AASP...12...18C} {12, 18}

\bibitem[\protect\citeauthoryear{{Conroy}, {Naidu}, {Garavito-Camargo},
  {Besla}, {Zaritsky}, {Bonaca}  \& {Johnson}}{{Conroy}
  et~al.}{2021}]{2021Natur.592..534C}
{Conroy} C.,  {Naidu} R.~P.,  {Garavito-Camargo} N.,  {Besla} G.,  {Zaritsky}
  D.,  {Bonaca} A.,   {Johnson} B.~D.,  2021, \mn@doi [\nat]
  {10.1038/s41586-021-03385-7}, \href
  {https://ui.adsabs.harvard.edu/abs/2021Natur.592..534C} {592, 534}

\bibitem[\protect\citeauthoryear{{Davies}, {Crain}, {Oppenheimer}  \&
  {Schaye}}{{Davies} et~al.}{2020}]{2020MNRAS.491.4462D}
{Davies} J.~J.,  {Crain} R.~A.,  {Oppenheimer} B.~D.,   {Schaye} J.,  2020,
  \mn@doi [\mnras] {10.1093/mnras/stz3201}, \href
  {https://ui.adsabs.harvard.edu/abs/2020MNRAS.491.4462D} {491, 4462}

\bibitem[\protect\citeauthoryear{{Donnari} et~al.,}{{Donnari}
  et~al.}{2021}]{2021MNRAS.500.4004D}
{Donnari} M.,  et~al., 2021, \mn@doi [\mnras] {10.1093/mnras/staa3006}, \href
  {https://ui.adsabs.harvard.edu/abs/2021MNRAS.500.4004D} {500, 4004}

\bibitem[\protect\citeauthoryear{{Drimmel} \& {Poggio}}{{Drimmel} \&
  {Poggio}}{2018}]{Drimmel2018}
{Drimmel} R.,  {Poggio} E.,  2018, \mn@doi [Research Notes of the American
  Astronomical Society] {10.3847/2515-5172/aaef8b}, \href
  {https://ui.adsabs.harvard.edu/abs/2018RNAAS...2..210D} {2, 210}

\bibitem[\protect\citeauthoryear{{Dwek} et~al.,}{{Dwek}
  et~al.}{1995}]{1995ApJ...445..716D}
{Dwek} E.,  et~al., 1995, \mn@doi [\apj] {10.1086/175734}, \href
  {https://ui.adsabs.harvard.edu/abs/1995ApJ...445..716D} {445, 716}

\bibitem[\protect\citeauthoryear{{Fattahi} et~al.,}{{Fattahi}
  et~al.}{2019}]{Fattahi2019}
{Fattahi} A.,  et~al., 2019, \mn@doi [\mnras] {10.1093/mnras/stz159}, \href
  {https://ui.adsabs.harvard.edu/abs/2019MNRAS.484.4471F} {484, 4471}

\bibitem[\protect\citeauthoryear{{Fern{\'a}ndez-Trincado}
  et~al.,}{{Fern{\'a}ndez-Trincado} et~al.}{2021}]{Fernandez-Trincado2021}
{Fern{\'a}ndez-Trincado} J.~G.,  et~al., 2021, \mn@doi [\apjl]
  {10.3847/2041-8213/abdf47}, \href
  {https://ui.adsabs.harvard.edu/abs/2021ApJ...908L..42F} {908, L42}

\bibitem[\protect\citeauthoryear{{Ferrone}, {Di Matteo},
  {Mastrobuono-Battisti}, {Haywood}, {Snaith}, {Montouri}, {Khoperskov}  \&
  {Valls-Gabaud}}{{Ferrone} et~al.}{2023}]{2023arXiv230105166F}
{Ferrone} S.,  {Di Matteo} P.,  {Mastrobuono-Battisti} A.,  {Haywood} M.,
  {Snaith} O.~N.,  {Montouri} M.,  {Khoperskov} S.,   {Valls-Gabaud} D.,  2023,
  \mn@doi [arXiv e-prints] {10.48550/arXiv.2301.05166}, \href
  {https://ui.adsabs.harvard.edu/abs/2023arXiv230105166F} {p. arXiv:2301.05166}

\bibitem[\protect\citeauthoryear{{Gaia Collaboration} et~al.,}{{Gaia
  Collaboration} et~al.}{2018}]{2018A&A...616A...1G}
{Gaia Collaboration} et~al., 2018, \mn@doi [\aap]
  {10.1051/0004-6361/201833051}, \href
  {https://ui.adsabs.harvard.edu/abs/2018A&A...616A...1G} {616, A1}

\bibitem[\protect\citeauthoryear{{Gaia Collaboration} et~al.,}{{Gaia
  Collaboration} et~al.}{2021}]{Gaia2021}
{Gaia Collaboration} et~al., 2021, \mn@doi [\aap]
  {10.1051/0004-6361/202039657}, \href
  {https://ui.adsabs.harvard.edu/abs/2021A&A...649A...1G} {649, A1}

\bibitem[\protect\citeauthoryear{{Garro}, {Minniti}, {G{\'o}mez},
  {Alonso-Garc{\'\i}a}, {Ripepi}, {Fern{\'a}ndez-Trincado}  \& {Vivanco
  C{\'a}diz}}{{Garro} et~al.}{2022}]{Garro2022}
{Garro} E.~R.,  {Minniti} D.,  {G{\'o}mez} M.,  {Alonso-Garc{\'\i}a} J.,
  {Ripepi} V.,  {Fern{\'a}ndez-Trincado} J.~G.,   {Vivanco C{\'a}diz} F.,
  2022, \mn@doi [\aap] {10.1051/0004-6361/202141819}, \href
  {https://ui.adsabs.harvard.edu/abs/2022A&A...658A.120G} {658, A120}

\bibitem[\protect\citeauthoryear{{Garrow}, {Webb}  \& {Bovy}}{{Garrow}
  et~al.}{2020}]{Garrow2020}
{Garrow} T.,  {Webb} J.~J.,   {Bovy} J.,  2020, \mn@doi [\mnras]
  {10.1093/mnras/staa2773}, \href
  {https://ui.adsabs.harvard.edu/abs/2020MNRAS.499..804G} {499, 804}

\bibitem[\protect\citeauthoryear{{Genel} et~al.,}{{Genel}
  et~al.}{2018}]{2018MNRAS.474.3976G}
{Genel} S.,  et~al., 2018, \mn@doi [\mnras] {10.1093/mnras/stx3078}, \href
  {https://ui.adsabs.harvard.edu/abs/2018MNRAS.474.3976G} {474, 3976}

\bibitem[\protect\citeauthoryear{{G{\'o}mez}, {Besla}, {Carpintero},
  {Villalobos}, {O'Shea}  \& {Bell}}{{G{\'o}mez}
  et~al.}{2015}]{2015ApJ...802..128G}
{G{\'o}mez} F.~A.,  {Besla} G.,  {Carpintero} D.~D.,  {Villalobos} {\'A}.,
  {O'Shea} B.~W.,   {Bell} E.~F.,  2015, \mn@doi [\apj]
  {10.1088/0004-637X/802/2/128}, \href
  {https://ui.adsabs.harvard.edu/abs/2015ApJ...802..128G} {802, 128}

\bibitem[\protect\citeauthoryear{{Grand}, {Deason}, {White}, {Simpson},
  {G{\'o}mez}, {Marinacci}  \& {Pakmor}}{{Grand}
  et~al.}{2019}]{2019MNRAS.487L..72G}
{Grand} R. J.~J.,  {Deason} A.~J.,  {White} S. D.~M.,  {Simpson} C.~M.,
  {G{\'o}mez} F.~A.,  {Marinacci} F.,   {Pakmor} R.,  2019, \mn@doi [\mnras]
  {10.1093/mnrasl/slz092}, \href
  {https://ui.adsabs.harvard.edu/abs/2019MNRAS.487L..72G} {487, L72}

\bibitem[\protect\citeauthoryear{{Gratton}, {Bragaglia}, {Carretta}, {D'Orazi},
  {Lucatello}  \& {Sollima}}{{Gratton} et~al.}{2019}]{2019A&ARv..27....8G}
{Gratton} R.,  {Bragaglia} A.,  {Carretta} E.,  {D'Orazi} V.,  {Lucatello} S.,
   {Sollima} A.,  2019, \mn@doi [\aapr] {10.1007/s00159-019-0119-3}, \href
  {https://ui.adsabs.harvard.edu/abs/2019A&ARv..27....8G} {27, 8}

\bibitem[\protect\citeauthoryear{{Gravity Collaboration} et~al.,}{{Gravity
  Collaboration} et~al.}{2019}]{Gravity2019}
{Gravity Collaboration} et~al., 2019, \mn@doi [\aap]
  {10.1051/0004-6361/201935656}, \href
  {https://ui.adsabs.harvard.edu/abs/2019A&A...625L..10G} {625, L10}

\bibitem[\protect\citeauthoryear{{Habouzit} et~al.,}{{Habouzit}
  et~al.}{2019}]{2019MNRAS.484.4413H}
{Habouzit} M.,  et~al., 2019, \mn@doi [\mnras] {10.1093/mnras/stz102}, \href
  {https://ui.adsabs.harvard.edu/abs/2019MNRAS.484.4413H} {484, 4413}

\bibitem[\protect\citeauthoryear{{Haghi}, {Zonoozi}  \& {Taghavi}}{{Haghi}
  et~al.}{2015}]{Haghi2015}
{Haghi} H.,  {Zonoozi} A.~H.,   {Taghavi} S.,  2015, \mn@doi [\mnras]
  {10.1093/mnras/stv827}, \href
  {https://ui.adsabs.harvard.edu/abs/2015MNRAS.450.2812H} {450, 2812}

\bibitem[\protect\citeauthoryear{{Harris}}{{Harris}}{2010}]{Harris2010}
{Harris} W.~E.,  2010, arXiv e-prints, \href
  {https://ui.adsabs.harvard.edu/abs/2010arXiv1012.3224H} {p. arXiv:1012.3224}

\bibitem[\protect\citeauthoryear{{Harris}, {Harris}  \& {Alessi}}{{Harris}
  et~al.}{2013}]{2013ApJ...772...82H}
{Harris} W.~E.,  {Harris} G. L.~H.,   {Alessi} M.,  2013, \mn@doi [\apj]
  {10.1088/0004-637X/772/2/82}, \href
  {https://ui.adsabs.harvard.edu/abs/2013ApJ...772...82H} {772, 82}

\bibitem[\protect\citeauthoryear{{Ibata} et~al.,}{{Ibata}
  et~al.}{2021}]{Ibata2021}
{Ibata} R.,  et~al., 2021, \mn@doi [\apj] {10.3847/1538-4357/abfcc2}, \href
  {https://ui.adsabs.harvard.edu/abs/2021ApJ...914..123I} {914, 123}

\bibitem[\protect\citeauthoryear{{Ishchenko}, {Sobolenko}, {Kalambay},
  {Shukirgaliyev}  \& {Berczik}}{{Ishchenko} et~al.}{2021}]{Ishchenko2021}
{Ishchenko} M.~V.,  {Sobolenko} M.~O.,  {Kalambay} M.~T.,  {Shukirgaliyev}
  B.~T.,   {Berczik} P.~P.,  2021, \mn@doi [Reports of National Academy of
  Sciences of the Republic of Kazakhstan] {10.32014/2021.2518-1483.116}, \href
  {https://ui.adsabs.harvard.edu/abs/2022arXiv220106891I} {6, 94}

\bibitem[\protect\citeauthoryear{{Johnson} \& {Soderblom}}{{Johnson} \&
  {Soderblom}}{1987}]{Johnson1987}
{Johnson} D. R.~H.,  {Soderblom} D.~R.,  1987, \mn@doi [\aj] {10.1086/114370},
  \href {https://ui.adsabs.harvard.edu/abs/1987AJ.....93..864J} {93, 864}

\bibitem[\protect\citeauthoryear{{Katz} \& {Gunn}}{{Katz} \&
  {Gunn}}{1991}]{1991ApJ...377..365K}
{Katz} N.,  {Gunn} J.~E.,  1991, \mn@doi [\apj] {10.1086/170367}, \href
  {https://ui.adsabs.harvard.edu/abs/1991ApJ...377..365K} {377, 365}

\bibitem[\protect\citeauthoryear{{Kere{\v{s}}}, {Katz}, {Weinberg}  \&
  {Dav{\'e}}}{{Kere{\v{s}}} et~al.}{2005}]{2005MNRAS.363....2K}
{Kere{\v{s}}} D.,  {Katz} N.,  {Weinberg} D.~H.,   {Dav{\'e}} R.,  2005,
  \mn@doi [\mnras] {10.1111/j.1365-2966.2005.09451.x}, \href
  {https://ui.adsabs.harvard.edu/abs/2005MNRAS.363....2K} {363, 2}

\bibitem[\protect\citeauthoryear{{Kharchenko}, {Piskunov}, {Schilbach},
  {R{\"o}ser}  \& {Scholz}}{{Kharchenko} et~al.}{2013}]{Kharchenko2013}
{Kharchenko} N.~V.,  {Piskunov} A.~E.,  {Schilbach} E.,  {R{\"o}ser} S.,
  {Scholz} R.~D.,  2013, \mn@doi [\aap] {10.1051/0004-6361/201322302}, \href
  {https://ui.adsabs.harvard.edu/abs/2013A&A...558A..53K} {558, A53}

\bibitem[\protect\citeauthoryear{{Khoperskov} et~al.,}{{Khoperskov}
  et~al.}{2022}]{2022arXiv220604522K}
{Khoperskov} S.,  et~al., 2022, arXiv e-prints, \href
  {https://ui.adsabs.harvard.edu/abs/2022arXiv220604522K} {p. arXiv:2206.04522}

\bibitem[\protect\citeauthoryear{{Laporte}, {Johnston}, {G{\'o}mez},
  {Garavito-Camargo}  \& {Besla}}{{Laporte} et~al.}{2018}]{2018MNRAS.481..286L}
{Laporte} C. F.~P.,  {Johnston} K.~V.,  {G{\'o}mez} F.~A.,  {Garavito-Camargo}
  N.,   {Besla} G.,  2018, \mn@doi [\mnras] {10.1093/mnras/sty1574}, \href
  {https://ui.adsabs.harvard.edu/abs/2018MNRAS.481..286L} {481, 286}

\bibitem[\protect\citeauthoryear{{Laporte}, {Belokurov}, {Koposov}, {Smith}  \&
  {Hill}}{{Laporte} et~al.}{2020}]{2020MNRAS.492L..61L}
{Laporte} C. F.~P.,  {Belokurov} V.,  {Koposov} S.~E.,  {Smith} M.~C.,   {Hill}
  V.,  2020, \mn@doi [\mnras] {10.1093/mnrasl/slz167}, \href
  {https://ui.adsabs.harvard.edu/abs/2020MNRAS.492L..61L} {492, L61}

\bibitem[\protect\citeauthoryear{{{\L}okas}}{{{\L}okas}}{2020}]{2020A&A...638A.133L}
{{\L}okas} E.~L.,  2020, \mn@doi [\aap] {10.1051/0004-6361/202037643}, \href
  {https://ui.adsabs.harvard.edu/abs/2020A&A...638A.133L} {638, A133}

\bibitem[\protect\citeauthoryear{{Lovell} et~al.,}{{Lovell}
  et~al.}{2018}]{2018MNRAS.481.1950L}
{Lovell} M.~R.,  et~al., 2018, \mn@doi [\mnras] {10.1093/mnras/sty2339}, \href
  {https://ui.adsabs.harvard.edu/abs/2018MNRAS.481.1950L} {481, 1950}

\bibitem[\protect\citeauthoryear{{Makino} \& {Aarseth}}{{Makino} \&
  {Aarseth}}{1992}]{MA1992}
{Makino} J.,  {Aarseth} S.~J.,  1992, \pasj, \href
  {https://ui.adsabs.harvard.edu/abs/1992PASJ...44..141M} {44, 141}

\bibitem[\protect\citeauthoryear{{Malhan} et~al.,}{{Malhan}
  et~al.}{2022}]{Malhan2022}
{Malhan} K.,  et~al., 2022, \mn@doi [\apj] {10.3847/1538-4357/ac4d2a}, \href
  {https://ui.adsabs.harvard.edu/abs/2022ApJ...926..107M} {926, 107}

\bibitem[\protect\citeauthoryear{{Mardini} et~al.,}{{Mardini}
  et~al.}{2020}]{Mardini2020}
{Mardini} M.~K.,  et~al., 2020, \mn@doi [\apj] {10.3847/1538-4357/abbc13},
  \href {https://ui.adsabs.harvard.edu/abs/2020ApJ...903...88M} {903, 88}

\bibitem[\protect\citeauthoryear{{Mar{\'\i}n-Franch}
  et~al.,}{{Mar{\'\i}n-Franch} et~al.}{2009}]{2009ApJ...694.1498M}
{Mar{\'\i}n-Franch} A.,  et~al., 2009, \mn@doi [\apj]
  {10.1088/0004-637X/694/2/1498}, \href
  {https://ui.adsabs.harvard.edu/abs/2009ApJ...694.1498M} {694, 1498}

\bibitem[\protect\citeauthoryear{{Marinacci} et~al.,}{{Marinacci}
  et~al.}{2018}]{2018MNRAS.480.5113M}
{Marinacci} F.,  et~al., 2018, \mn@doi [\mnras] {10.1093/mnras/sty2206}, \href
  {https://ui.adsabs.harvard.edu/abs/2018MNRAS.480.5113M} {480, 5113}

\bibitem[\protect\citeauthoryear{{Massari}, {Koppelman}  \& {Helmi}}{{Massari}
  et~al.}{2019}]{Massari2019}
{Massari} D.,  {Koppelman} H.~H.,   {Helmi} A.,  2019, \mn@doi [\aap]
  {10.1051/0004-6361/201936135}, \href
  {https://ui.adsabs.harvard.edu/abs/2019A&A...630L...4M} {630, L4}

\bibitem[\protect\citeauthoryear{{Mateu}}{{Mateu}}{2023}]{Mateu2023}
{Mateu} C.,  2023, \mn@doi [\mnras] {10.1093/mnras/stad321}, \href
  {https://ui.adsabs.harvard.edu/abs/2023MNRAS.520.5225M} {520, 5225}

\bibitem[\protect\citeauthoryear{{McConnachie}}{{McConnachie}}{2012}]{2012AJ....144....4M}
{McConnachie} A.~W.,  2012, \mn@doi [\aj] {10.1088/0004-6256/144/1/4}, \href
  {https://ui.adsabs.harvard.edu/abs/2012AJ....144....4M} {144, 4}

\bibitem[\protect\citeauthoryear{{McConnachie}, {Higgs}, {Thomas}, {Venn},
  {C{\^o}t{\'e}}, {Battaglia}  \& {Lewis}}{{McConnachie}
  et~al.}{2021}]{McConnachie2021}
{McConnachie} A.~W.,  {Higgs} C.~R.,  {Thomas} G.~F.,  {Venn} K.~A.,
  {C{\^o}t{\'e}} P.,  {Battaglia} G.,   {Lewis} G.~F.,  2021, \mn@doi [\mnras]
  {10.1093/mnras/staa3740}, \href
  {https://ui.adsabs.harvard.edu/abs/2021MNRAS.501.2363M} {501, 2363}

\bibitem[\protect\citeauthoryear{Merritt}{Merritt}{1999}]{merritt_1999}
Merritt D.,  1999, \mn@doi [Publications of the Astronomical Society of the
  Pacific] {10.1086/316307}, 111, 129

\bibitem[\protect\citeauthoryear{{Miyamoto} \& {Nagai}}{{Miyamoto} \&
  {Nagai}}{1975}]{Miyamoto1975}
{Miyamoto} M.,  {Nagai} R.,  1975, \pasj, \href
  {https://ui.adsabs.harvard.edu/abs/1975PASJ...27..533M} {27, 533}

\bibitem[\protect\citeauthoryear{{Myeong}, {Vasiliev}, {Iorio}, {Evans}  \&
  {Belokurov}}{{Myeong} et~al.}{2019}]{Myeong2019}
{Myeong} G.~C.,  {Vasiliev} E.,  {Iorio} G.,  {Evans} N.~W.,   {Belokurov} V.,
  2019, \mn@doi [\mnras] {10.1093/mnras/stz1770}, \href
  {https://ui.adsabs.harvard.edu/abs/2019MNRAS.488.1235M} {488, 1235}

\bibitem[\protect\citeauthoryear{{Naiman} et~al.,}{{Naiman}
  et~al.}{2018}]{2018MNRAS.477.1206N}
{Naiman} J.~P.,  et~al., 2018, \mn@doi [\mnras] {10.1093/mnras/sty618}, \href
  {https://ui.adsabs.harvard.edu/abs/2018MNRAS.477.1206N} {477, 1206}

\bibitem[\protect\citeauthoryear{{Navarro}, {Frenk}  \& {White}}{{Navarro}
  et~al.}{1997}]{NFW1997}
{Navarro} J.~F.,  {Frenk} C.~S.,   {White} S. D.~M.,  1997, \mn@doi [\apj]
  {10.1086/304888}, \href
  {https://ui.adsabs.harvard.edu/abs/1997ApJ...490..493N} {490, 493}

\bibitem[\protect\citeauthoryear{{Nelson} et~al.,}{{Nelson}
  et~al.}{2018}]{2018MNRAS.475..624N}
{Nelson} D.,  et~al., 2018, \mn@doi [\mnras] {10.1093/mnras/stx3040}, \href
  {https://ui.adsabs.harvard.edu/abs/2018MNRAS.475..624N} {475, 624}

\bibitem[\protect\citeauthoryear{{Nelson} et~al.,}{{Nelson}
  et~al.}{2019}]{2019ComAC...6....2N}
{Nelson} D.,  et~al., 2019, \mn@doi [Computational Astrophysics and Cosmology]
  {10.1186/s40668-019-0028-x}, \href
  {https://ui.adsabs.harvard.edu/abs/2019ComAC...6....2N} {6, 2}

\bibitem[\protect\citeauthoryear{{Ness} \& {Lang}}{{Ness} \&
  {Lang}}{2016}]{2016AJ....152...14N}
{Ness} M.,  {Lang} D.,  2016, \mn@doi [\aj] {10.3847/0004-6256/152/1/14}, \href
  {https://ui.adsabs.harvard.edu/abs/2016AJ....152...14N} {152, 14}

\bibitem[\protect\citeauthoryear{{Pagnini}, {Di Matteo}, {Khoperskov},
  {Mastrobuono-Battisti}, {Haywood}, {Renaud}  \& {Combes}}{{Pagnini}
  et~al.}{2022}]{2022arXiv221004245P}
{Pagnini} G.,  {Di Matteo} P.,  {Khoperskov} S.,  {Mastrobuono-Battisti} A.,
  {Haywood} M.,  {Renaud} F.,   {Combes} F.,  2022, \mn@doi [arXiv e-prints]
  {10.48550/arXiv.2210.04245}, \href
  {https://ui.adsabs.harvard.edu/abs/2022arXiv221004245P} {p. arXiv:2210.04245}

\bibitem[\protect\citeauthoryear{{Panithanpaisal}, {Sanderson}, {Wetzel},
  {Cunningham}, {Bailin}  \& {Faucher-Gigu{\`e}re}}{{Panithanpaisal}
  et~al.}{2021}]{2021ApJ...920...10P}
{Panithanpaisal} N.,  {Sanderson} R.~E.,  {Wetzel} A.,  {Cunningham} E.~C.,
  {Bailin} J.,   {Faucher-Gigu{\`e}re} C.-A.,  2021, \mn@doi [\apj]
  {10.3847/1538-4357/ac1109}, \href
  {https://ui.adsabs.harvard.edu/abs/2021ApJ...920...10P} {920, 10}

\bibitem[\protect\citeauthoryear{{Petersen} \& {Pe{\~n}arrubia}}{{Petersen} \&
  {Pe{\~n}arrubia}}{2021}]{2021NatAs...5..251P}
{Petersen} M.~S.,  {Pe{\~n}arrubia} J.,  2021, \mn@doi [Nature Astronomy]
  {10.1038/s41550-020-01254-3}, \href
  {https://ui.adsabs.harvard.edu/abs/2021NatAs...5..251P} {5, 251}

\bibitem[\protect\citeauthoryear{{Pillepich} et~al.,}{{Pillepich}
  et~al.}{2018}]{2018MNRAS.473.4077P}
{Pillepich} A.,  et~al., 2018, \mn@doi [\mnras] {10.1093/mnras/stx2656}, \href
  {https://ui.adsabs.harvard.edu/abs/2018MNRAS.473.4077P} {473, 4077}

\bibitem[\protect\citeauthoryear{{Pillepich} et~al.,}{{Pillepich}
  et~al.}{2019}]{2019MNRAS.490.3196P}
{Pillepich} A.,  et~al., 2019, \mn@doi [\mnras] {10.1093/mnras/stz2338}, \href
  {https://ui.adsabs.harvard.edu/abs/2019MNRAS.490.3196P} {490, 3196}

\bibitem[\protect\citeauthoryear{{Reid} \& {Brunthaler}}{{Reid} \&
  {Brunthaler}}{2004}]{Reid2004}
{Reid} M.~J.,  {Brunthaler} A.,  2004, \mn@doi [\apj] {10.1086/424960}, \href
  {https://ui.adsabs.harvard.edu/abs/2004ApJ...616..872R} {616, 872}

\bibitem[\protect\citeauthoryear{{Sch{\"o}nrich}, {Binney}  \&
  {Dehnen}}{{Sch{\"o}nrich} et~al.}{2010}]{Schonrich2010}
{Sch{\"o}nrich} R.,  {Binney} J.,   {Dehnen} W.,  2010, \mn@doi [\mnras]
  {10.1111/j.1365-2966.2010.16253.x}, \href
  {https://ui.adsabs.harvard.edu/abs/2010MNRAS.403.1829S} {403, 1829}

\bibitem[\protect\citeauthoryear{{Snaith}, {Haywood}, {Di Matteo}, {Lehnert},
  {Combes}, {Katz}  \& {G{\'o}mez}}{{Snaith}
  et~al.}{2014}]{2014ApJ...781L..31S}
{Snaith} O.~N.,  {Haywood} M.,  {Di Matteo} P.,  {Lehnert} M.~D.,  {Combes} F.,
   {Katz} D.,   {G{\'o}mez} A.,  2014, \mn@doi [\apjl]
  {10.1088/2041-8205/781/2/L31}, \href
  {https://ui.adsabs.harvard.edu/abs/2014ApJ...781L..31S} {781, L31}

\bibitem[\protect\citeauthoryear{{Springel} et~al.,}{{Springel}
  et~al.}{2018}]{2018MNRAS.475..676S}
{Springel} V.,  et~al., 2018, \mn@doi [\mnras] {10.1093/mnras/stx3304}, \href
  {https://ui.adsabs.harvard.edu/abs/2018MNRAS.475..676S} {475, 676}

\bibitem[\protect\citeauthoryear{{Sun}, {Wang}, {Liu}, {Long}, {Chen}  \&
  {Gao}}{{Sun} et~al.}{2023}]{Sun2023}
{Sun} G.,  {Wang} Y.,  {Liu} C.,  {Long} R.~J.,  {Chen} X.,   {Gao} Q.,  2023,
  \mn@doi [Research in Astronomy and Astrophysics] {10.1088/1674-4527/ac9e91},
  \href {https://ui.adsabs.harvard.edu/abs/2023RAA....23a5013S} {23, 015013}

\bibitem[\protect\citeauthoryear{{Torrey} et~al.,}{{Torrey}
  et~al.}{2019}]{2019MNRAS.484.5587T}
{Torrey} P.,  et~al., 2019, \mn@doi [\mnras] {10.1093/mnras/stz243}, \href
  {https://ui.adsabs.harvard.edu/abs/2019MNRAS.484.5587T} {484, 5587}

\bibitem[\protect\citeauthoryear{{Valcin}, {Bernal}, {Jimenez}, {Verde}  \&
  {Wandelt}}{{Valcin} et~al.}{2020}]{Valcin2020}
{Valcin} D.,  {Bernal} J.~L.,  {Jimenez} R.,  {Verde} L.,   {Wandelt} B.~D.,
  2020, \mn@doi [\jcap] {10.1088/1475-7516/2020/12/002}, \href
  {https://ui.adsabs.harvard.edu/abs/2020JCAP...12..002V} {2020, 002}

\bibitem[\protect\citeauthoryear{{VandenBerg}, {Brogaard}, {Leaman}  \&
  {Casagrande}}{{VandenBerg} et~al.}{2013}]{VandenBerg2013}
{VandenBerg} D.~A.,  {Brogaard} K.,  {Leaman} R.,   {Casagrande} L.,  2013,
  \mn@doi [\apj] {10.1088/0004-637X/775/2/134}, \href
  {https://ui.adsabs.harvard.edu/abs/2013ApJ...775..134V} {775, 134}

\bibitem[\protect\citeauthoryear{{Vasiliev}}{{Vasiliev}}{2019}]{Vasiliev2019}
{Vasiliev} E.,  2019, \mn@doi [\mnras] {10.1093/mnras/stz171}, \href
  {https://ui.adsabs.harvard.edu/abs/2019MNRAS.484.2832V} {484, 2832}

\bibitem[\protect\citeauthoryear{{Vasiliev} \& {Baumgardt}}{{Vasiliev} \&
  {Baumgardt}}{2021}]{VasBaum2021}
{Vasiliev} E.,  {Baumgardt} H.,  2021, \mn@doi [\mnras]
  {10.1093/mnras/stab1475}, \href
  {https://ui.adsabs.harvard.edu/abs/2021MNRAS.505.5978V} {505, 5978}

\bibitem[\protect\citeauthoryear{{Wegg} \& {Gerhard}}{{Wegg} \&
  {Gerhard}}{2013}]{2013MNRAS.435.1874W}
{Wegg} C.,  {Gerhard} O.,  2013, \mn@doi [\mnras] {10.1093/mnras/stt1376},
  \href {https://ui.adsabs.harvard.edu/abs/2013MNRAS.435.1874W} {435, 1874}

\bibitem[\protect\citeauthoryear{{Xiang} \& {Rix}}{{Xiang} \&
  {Rix}}{2022}]{Xiang2022}
{Xiang} M.,  {Rix} H.-W.,  2022, \mn@doi [\nat] {10.1038/s41586-022-04496-5},
  \href {https://ui.adsabs.harvard.edu/abs/2022Natur.603..599X} {603, 599}

\bibitem[\protect\citeauthoryear{{Yun} et~al.,}{{Yun}
  et~al.}{2019}]{2019MNRAS.483.1042Y}
{Yun} K.,  et~al., 2019, \mn@doi [\mnras] {10.1093/mnras/sty3156}, \href
  {https://ui.adsabs.harvard.edu/abs/2019MNRAS.483.1042Y} {483, 1042}

\bibitem[\protect\citeauthoryear{{van Dokkum} et~al.,}{{van Dokkum}
  et~al.}{2013}]{2013ApJ...771L..35V}
{van Dokkum} P.~G.,  et~al., 2013, \mn@doi [\apjl]
  {10.1088/2041-8205/771/2/L35}, \href
  {https://ui.adsabs.harvard.edu/abs/2013ApJ...771L..35V} {771, L35}

\makeatother
\end{thebibliography}
%%%%%%%%%%%%%%%%%%%%%%%%%%%%%%%%%%%%%%%%%%%%%%%%%%%%%%%%%%%%%%%%%%%%%

%%%%%%%%%%%%%%%%%%%%%%%%%%%%%%%%%%%%%%%%%%%%%%%%%%%%%%%%%%%%%%%%%%%%%
\begin{appendix}
%%%%%%%%%%%%%%%%%%%%%%%%%%%%%%%%%%%%%%%%%%%%%%%%%%%%%%%%%%%%%%%%%%%%%

%%%%%%%%%%%%%%%%%%%%%%%%%%%%%%%%%%%%%%%%%%%%%%%%%%%%%%%%%%%%%%%%%%%%%
\section{Evolution of the circular velocity at the distance of the Sun for TNG-TVPs.}\label{app:tng-pot-vel}
%%%%%%%%%%%%%%%%%%%%%%%%%%%%%%%%%%%%%%%%%%%%%%%%%%%%%%%%%%%%%%%%%%%%%

%-------------------------------------------------------------------------%
\begin{figure*}[ht]
\centering
\includegraphics[width=0.31\linewidth]{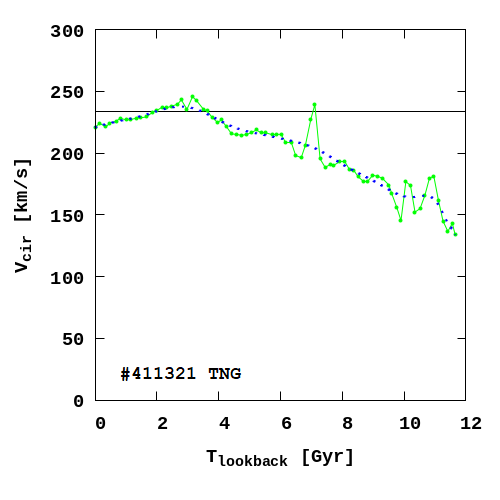}
\includegraphics[width=0.31\linewidth]{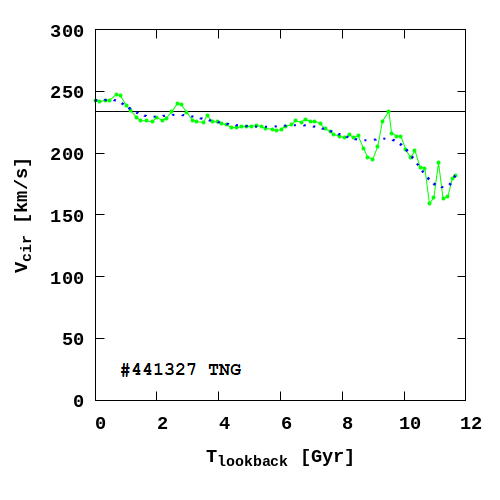}
\includegraphics[width=0.31\linewidth]{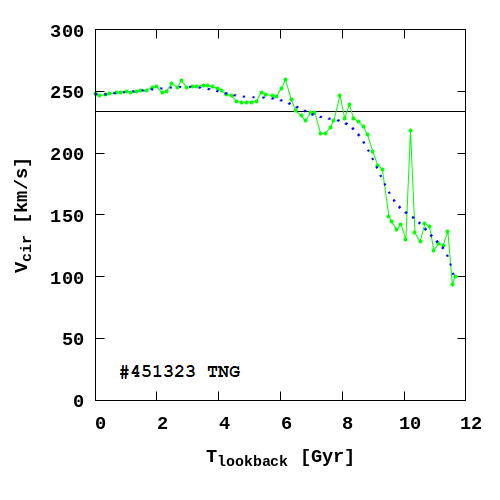}
\includegraphics[width=0.31\linewidth]{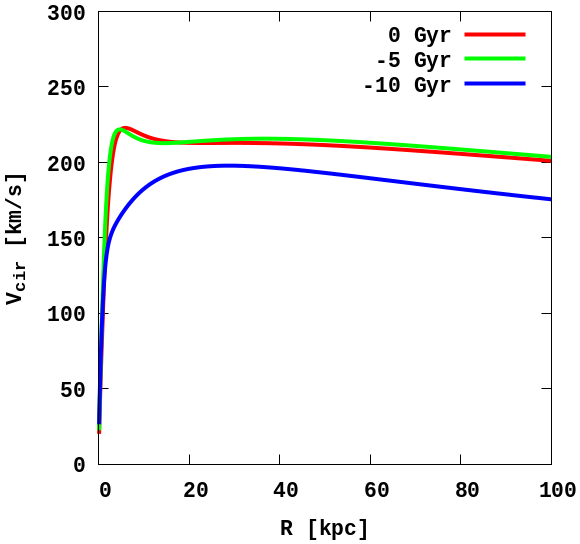}
\includegraphics[width=0.31\linewidth]{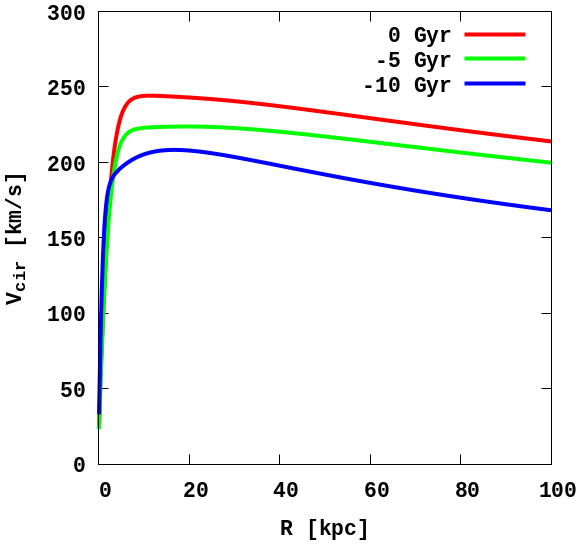}
\includegraphics[width=0.31\linewidth]{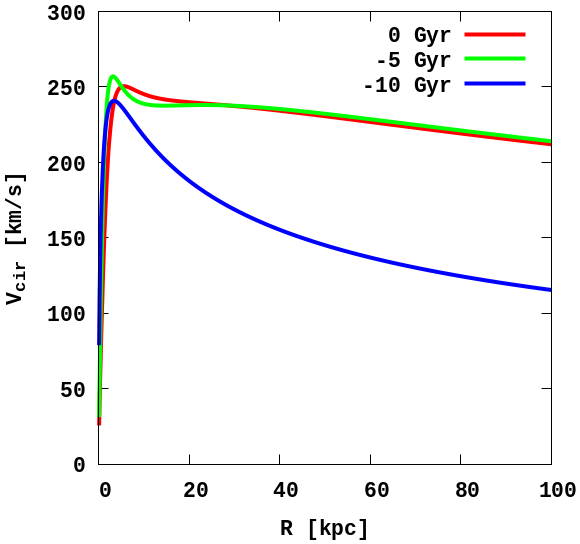}
\includegraphics[width=0.31\linewidth]{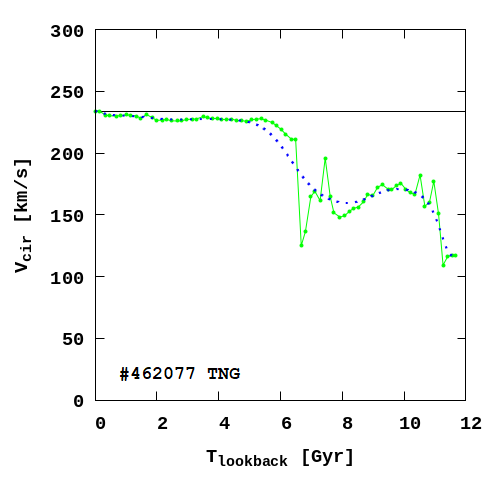}
\includegraphics[width=0.31\linewidth]{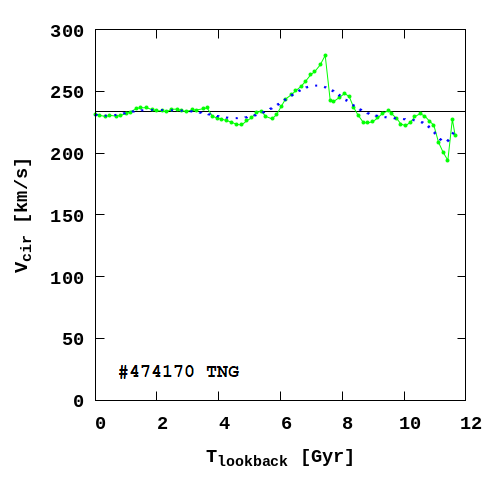}
\hspace{10em}
\includegraphics[width=0.31\linewidth]{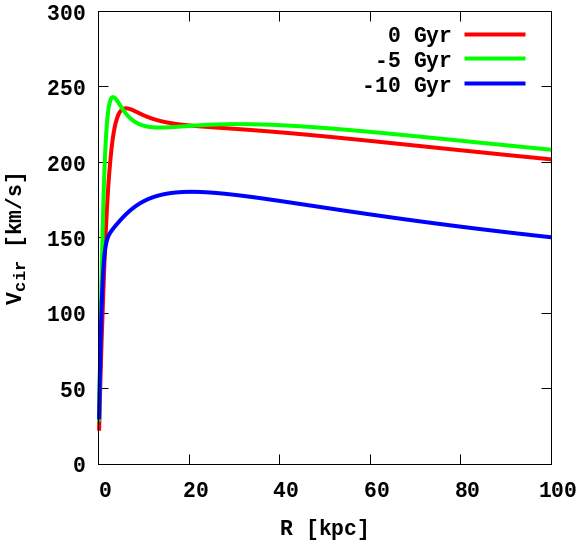}
\includegraphics[width=0.31\linewidth]{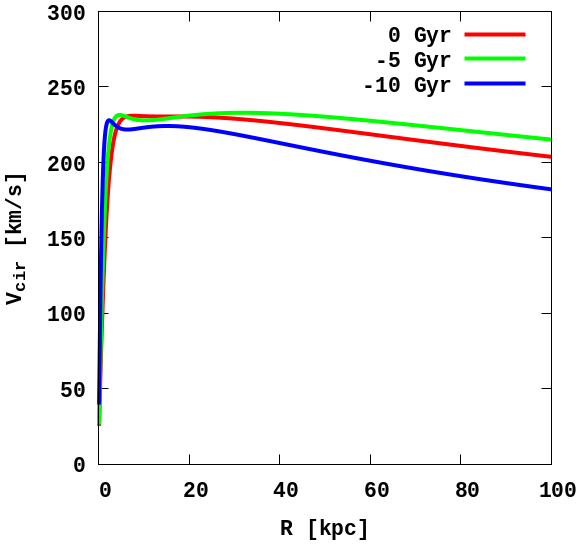}
\caption{Evolution of the circular velocity at the distance of the Sun \citep[$R_{\odot}\approx8.12$ kpc;][]{Gravity2019} in the Galactic disk as a function of backward integration time in five TNG-TVPs, {\textit{first and third panels}. A straight line indicates the current velocity value of the Sun rotation \citep[$V_{\odot}\approx235$~km~s$^{-1}$;][]{Mardini2020}. Green dotted lines represent the original data from IllustrisTNG-100. Dotted blue lines represent the interpolation and smoothing with a 1~Myr time step. {Second and fourth panels}: evolution of the TNG-TVP rotation curves to the initial (at present -- red colour), average (5 Gyr -- green) and final moments of time (10 Gyr -- blue)}. \textit{From left to right from top to bottom}: {\tt \#411321}, {\tt \#441327}, {\tt \#451323}, {\tt \#462077}, and {\tt \#474170}.}
\label{fig:v_cir}
\end{figure*}
%-------------------------------------------------------------------------%

%%%%%%%%%%%%%%%%%%%%%%%%%%%%%%%%%%%%%%%%%%%%%%%%%%%%%%%%%%%%%%%%%%%%%
\section{Types of orbits.}\label{app:types-orb}
%%%%%%%%%%%%%%%%%%%%%%%%%%%%%%%%%%%%%%%%%%%%%%%%%%%%%%%%%%%%%%%%%%%%%
%-------------------------------------------------------------------------%
\begin{figure*}[htbp]
\centering
\includegraphics[width=0.9\linewidth]{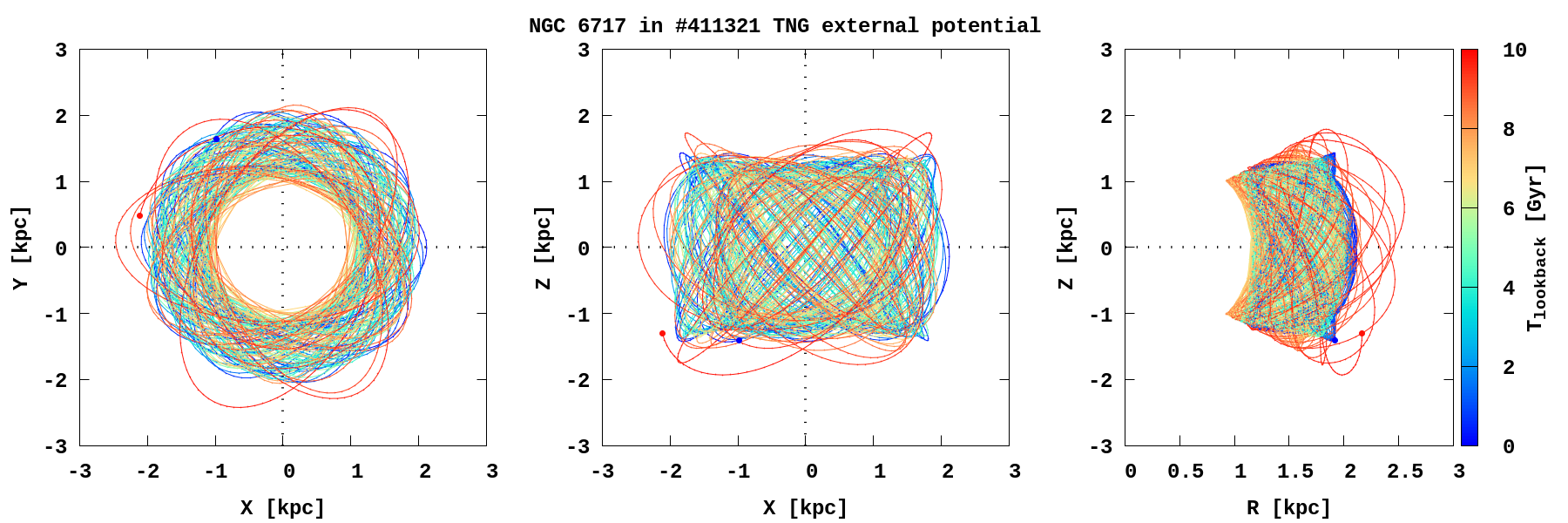}
\includegraphics[width=0.9\linewidth]{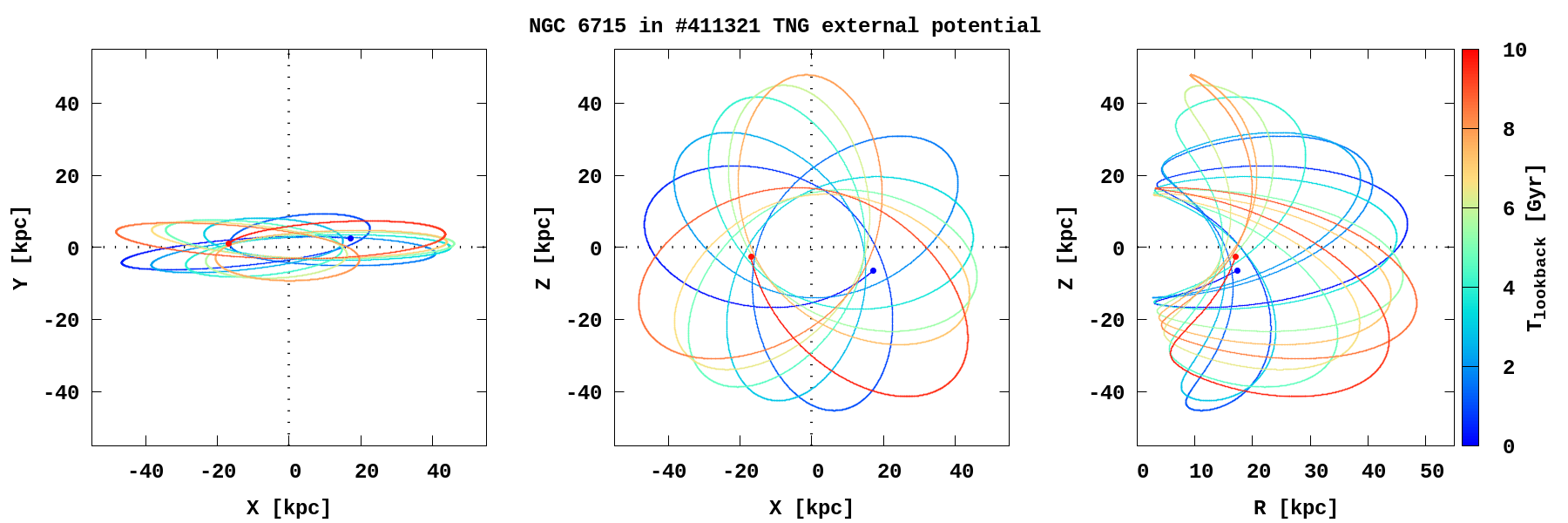}
\includegraphics[width=0.9\linewidth]{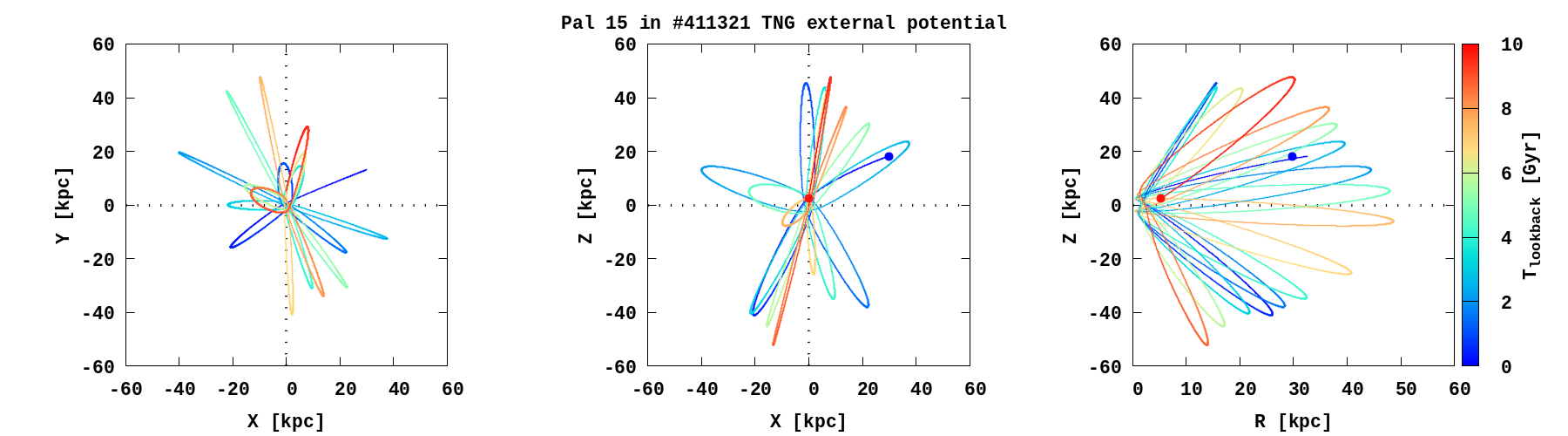}
\includegraphics[width=0.9\linewidth]{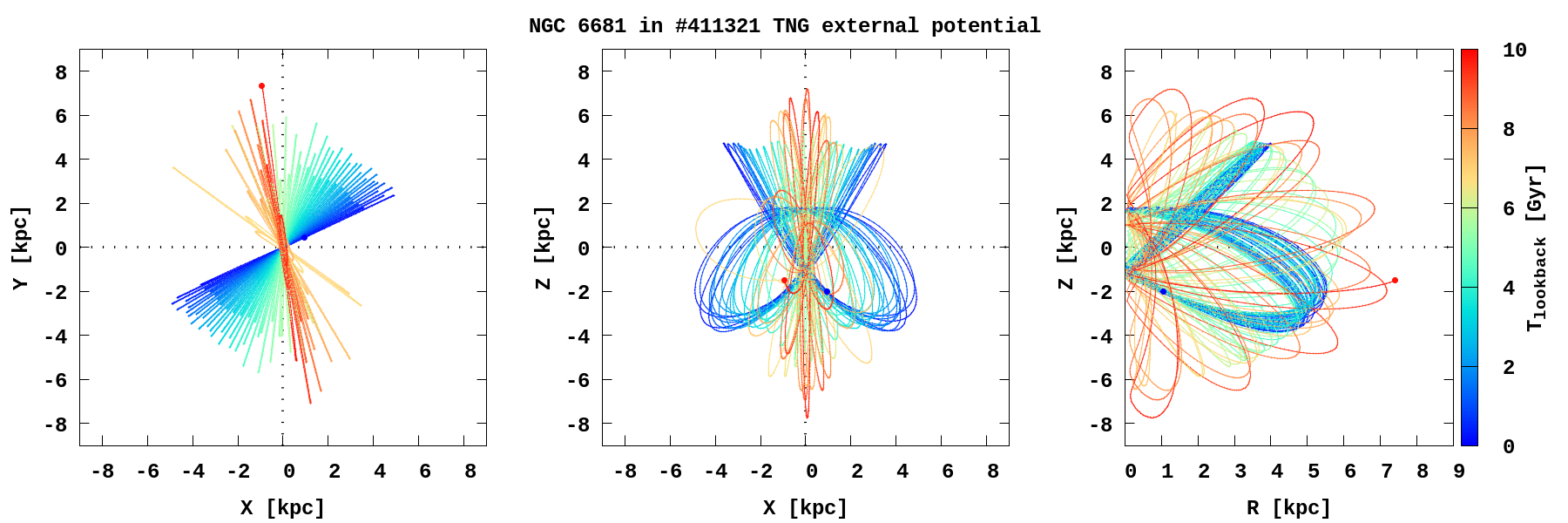}
\caption{Examples of orbits of different types in three coordinate planes ($X$, $Y$), ($X$, $Z$) and ($R$, $Z$), where $R$ is the planar Galactocentric radius. Tube orbit (TB) of NGC~6717, perpendicular tube orbit (PT) of NGC~6715, long radial orbit (LR) of Palomar~14, and irregular orbit (IR) of NGC~6681, \textit{from top to bottom}. The colour shows the evolution up to 10 Gyr, where the blue dot is the initial location and the red dot is the end location.}
\label{fig:to}
\end{figure*}
%-------------------------------------------------------------------------%

%%%%%%%%%%%%%%%%%%%%%%%%%%%%%%%%%%%%%%%%%%%%%%%%%%%%%%%%%%%%%%%%%%%%%
\section{Globular clusters from \textit{Gaia}~DR3 catalogue.}\label{app:table}
%%%%%%%%%%%%%%%%%%%%%%%%%%%%%%%%%%%%%%%%%%%%%%%%%%%%%%%%%%%%%%%%%%%%%

%-------------------------------------------------------------------------%
\begin{table*}[!b]
\caption{Globular clusters from \textit{Gaia}~DR3 catalogue.}
\centering
%\resizebox{0.98\textwidth}{!}{
\begin{tabular}{
r@{\hspace{0.95\tabcolsep}}l@{\hspace{0.95\tabcolsep}}c@{\hspace{0.95\tabcolsep}}c@{\hspace{0.95\tabcolsep}}|
r@{\hspace{0.95\tabcolsep}}l@{\hspace{0.95\tabcolsep}}c@{\hspace{0.95\tabcolsep}}c@{\hspace{0.95\tabcolsep}}|
r@{\hspace{0.95\tabcolsep}}l@{\hspace{0.95\tabcolsep}}c@{\hspace{0.95\tabcolsep}}c@{\hspace{0.95\tabcolsep}}|
r@{\hspace{0.95\tabcolsep}}l@{\hspace{0.95\tabcolsep}}c@{\hspace{0.95\tabcolsep}}c@{\hspace{0.95\tabcolsep}}}
\hline
\hline
\multicolumn{1}{c}{ID} & Name & TO & GR & 
\multicolumn{1}{c}{ID} & Name & TO & GR & 
\multicolumn{1}{c}{ID} & Name & TO & GR & 
\multicolumn{1}{c}{ID} & Name & TO & GR \\
\multicolumn{1}{c}{(1)} & (2) & (3) & (4) & 
\multicolumn{1}{c}{(5)} & (6) & (7) & (8) & 
\multicolumn{1}{c}{(9)} & (10) & (11) & (12) & 
\multicolumn{1}{c}{(13)} & (14) & (15) & (16) \\
\hline
\hline
1  & NGC~104        & TB & HL &      41 & NGC~5927          & TB & TH &       81  & HP~1$^{\rm c}$    & IR & HL &       121 & NGC~6569       & TB & TH  \\ 
2  & NGC~288        & IR & HL &      42 & NGC~5946          & TB & TH &       82  & FSR~1758          & TB & HL &       122 & BH~261         & TB & TH  \\ 
3  & NGC~362        & LR & HL &      43 & NGC~5986          & TB & TH &       83  & NGC~6362          & TB & HL &       123 & NGC~6584       & IR & HL  \\ 
4  & Whiting~1      & PT & HL &      44 & FSR~1716          & TB & TH &       84  & Liller~1          & TB & BL &       124 & NGC~6624       & TB & TH  \\ 
5  & NGC~1261       & IR & HL &      45 & Pal~14            & LR & HL &       85  & NGC~6380          & TB & BL &       125 & NGC~6626       & TB & TH  \\ 
6  & Pal~1          & TB & HL &      46 & Lynga~7$^{\rm b}$ & TB & TH &       86  & Terzan~1          & TB & TN &       126 & NGC~6638       & TB & TH  \\ 
7  & AM~1$^{\rm a}$ & LR & HL &      47 & NGC~6093          & TB & HL &       87  & Ton~2             & TB & TH &       127 & NGC~6637       & TB & TH  \\ 
8  & Eridanus       & IR & HL &      48 & NGC~6121          & TB & TH &       88  & NGC~6388          & TB & TH &       128 & NGC~6642       & TB & TN  \\ 
9  & Pal~2          & LR & HL &      49 & NGC~6101          & IR & HL &       89  & NGC~6402          & TB & TH &       129 & NGC~6652       & TB & HL  \\
10 & NGC~1851       & LR & HL &      50 & NGC~6144          & TB & HL &       90  & NGC~6401          & TB & TH &       130 & NGC~6656       & TB & HL \\
11 & NGC~1904       & LR & HL &      51 & NGC~6139          & TB & TH &       91  & NGC~6397          & TB & HL &       131 & Pal~8          & TB & TH  \\
12 & NGC~2298       & IR & HL &      52 & Terzan~3          & TB & TH &       92  & Pal~6             & TB & TH &       132 & NGC~6681       & IR & HL  \\
13 & NGC~2419       & IR & HL &      53 & NGC~6171          & TB & TH &       93  & NGC~6426          & TB & HL &       133 & NGC~6712       & TB & HL  \\
14 & Pyxis          & IR & HL &      54 & ESO~452-11        & TB & TH &       94  & Djorg~1           & TB & HL &       134 & NGC~6715       & PT & HL  \\
15 & NGC~2808       & TB & HL &      55 & NGC~6205          & TB & HL &       95  & Terzan~5          & TB & BL &       135 & NGC~6717       & TB & TH  \\
16 & E~3            & TB & HL &      56 & NGC~6229          & LR & HL &       96  & NGC~6440          & TB & BL &       136 & NGC~6723       & TB & HL  \\
17 & Pal~3          & IR & HL &      57 & NGC~6218          & TB & TH &       97  & NGC~6441          & TB & TH &       137 & NGC~6749       & TB & TH  \\
18 & NGC~3201       & TB & HL &      58 & FSR~1735          & TB & TH &       98  & Terzan~6          & TB & BL &       138 & NGC~6752       & TB & TH  \\
19 & Pal~4          & LR & HL &      59 & NGC~6235          & TB & HL &       99  & NGC~6453          & TB & TH &       139 & NGC~6760       & TB & HL  \\
20 & Crater         & IR & HL &      60 & NGC~6254          & TB & TH &       100 & UKS~1             & TB & HL &       140 & NGC~6779       & TB & HL  \\
21 & NGC~4147       & LR & HL &      61 & NGC~6256          & TB & TN &       101 & VVV~CL001         & TB & TN &       141 & Terzan~7       & PT & HL  \\
22 & NGC~4372       & TB & TH &      62 & Pal~15            & LR & HL &       102 & NGC~6496          & TB & TH &       142 & Pal~10         & TB & HL  \\
23 & Rup~106        & TB & HL &      63 & NGC~6266          & TB & TN &       103 & Terzan~9          & TB & BL &       143 & Arp~2          & PT & HL  \\
24 & NGC~4590       & TB & HL &      64 & NGC~6273          & TB & HL &       104 & Djorg~2           & TB & BL &       144 & NGC~6809       & TB & HL  \\
25 & BH~140         & TB & HL &      65 & NGC~6284          & TB & HL &       105 & NGC~6517          & TB & TN &       145 & Terzan~8       & PT & HL  \\
26 & NGC~4833       & TB & HL &      66 & NGC~6287          & TB & HL &       106 & Terzan~10         & TB & HL &       146 & Pal~11         & TB & HL  \\
27 & NGC~5024       & IR & HL &      67 & NGC~6293          & TB & HL &       107 & NGC~6522          & TB & TH &       147 & Sagittarius~II & PT & HL  \\
28 & NGC~5053       & IR & HL &      68 & NGC~6304          & TB & TN &       108 & NGC~6535          & TB & TH &       148 & NGC~6838       & TB & HL  \\
29 & NGC~5139       & TB & HL &      69 & NGC~6316          & TB & TH &       109 & NGC~6528          & TB & TN &       149 & NGC~6864       & TB & HL \\
30 & NGC~5272       & IR & HL &      70 & NGC~6341          & LR & HL &       110 & NGC~6539          & TB & TH &       150 & NGC~6934       & LR & HL  \\
31 & NGC~5286       & TB & HL &      71 & NGC~6325          & TB & TH &       111 & NGC~6540          & TB & TH &       151 & NGC~6981       & LR & HL  \\
32 & AM~4           & PT & HL &      72 & NGC~6333          & TB & HL &       112 & NGC~6544          & TB & TH &       152 & NGC~7006       & LR & HL  \\
33 & NGC~5466       & IR & HL &      73 & NGC~6342          & TB & TH &       113 & NGC~6541          & TB & TH &       153 & Laevens~3      & IR & HL  \\
34 & NGC~5634       & IR & HL &      74 & NGC~6356          & TB & HL &       114 & 2MASS-GC01        & TB & TN &       154 & NGC~7078       & TB & HL  \\
35 & NGC~5694       & LR & HL &      75 & NGC~6355          & TB & TH &       115 & ESO~280-06        & TB & HL &       155 & NGC~7089       & LR & HL  \\
36 & IC~4499        & IR & HL &      76 & NGC~6352          & TB & TH &       116 & NGC~6553          & TB & TH &       156 & NGC~7099       & TB & HL  \\
37 & NGC~5824       & IR & HL &      77 & IC~1257           & LR & HL &       117 & 2MASS-GC02        & TB & TH &       157 & Pal~12         & PT & HL  \\
38 & Pal~5          & IR & HL &      78 & Terzan~2          & TB & BL &       118 & NGC~6558          & TB & TN &       158 & Pal~13         & IR & HL  \\
39 & NGC~5897       & TB & HL &      79 & NGC~6366          & TB & TH &       119 & IC~1276$^{\rm d}$ & TB & HL &       159 & NGC~7492       & LR & HL  \\
40 & NGC~5904       & LR & HL &      80 & Terzan~4          & TB & BL &       120 & Terzan~12         & TB & TH &           &                &    &     \\
\hline
\end{tabular}
%}
\tablefoot{Columns (1), (5), (9) and (13) represent the Globular Cluster index number. Columns (2), (6), (10) and (14) represent the names of all GCs taken from \citep{VasBaum2021}. The alternative names of the GCs, which were used in our previous works \citep{Ishchenko2021,Chemerynska2022}: $^{\rm a}$E~1, $^{\rm b}$BH~184, $^{\rm c}$BH 229, and $^{\rm d}$Pal 7. Columns (3), (7), (11), and (15) show the type of GCs orbits (TO), where tube orbit -- TB, perpendicular tube orbit -- PT, long radial orbit -- LR, and irregular orbit -- IR. Examples for the each type are given in Appendix~\ref{app:tng-pot-vel}. Columns (4), (8), (12), and (16) show GCs belonging to the different regions of the Galaxy (GR) at present, according to the classification from \cite{Bland-Hawthorn2016}, where bulge -- BL, thin disk -- TN, thick disk -- TH, and halo -- HL.}
\label{tab:GC3}
\end{table*}
%-------------------------------------------------------------------------%

\end{appendix}

%%%%%%%%%%%%%%%%%%%%%%%%%%%%%%%%%%%%%%%%%%%%%%%%%%%%%%%%%%%%%%%%%%%%%
\end{document}